\documentclass[aps,pra,twocolumn,superscriptaddress,showpacs,nofootinbibfloatfix,amsmath,amsfonts,amssymb]{revtex4-2}%

\usepackage{graphicx}
\usepackage{float} 
\usepackage{subfigure}
\usepackage{amsmath}
\usepackage{amsfonts,amscd,amssymb,amsthm}
\usepackage{booktabs}  
\usepackage{verbatim}
\usepackage{bm}
\usepackage{amstext}

\usepackage{amsmath,amsfonts,amssymb,color}
\usepackage{amsthm}
\usepackage{leftidx}
\usepackage{graphicx}
\usepackage{xcolor}
\usepackage{dcolumn}
\usepackage{bm}
\usepackage{epstopdf}
\usepackage{epsfig}
\usepackage{environ}
\usepackage{pdfcomment}

\usepackage{multirow}
\usepackage{setspace}
\usepackage{color}

\usepackage{float}
\usepackage[T1]{fontenc}
\usepackage{setspace}
\usepackage{esint}

\begin{document}
	
	\title{Correlation-induced phase transitions and mobility edges in an interacting non-Hermitian quasicrystal}
	
	\author{Tian Qian}
\affiliation{%
	College of Physics and Optoelectronic Engineering, Ocean University of China, Qingdao, China 266100
}

\author{Yongjian Gu}
\affiliation{%
	College of Physics and Optoelectronic Engineering, Ocean University of China, Qingdao, China 266100
}
\affiliation{%
	Key Laboratory of Optics and Optoelectronics, Qingdao, China 266100
}

\author{Longwen Zhou}
\email{zhoulw13@u.nus.edu}
\affiliation{%
	College of Physics and Optoelectronic Engineering, Ocean University of China, Qingdao, China 266100
}
\affiliation{%
	Key Laboratory of Optics and Optoelectronics, Qingdao, China 266100
}
\affiliation{%
	Engineering Research Center of Advanced Marine Physical Instruments and Equipment of MOE, Qingdao, China 266100
}
 
	\date{\today}
	
	
	\begin{abstract}
        Non-Hermitian quasicrystal constitutes a unique class of disordered open system with PT-symmetry breaking, localization and topological triple phase transitions. In this work, we uncover the effect of quantum correlation on phase transitions and entanglement dynamics in non-Hermitian quasicrystals. Focusing on two interacting bosons in a Bose-Hubbard lattice with quasiperiodically modulated gain and loss, we find that the onsite interaction between bosons could drag the PT and localization transition thresholds towards weaker disorder regions compared with the noninteracting case. Moreover, the interaction facilitates the expansion of the critical point of a triple phase transition in the noninteracting system into a critical phase with mobility edges, whose domain could be flexibly controlled by tuning the interaction strength. Systematic analyses of the spectrum, inverse participation ratio, topological winding number, wavepacket dynamics and entanglement entropy lead to consistent predictions about the correlation-driven phases and transitions in our system. Our findings pave the way for further studies of the interplay between disorder and interaction in non-Hermitian quantum matter.
	\end{abstract}

	\maketitle
	
	\section{Introduction}\label{sec.int}
	Over the past decade, non-Hermitian systems have garnered increasing attention due to their rich dynamics, topological features and transport properties
	\cite{Ashida_2020,El-Ganainy_2018,Gong_2018,Bergholtz_2021,MartinezAlvarez_2018,Ghatak_2019}.
 	Theoretical studies have revealed distinctive phenomena in non-Hermitian physics, such as PT-symmetry breaking \cite{Bender_1998}, exceptional points \cite{Berry_2004,Heiss_2012,Miri_2019}, non-Hermitian skin effects \cite{Yao_2018,Kunst_2018,MartinezAlvarez_2018a,Lee_2019}, anomalous localization transitions \cite{Hatano_1996,Feinberg_1999,Feinberg_1999a,Hatano_2016}, and enlarged symmetry classifications of topological matter \cite{Kawabata_2019b,Zhou_2019,Wojcik_2020,Shiozaki_2021}. Experiments have enabled us to observe novel non-Hermitian dynamics and topological phases on various platforms, including cold atoms \cite{Gou_2020,Li_2019,Xu_2017,Lapp_2019,Ren_2022}, photonics \cite{Zeuner_2015,Weimann_2017,Wang_2019,Xiao_2020,Lukin_2019,Lin_2022,Weidemann_2022}, acoustics \cite{Zhu_2018,Shen_2018,Gao_2020}, electrical circuits \cite{Hofmann_2020,Helbig_2020,Liu_2020b}, and nitrogen-vacancy center in diamonds \cite{Wu_2019}.
	
	The non-Hermitian quasicrystal (NHQC) forms an intriguing class of matter with rich physical properties. The interplay between non-Hermitian effects (such as gain and loss or nonreciprocity) and correlated disorder within an NHQC could induce parity and time-reversal (PT) symmetry breaking, localization, and topological phase transitions  \cite{Sarnak_1982,Jazaeri_2001,Zeng_2017,Jiang_2019,Longhi_2019,Longhi_2019a,Liu_2020a,Liu_2020,Zeng_2020,Zeng_2020b,Zhai_2020,Liu_2021a,Liu_2021b,Xu_2021,Cai_2021,Tang_2021,Liu_2021c,Zhai_2021,Zhou_2021b,Wang_2021c,Liu_2021,Cai_2021a,Longhi_2021,Longhi_2021a,Zhou_2021a,Acharya_2022,Zhou_2022,Xia_2022,Zeng_2020a,Cai_2022,Zhai_2022,Chakrabarty_2023}.
	The addition of long-range hopping \cite{Liu_2020a}, lattice dimerization \cite{Zhou_2021b}, non-Abelian potential \cite{Zhou_2023} and time-periodic driving \cite{Zhou_2022a} could further create reentrant and alternating localization transitions between different phases in non-Hermitian quasicrystals (NHQCs). Despite great theoretical progress, NHQCs have also been simulated experimentally by nonunitary photonic quantum walks \cite{Lin_2022,Weidemann_2022}, motivating further considerations of their usage in phase-change and nonreciprocal quantum devices.
	
	\begin{figure}
		\centering
		\includegraphics[scale=0.287]{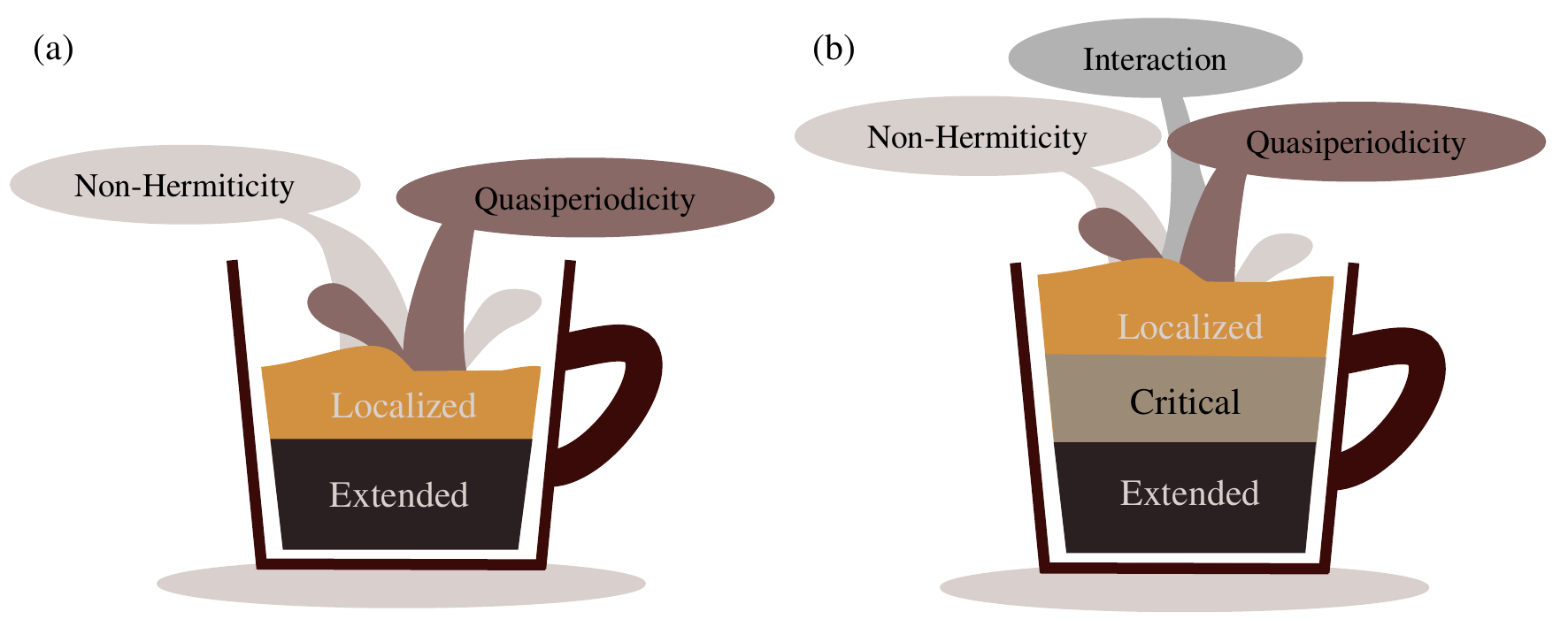}
		\caption{
		A schematic illustration of the (a) noninteracting and (b) interacting NHQCs. Correlations induced by interparticle interactions are expected to yield phases and transitions beyond the single-particle case, such as two-body non-Hermitian critical phases with mobility edges.}\label{fig.coffee.2}
	\end{figure}
	
	Up to now, a great deal of knowledge about NHQCs has been accumulated on the single-particle level. However, much less is known regarding the effect of interparticle interactions (see a schematic illustration in Fig.~\ref{fig.coffee.2}). Specially, could the quantum correlation between interacting particles be able to modify the critical points of single-particle phase transitions or even generate new phases that are absent in the noninteracting limit of an NHQC? Moreover, which set of unique features originated from the interplay among non-Hermiticity, disorder, and many-body interactions could be established for the spectral, localization and topological transitions in NHQCs? Resolving these issues is essential for our further understanding and realization of NHQC phases in quantum many-body systems.
	In this work, we make progress along this line of thought. Focusing on two interacting bosons in a typical non-Hermitian variant of the Aubry-Andr\'e-Harper (AAH) \cite{Aubry_1980,Harper_1955,Sokoloff_1985} quasicrystal, we reveal that the onsite interaction between bosons could generate a lower threshold for localization phase transitions to happen compared with the single-particle case. Moreover, the interaction expands an original PT, localization and topological triple phase transition point of the noninteracting system into a whole critical phase, in which extended and localized two-particle states are energetically separated by mobility edges. In Sec.~\ref{sec.model}, we introduce our non-Hermitian AAH model with quasiperiodically modulated gain/loss and onsite interactions. In Sec.~\ref{sec.static}, we uncover disorder and correlation induced phases and transitions in our system by systematically investigating the two-body spectrum, inverse participation ratio (IPR) and topological winding number of the interacting bosons. In Sec.~\ref{sec.dynamic}, we further investigate the wavepacket and entanglement dynamics and establish their connections with the found phases and transitions. In Sec.~\ref{sec.sum}, we summarize our results, discuss their physical implications and point out potential future directions.
	
	\section{Model}\label{sec.model}
	In this work, we focus on interaction-induced phenomena in NHQCs.
	We start with a one-dimensional (1D) model of noninteracting bosons, which can be viewed as a minimal non-Hermitian extension of the AAH model. The Hamiltonian of the model reads \cite{Longhi_2019a}
	\begin{equation}
		\hat{H}_{0}=-J\sum_{l}(\hat{b}_{l}^{\dagger}\hat{b}_{l+1}+{\rm H.c.})-\mu\sum_{l}e^{i2\pi\alpha l}\hat{n}_{l}.\label{eq.NHAAH}
	\end{equation}
	Here $\hat{b}_{l}^{\dagger}$ ($\hat{b}_{l}$) creates (annihilates) a boson on the lattice site $l$, and the particle number operator $\hat{n}_{l}=\hat{b}_{l}^{\dagger}\hat{b}_{l}$. The lattice constant has been set to one. $J$ is the nearest-neighbor hopping amplitude. $\mu$ is the amplitude of chemical potential. The potential $\mu_{l}=\mu e^{i2\pi\alpha l}$ is quasiperiodic for any irrational $\alpha$, and it is further non-Hermitian for any $\mu\neq0$. The imaginary part of the potential yields a quasiperiodically modulated gain and loss in space. We will assume $\alpha=(\sqrt{5}-1)/2$ in all our discussions without loss of generality.

	The system described by $\hat{H}_{0}$ possesses the PT symmetry, where the parity ${\cal P}:l\rightarrow-l$ and time-reversal ${\cal T}={\cal K}$, with ${\cal K}$ performing the complex conjugation. The spectrum of $\hat{H}_{0}$ could thus be real in certain parameter regions even though $\hat{H}_{0}\neq\hat{H}_{0}^{\dagger}$. In the meantime, we notice that all the eigenstates of the system are extended when $\mu=0$. With the increase of $\mu$ from zero, the correlated disorder in the system becomes stronger, and we expect localized eigenstates to appear at large $\mu$. Indeed, it was found that under the periodic boundary condition (PBC), the single-particle spectrum of $\hat{H}_{0}$ has the form $E=2J\cos k$ for $|\mu|<|J|$ and $E=2J\cos(k-ih)$ with $h=\ln(\mu/J)$ for $|\mu|>|J|$ ($k\in[-\pi,\pi]$), corresponding to a line segment along the real axis and an ellipse on the complex energy plane, respectively. There is thus a PT transition at $|\mu|=|J|$, where the spectrum of the system changes from real ($|\mu|<|J|$) to complex ($|\mu|>|J|$) \cite{Longhi_2019a}. Moreover, it was identified that all the eigenstates of $\hat{H}_{0}$ are extended (localized) for $|\mu|<|J|$ ($|\mu|>|J|$). The PT transition at $|\mu|=|J|$ thus also accompanies a metal-to-insulator transition with the increasing of $|\mu|$ from below to above $|J|$. These PT and localization transitions could be further characterized by the quantized jump of a winding number of $E$ around the origin of the complex plane, which goes from zero to one following the PT symmetry breaking of the system \cite{Zhou_2021a}. Therefore, for $|\mu|<|J|$ ($|\mu|>|J|$), $\hat{H}_{0}$ has a real (complex) spectrum with only extended (localized) eigenstates and a vanishing (unit-quantized) winding number. We will thus encounter a spectral, localization and topological triple phase transition at $|\mu|=|J|$ in the single-particle system $\hat{H}_{0}$.

  	We now add onsite interactions between bosons to $\hat{H}_{0}$. The resulting system is described by the Hamiltonian
    \begin{alignat}{1}
    \hat{H}= & -J\sum_{l}(\hat{b}_{l}^{\dagger}\hat{b}_{l+1}+{\rm H.c.})-\mu\sum_{l}e^{i2\pi\alpha l}\hat{n}_{l}\nonumber \\
    & +\frac{U}{2}\sum_{l}\hat{n}_{l}(\hat{n}_{l}-1),\label{eq.NHBHHM}
    \end{alignat}
	where $U$ measures the interaction strength. In the limit $\mu\rightarrow0$, $\hat{H}$ reduces to the Bose-Hubbard model, which is a paradigm in the study of quantum phase transitions between superfluids and Mott insulators \cite{Ma_1986,Fisher_1989}. The system described by $\hat{H}$ can thus be viewed also as a Bose-Hubbard model subject to a quasiperiodic non-Hermitian superlattice potential. Since both disorders and interactions could induce localization, their collaboration may allow localization transitions to happen at lower thresholds of the non-Hermitian quasiperiodic potential. Nevertheless, with $U\neq0$, it is unclear whether all the eigenstates of $\hat{H}$ will become localized following the localization transition, just as what happens in $\hat{H}_{0}$. Moreover, since the interaction term in $\hat{H}$ preserves the original PT symmetry of the single-particle model, it is curious to know whether the interaction itself could induce PT symmetry breaking and how would such PT breaking transitions accompany localization transitions in the system. The robustness of the triple phase transition point of $\hat{H}_{0}$ to interactions also deserves to be tested. In the following sections, we address these issues by investigating the spectrum, eigenstates, topology, dynamics and entanglement of our Bose-Hubbard non-Hermitian AAH (NHAAH) model $\hat{H}$. We will focus on the case of two interacting bosons, which serves as a minimal situation of revealing nontrivial physics due to quantum correlations in NHQCs.
	
	\section{Interaction-induced phases and phase transitions}\label{sec.static}

	\begin{figure*}
		\centering  
		\includegraphics[width=0.8\textwidth]{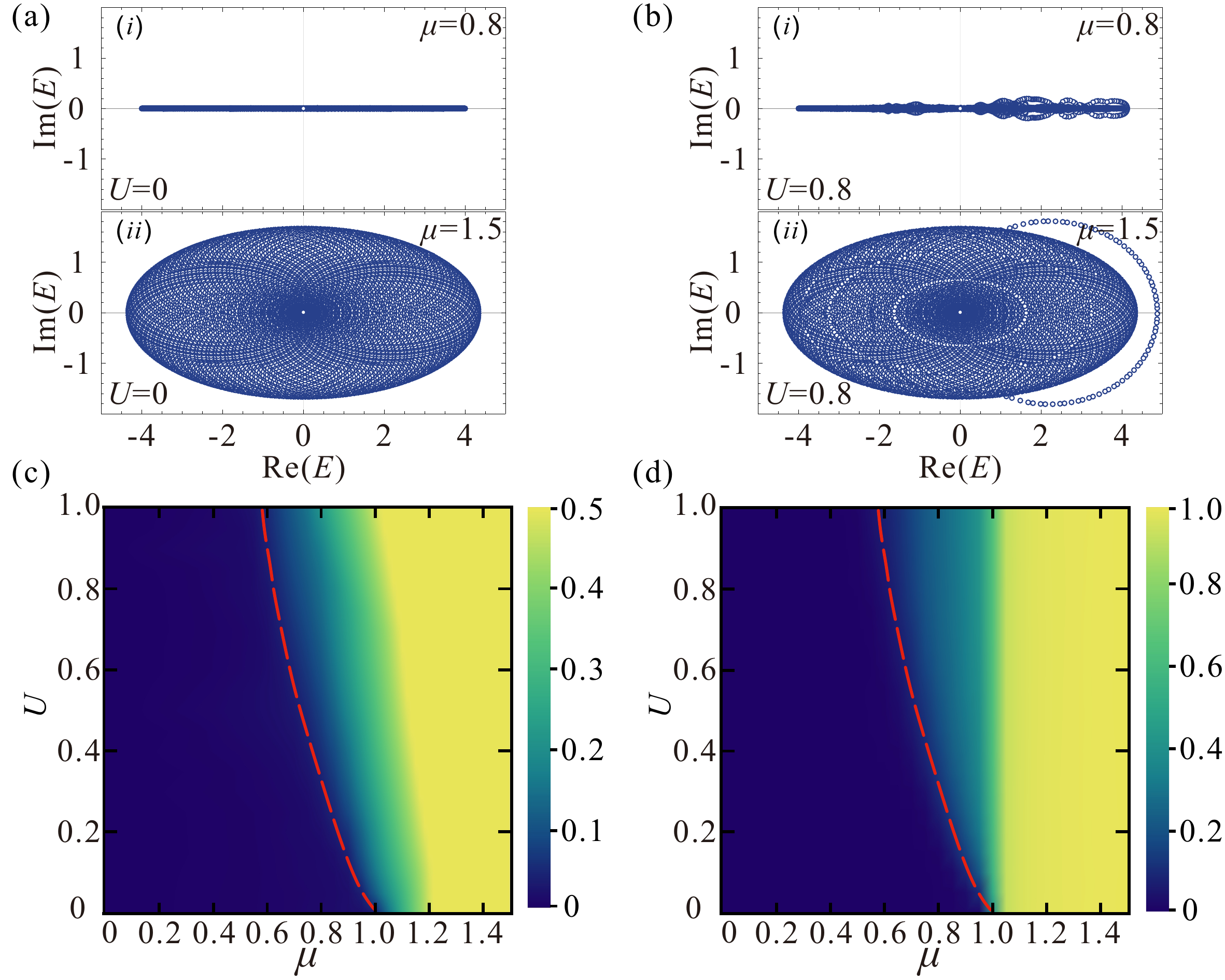}
		\caption{Spectrum properties of the Bose-Hubbard NHAAH model under PBC, computed with $J=1$ and the lattice size $L = 144$. 
		(a) and (b) show the energy spectra $E$ of $\hat{H}$ for $U=0$ and $0.8$ with two typical values of $\mu$. 
		(c) Maximal imaginary parts of $E$, $|{\rm Im}(E)|_{\max}$ [Eq.~(\ref{eq:EImax})], vs $\mu$ and $U$. The dark and light regions have $|{\rm Im}(E)|_{\max}\simeq0$ and $|{\rm Im}(E)|_{\max}>0$. Each $|{\rm Im}(E)|_{\max}$ is rescaled by its maximum over the considered parameter space.
		(d) DOSs $\rho_{\rm Im}$ [Eq.~(\ref{eq:RIm})] with nonzero imaginary parts of $E$.
		Red dashed lines in (c) and (d) represent the boundary of PT transition, numerically extracted from (c).}\label{fig.energy}
	\end{figure*}

	In this section, we uncover phases and transitions in our Bose-Hubbard NHAAH model from an integrated perspective. In Sec.~\ref{sec.spectrum}, we obtain the spectrum of our system by exact diagonalization and analyze the real-to-complex spectral transitions due to the collaborated efforts of disorder and interactions. With the established PT phase diagram, we further identify localization transitions and characterize the critical mobility edge phases in our system in Sec.~\ref{sec.states}. In Sec.~\ref{sec.topo}, we show that the PT and localization transitions can be associated with a spectral winding number, whose values get quantized jumps at the transition borders between extended/critical and critical/localized phases, yielding a topological phase diagram that is consistent with the spectral and localization patterns of our system.
	
	\subsection{Spectrum and PT transitions}\label{sec.spectrum}
	 We first analyze the energy spectrum and reveal the PT transitions in our Bose-Hubbard NHAAH model. We obtain the spectrum of our system under PBC by diagonalizing $\hat{H}$ {[}Eq.~(\ref{eq.NHBHHM}){]} exactly in the Fock space of two bosons $\{|1_{l},1_{l'}\rangle,|2_{l}\rangle\}$. The Fock basis $|1_{l},1_{l'}\rangle$ represents the state with one boson on the lattice site $l$ and the other one on the site $l'$. The Fock basis $|2_{l}\rangle$ denotes the state with two bosons on the same lattice site $l$. Here $l\neq l'$ and $l,l'=1,...,L$, with $L$ being the length of lattice.

 	In the absence of interactions ($U=0$), two typical examples of the spectrum are shown in Figs.~\ref{fig.energy}(a)$(i)$ and \ref{fig.energy}(a)$(ii)$. Since the two bosons are decoupled when $U=0$, we expect that in the limit $L\rightarrow\infty$, the spectrum in the noninteracting case takes the form 
	\begin{equation}
		E=\begin{cases}
		2J(\cos k+\cos k') & |\mu|\leq|J|\\
		2J[\cos(k-ih)+\cos(k'-ih)] & |\mu|>|J|
		\end{cases},\label{eq:E2}
	\end{equation}
	where $k,k'\in[-\pi,\pi]$ and $h=\ln(\mu/J)$. Our numerical results in Figs.~\ref{fig.energy}(a)$(i)$ and \ref{fig.energy}(a)$(ii)$ are coincident with this theoretical prediction. Therefore, our system with two free bosons resides in a PT-unbroken phase when $|\mu|<|J|$, moves into a PT-broken phase when $|\mu|>|J|$, and undergoes a PT transition at $|\mu|=|J|$, which are all similar to the single-particle case. Nevertheless, when $|\mu|>|J|$, the two-body spectrum forms multiple nested loops around the origin ($E=0$) of the complex energy plane, which is rather different from the one-loop spectrum of the single-particle case \cite{Longhi_2019a}. This observation also implies the presence of a much larger spectral winding number in the PT-broken phase of the two-body system compared with the single-particle case, as will be confirmed explicitly in Sec.~\ref{sec.topo}.

	After the switching on of interactions, we observe notable changes in the energy spectrum, as showcased in Figs.~\ref{fig.energy}(b)$(i)$ and \ref{fig.energy}(b)$(ii)$. First, complex eigenenergies emerge even when $|\mu|<|J|$ {[}Fig.~\ref{fig.energy}(b)$(i)${]}. It indicates that compared with single-particle and noninteracting cases, a PT-broken transition has happened at a lower threshold of correlated disorder strength $|\mu|$ when $U\neq0$. Interactions could thus control and facilitate PT transitions in NHQCs. Second, when $|\mu|>|J|$, the energy spectrum separates into two distinguishable portions {[}Fig.~\ref{fig.energy}(b)$(ii)${]}. One of them looks rather similar to the noninteracting two-body spectrum of the system {[}Fig.~\ref{fig.energy}(a)$(ii)${]} at the same parameters in both its range and shape. Another portion constitutes a loop centred around some $E\neq0$, which can be completely separated from the first portion by a line-gap at larger $U$. As will be confirmed in Appendix~\ref{app.doublon}, two-body states with energies lying along this loop are doublons formed by interaction-induced bounded boson pairs. Their emergence provides a key reason for the PT transition to happen in advance in our interacting NHQCs.

	To generate an all-round view of spectral transitions in the Bose-Hubbard NHAAH model, we introduce the quantities
	\begin{equation}
		|{\rm Im}(E)|_{\max}\equiv\max_{j\in\{1,...,D\}}|{\rm Im}E_{j}|,\label{eq:EImax}
	\end{equation}
	\begin{equation}
		\rho_{{\rm Im}}\equiv D_{{\rm Im}}/D.\label{eq:RIm}
	\end{equation}
	Here $E_{j}$ is the $j$th eigenvalue of $\hat{H}$ {[}Eq.~(\ref{eq.NHBHHM}){]}. $D$ represents the Fock-space dimension of our two bosons. $D_{{\rm Im}}$ counts the number of eigenvalues of $\hat{H}$ whose imaginary parts are nonzero. $|{\rm Im}(E)|_{\max}$ and $\rho_{{\rm Im}}$ thus give the maximal imaginary part of $E$ and the density of states (DOSs) with complex energies over the spectrum at a given set of system parameters. In the PT-invariant phase, we expect $|{\rm Im}(E)|_{\max}=0$ and $\rho_{{\rm Im}}=0$. In the PT-broken phase, we would instead have $|{\rm Im}(E)|_{\max}>0$ and $0<\rho_{{\rm Im}}\leq1$. If all the energies of $\hat{H}$ have nonzero imaginary parts, we will have $\rho_{{\rm Im}}=1$. In Figs.~\ref{fig.energy}(c) and \ref{fig.energy}(d), we present the $|{\rm Im}(E)|_{\max}$ and $\rho_{{\rm Im}}$ of $\hat{H}$ vs the disorder strength $\mu$ and interaction strength $U$ for a sufficiently large system size, yielding the PT phase diagram of our Bose-Hubbard NHAAH model. We observe that the PT transition threshold of disorder $|\mu|$ indeed decreases with the raise of $U$ {[}Fig.~\ref{fig.energy}(c){]}. 
	In the Appendix~\ref{app.doublon}, we further show that the PT phase boundary can be approximately described by $J^2=|U\mu|$ for large $U$.
	Meanwhile, the increasing of interaction strength $U$ could also induce PT transitions even if $|\mu|<|J|$ {[}Figs.~\ref{fig.energy}(c) and \ref{fig.energy}(d){]}. Besides, not all the eigenenergies possess nonzero imaginary parts after the PT-breaking transition happens, as reflected by the region with $0<\rho_{{\rm Im}}<1$ in Fig.~\ref{fig.energy}(d). Nevertheless, almost all the eigenenergies finally become complex when $|\mu|>|J|$ as in the noninteracting case. Hence, one pivotal role played by interaction here is to widen the PT transition point of the free-boson system into a critical region with coexisting real and complex eigenenergies. Further roles of interactions in our Bose-Hubbard NHAAH model will be uncovered in the following subsections.
	
    \subsection{Localization transitions}\label{sec.states}
   We next investigate the spatial distributions of eigenstates and unveil the phases with different localization nature in our system. Let $|\psi_{j}\rangle$ be the $j$th two-body eigenstate of $\hat{H}$ {[}Eq.~(\ref{eq.NHBHHM}){]} with energy $E_{j}$. The symmetrized position eigenbasis takes the form 
	\begin{equation}
		\{|l,l\rangle|l=1,...,L\}\cup\{(|l,l'\rangle+|l',l\rangle)/\sqrt{2}\}.\label{eq.Bas}
	\end{equation}
	Here $l$ and $l'$ denote lattice site indices of the two bosons with $l,l'=1,...,L$ and $l\neq l'$. $L$ is the length of the lattice. Projecting $|\psi_{j}\rangle$ onto each position basis and summing over the fourth power of the resulting absolute amplitudes, we arrive at the IPR of $|\psi_{j}\rangle$ in space, i.e.,
	\begin{equation}
		{\rm IPR}_{j}=\sum_{n=1}^{D}|\psi_{n}^{(j)}|^{4},\label{eq.IPR}
	\end{equation}
	where $\psi_{n}^{(j)}$ denotes the overlap between $|\psi_{j}\rangle$ and the $n$th basis in Eq.~(\ref{eq.Bas}). A localized (an extended) eigenstate $|\psi_{j}\rangle$ is expected to have a finite (a vanishing) ${\rm IPR}_{j}$ in the limit $L\rightarrow\infty$. Similarly, we can define the normalized participation ratio (NPR) of $|\psi_{j}\rangle$ in real space as
	\begin{equation}
		{\rm NPR}_{j}={\rm IPR}_{j}^{-1}/D.\label{eq.NPR}
	\end{equation}
	The behavior of ${\rm NPR}_{j}$ is opposite to ${\rm IPR}_{j}$, i.e., it takes a finite value if $|\psi_{j}\rangle$ is spatially extended and vanishes in the limit $L\rightarrow\infty$ if $|\psi_{j}\rangle$ is localized.

	\begin{figure*}
		\centering 
		\includegraphics[width=1.0\textwidth]{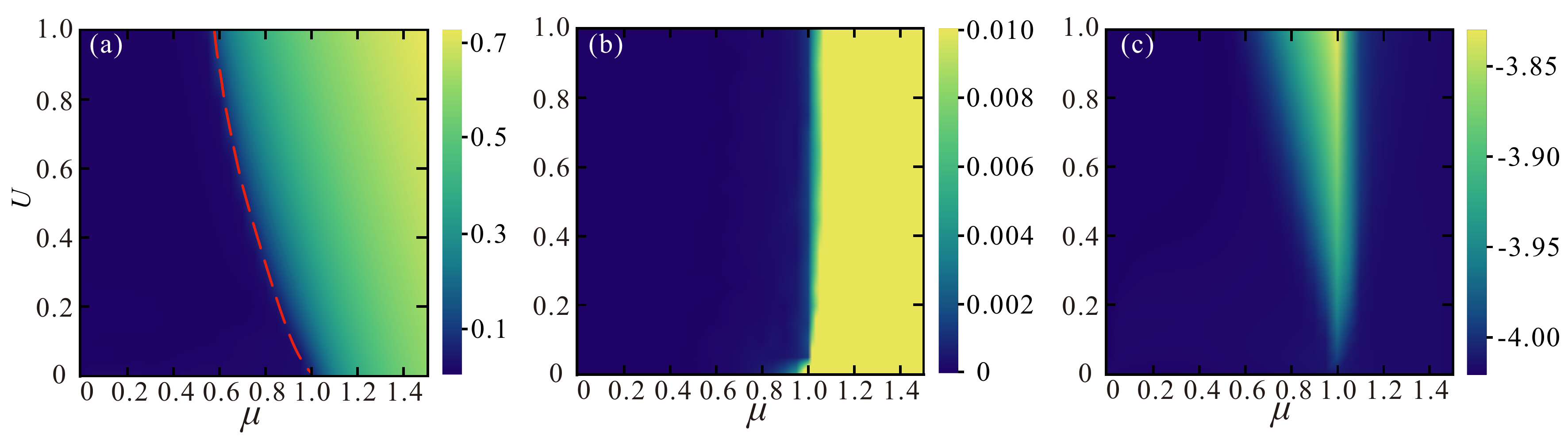}
		\caption{ 
		State properties of the Bose-Hubbard NHAAH model under PBC, computed with $J=1$ and the lattice size $L = 144$. 
			(a) and (b) show the ${\rm IPR}_{\max}$ [Eq.~(\ref{eq.IPRmax})] and ${\rm IPR}_{\min}$ [Eq.~(\ref{eq.IPRmin})].
		    (c) shows the smoking-gun function of mobility edge $\zeta$ [Eq.~(\ref{eq.Zeta})].
		    The red dashed line in (a) represents the boundary of PT transition, numerically extracted from Fig.~\ref{fig.energy}(c).}\label{fig.IPR}
	\end{figure*}

 	Collecting the information of IPR and NPR, we could introduce the following quantities to characterize the localization features of each possible phase in our two-body system, i.e., 
	\begin{equation}
		{\rm IPR}_{\max}=\max_{j\in\{1,...,D\}}({\rm IPR}_{j}),\label{eq.IPRmax}
	\end{equation}
	\begin{equation}
		{\rm IPR}_{\min}=\min_{j\in\{1,...,D\}}({\rm IPR}_{j}),\label{eq.IPRmin}
	\end{equation}
	\begin{equation}
		\zeta=\log_{10}({\rm IPR}_{{\rm ave}}\cdot{\rm NPR}_{{\rm ave}}).\label{eq.Zeta}
	\end{equation}
	Here 
	\begin{equation}
	{\rm IPR}_{{\rm ave}}\equiv\frac{1}{D}\sum_{j=1}^{D}{\rm IPR}_{j}
	\label{eq.IPRave}
	\end{equation}
	and 
	\begin{equation}
	{\rm NPR}_{{\rm ave}}\equiv\frac{1}{D}\sum_{j=1}^{D}{\rm NPR}_{j}
	\label{eq.NPRave}
	\end{equation}
	are the averages of IPR and NPR over all the eigenstates, respectively, at a given set of system parameters. By definition, it is clear that in a metallic phase where all the eigenstates are extended, we would have both ${\rm IPR}_{\max}\rightarrow0$ and ${\rm IPR}_{\min}\rightarrow0$ in the limit $L\rightarrow\infty$. On the contrary, both ${\rm IPR}_{\max}$ and ${\rm IPR}_{\min}$ would take finite values in an insulator phase with only localized eigenstates. In a critical phase with coexisting localized and extended eigenstates, we will have a finite ${\rm IPR}_{\max}$ and a vanishing ${\rm IPR}_{\min}$. The presence of such a critical phase can be further characterized by the smoking-gun function $\zeta$, which in the limit $L\rightarrow\infty$ takes finite values in the critical phase but goes to $-\infty$ in both the extended and localized phases \cite{Han_2022,Li_2020a}. The ${\rm IPR}_{\max}$, ${\rm IPR}_{\min}$ and $\zeta$ defined here thus provide us with a complete set of quantities to distinguish phases with different transport properties in our system.
	
	 In Fig.~\ref{fig.IPR}, we present the ${\rm IPR}_{\max}$ [\ref{fig.IPR}(a)], ${\rm IPR}_{\min}$ [\ref{fig.IPR}(b)] and $\zeta$ [\ref{fig.IPR}(c)] versus the quasiperiodic potential $\mu$ and interaction strength $U$. Their combination forms the localization phase diagram of our Bose-Hubbard NHAAH model with two bosons under the PBC. 
	 In Appendix~\ref{app.mobility.edge}, we further demonstrate the characterization of extended, critical and localized phases using ${\rm IPR_{ave}}$ and ${\rm NPR_{ave}}$, yielding consistent results.
	 In Fig.~\ref{fig.IPR}(a), we observe a clear boundary separating an extended phase (${\rm IPR}_{\max}\simeq0$) from a region with localized eigenstates (${\rm IPR}_{\max}>0$), which is the borderline of localization transitions in our system. These transitions happen at weaker disorder strengths $\mu$ with the increase of $U$. Moreover, even with $|\mu|<1$, a localization transition could still be induced by raising the strength of interaction $U$. The interparticle interaction thus plays an active role in controlling localization transitions. Notably, the localization phase boundary in Fig.~\ref{fig.IPR}(a) is found to be coincident with the PT phase boundary observed in Figs.~\ref{fig.energy}(c) and ~\ref{fig.energy}(d),
	 as illustrated by the red dashed line. Therefore, the PT breaking transitions of the spectra go hand-in-hand with localization transitions of the states in the regime $|\mu|\leq1$ of our system.

	A possible reason behind the lowering of PT and localization transition thresholds of $\mu$ in the presence of interactions may be understood as follows. When $U\neq0$, the two bosons may form a doublon, i.e., a spatially bounded pair of particles. The effective co-hopping rate of such a doublon state could be much smaller than a state in which the two bosons are well-separated in space and can thus undergo almost uncorrelated single-particle hopping processes. The hopping rate of doublons may further decrease with the increase of $U$ (see also Appendix \ref{app.doublon}). Therefore, before reaching the single-particle localization transition point ($|\mu|=1$), the doublon states first become localized with the increase of $|\mu|$ up to some $|\mu_{c}|<1$ when $U\neq0$, and the value of $|\mu_{c}|$ further decreases with the raise of interaction strength. Furthermore, when the doublon states become localized, their eigenenergies also acquire nonzero imaginary parts, triggering the PT-symmetry breaking transition at $|\mu_{c}|$ before reaching the single-particle PT transition point ($|\mu|=1$).

	In Fig.~\ref{fig.IPR}(b), we observe a second localization transition around $|\mu|=1$, after which all the eigenstates of $\hat{H}$ become localized (${\rm IPR}_{\min}>0$) for every $U$. This transition boundary is consistent with the border in Fig.~\ref{fig.energy}(d) that separates the region with coexisting real and complex eigenenergies {[}$\rho_{{\rm Im}}\in(0,1)${]} from the region where almost all eigenenergies have nonzero imaginary parts ($\rho_{{\rm Im}}\simeq1$). This phase boundary tends out to be independent of the interaction strength, implying that it is mainly associated with the localization of weakly correlated two-body states, in which the center of two bosons are spatially well-separated.

	Between the two phase boundaries in Figs.~\ref{fig.IPR}(a) and \ref{fig.IPR}(b), we observe a third phase in which extended and localized eigenstates are coexistent (${\rm IPR}_{\max}>0$, ${\rm IPR}_{\min}\simeq0$), as highlighted by the region with a finite $\zeta$ in Fig.~\ref{fig.IPR}(c). This region thus corresponds to a critical mobility edge phase.
	Details of the mobility edge are illustrated in Appendix \ref{app.mobility.edge}.
	Notably, its domain expands with the increase of $U$ and yet shrinks to a point at $|\mu|=1$ in the limit $U\rightarrow0$. Therefore, the critical mobility edge phase in our Bose-Hubbard NHAAH model could not appear in the free-particle case. It is made possible by interactions and hence unique to strongly correlated NHQCs.

	The interplay between correlated disorder and interaction now allows us to have three distinct phases in our two-body system, i.e., a PT-invariant extended phase, a PT-broken localized phase, and a PT-broken mixed phase in which extended states (with real energies) and localized states (with complex energies) coexist. The domain of the extended phase decreases with the increase of $U$. The domain of the mixed phase increases with the increase of $U$. While the domain of the localized phase is almost not affected by $U$. We will further discriminate these different phases by topological invariants in the following subsection.
	
	\subsection{Topological transitions}\label{sec.topo}
	In previous studies, a topological winding number has been introduced to characterize spectral and localization transitions in single-particle NHQCs \cite{Longhi_2019,Jiang_2019}. For our noninteracting model $\hat{H}_{0}$, such a winding number can be defined under the PBC as 
	\begin{equation}
		w=\int_{0}^{2\pi}\frac{d\theta}{2\pi i}\partial_{\theta}\ln\{\det[\hat{H}_{0}(\theta/L)-E_{0}]\}.\label{eq.w}
	\end{equation}
	Here $E_{0}$ is a base energy that can be chosen rather freely on the complex energy plane, so long as it does not belong to the spectrum of $\hat{H}_{0}$. $\hat{H}_{0}(\theta/L)$ is obtained from $\hat{H}_{0}$ by replacing the onsite potential $\mu_{l}\equiv\mu e^{i2\pi\alpha l}$ with $\mu e^{i(2\pi\alpha l+\theta/L)}$ for all $l=1,...,L$, with $L$ being the length of lattice. Thus defined, the winding number $w$ in Eq.~(\ref{eq.w}) counts the number of times that the spectrum of $\hat{H}_{0}$ encircles $E_{0}$ when the phase shift $\theta$ is varied over a cycle.

	In the single-particle case, it was found that for $\hat{H}_{0}$, we have $w=0$ in the PT-invariant extended phase ($|\mu|<|J|$) and $w=1$ in the PT-broken localized phase ($|\mu|>|J|$) \cite{Zhou_2021a}. The value of $w$ thus experiences a quantized jump following the PT and localization transitions of $\hat{H}_{0}$, yielding a topological signature for these transitions. In cases with two free bosons, we still have $w=0$ in the PT-invariant extended phase ($|\mu|<|J|$) of $\hat{H}_{0}$, which can be directly inspected from the expression of spectrum in Eq.~(\ref{eq:E2}). Meanwhile, in the PT-broken localized phase ($|\mu|>|J|$) of $\hat{H}_{0}$, $w$ is found to be quantized as a nonzero integer, whose precise value depends on the size of the lattice. This may also be deduced from Eq.~(\ref{eq:E2}), where the spectrum forms many loops on the complex energy plane for $|\mu|>|J|$, with each loop contributing a winding $+1$ (counterclockwise) or $-1$ (clockwise) to $w$. The number of such spectral loops increases with the growth of $L$. We have verified these conclusions for two noninteracting bosons in our $\hat{H}_{0}$ by numerical calculations.

	\begin{figure}
		\centering 
		\includegraphics[width=0.48\textwidth]{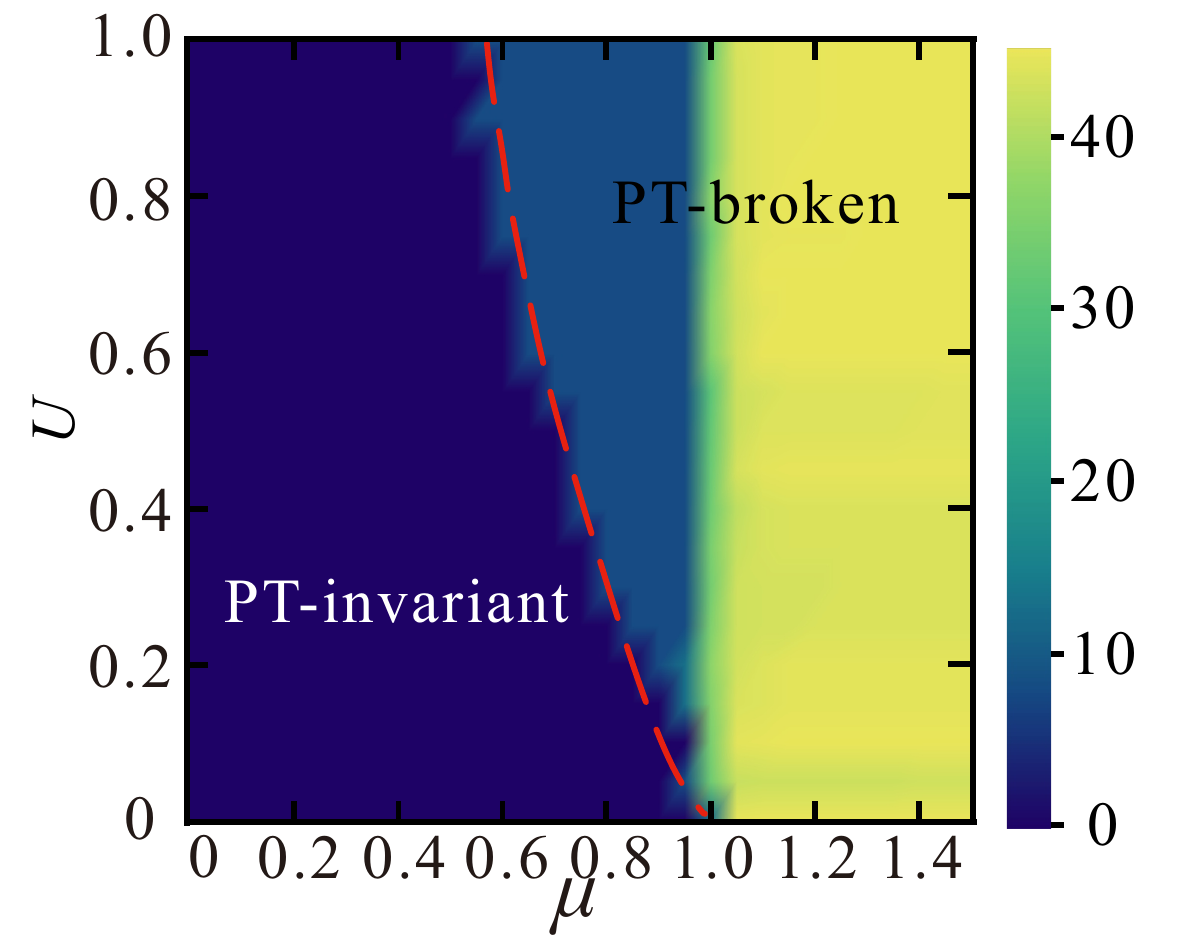}
		\caption{
			Topological phase diagram of the Bose-Hubbard NHAAH model under PBC, with the lattice size $L = 89$.  
			Regions in blue, green, and yellow correspond to extended, critical and localized phases with different winding numbers $w_{1}+w_{2}$, whose values can be read out from the color bar.
			The red dashed line represents the boundary of PT transition, numerically extracted from Fig.~\ref{fig.energy}(c).
		}\label{fig.WN.phase}
	\end{figure}

	In the presence of interactions, we have observed two sets of spectrum and localization transitions plus a critical mobility edge phase in between in our Bose-Hubbard NHAAH model $\hat{H}$ {[}Eq.~(\ref{eq.NHBHHM}) with $U\neq0${]}, as discussed in the last two subsections. A single winding number $w$ as defined in Eq.~(\ref{eq.w}), however, may not be able to fully capture all these transitions and thus distinguish the three different phases of two interacting bosons.
	
	\begin{table*}
		\begin{center}
		\caption{
				Summary of the results for the Bose-Hubbard NHAAH model, including the spectra, DOSs, IPRs and topological winding numbers of the extended, critical and localized phases.}\label{table.1}
			\begin{tabular}{cccc}
				\toprule[1pt]
				Phase& Extended& Critical& Localized \\
				\midrule
				Spectrum& Real& \multicolumn{2}{c}{Complex}\\
				$\rho_{\rm Im}$& $\rho_{\rm Im}\sim 0$& \multicolumn{2}{c}{0<$\rho_{\rm Im}\leq1$}\\
				IPR& $\sim 0$ for all states & \quad $>0 \& \sim 0$ coexist& \quad $>0$ for all states\\
				Winding number & \quad $(w_{1},w_{2})=(0,0)$ & \quad $(w_{1},w_{2})=(1,0)$&$(w_{1},w_{2})\neq(0,0)$\\
				\bottomrule[1pt]
			\end{tabular}
		\end{center}
	\end{table*}

	To address this issue, we introduce a pair of winding numbers $(w_{1},w_{2})$ for $\hat{H}$, which are defined under the PBC as \cite{Zhou_2021b} 
	\begin{equation}
		w_{\ell}=\int_{0}^{2\pi}\frac{d\theta}{2\pi i}\partial_{\theta}\ln\{\det[\hat{H}(\theta/L)-E_{{\rm B}\ell}]\},\quad\ell=1,2.\label{eq.w12}
	\end{equation}
	Here $\hat{H}(\theta/L)$ is obtained by setting the onsite potential $\mu e^{i2\pi\alpha l}\rightarrow\mu e^{i(2\pi\alpha l+\theta/L)}$ for $\hat{H}$. At a given $U$, we choose $E_{{\rm B1}}={\rm Re}E_{j}$ if $|\psi_{j}\rangle$ is the first eigenstate of $\hat{H}$ whose ${\rm IPR}_{j}$ starts to deviate from zero (${\rm IPR}_{\max}\simeq0\rightarrow>0$) and then becoming localized with a complex energy $E_{j}$. The winding number $w_{1}$ is hence expected to be a marker of the PT-breaking transition and the localization transition between extended and critical phases in our two-body system with the increase of $\mu$. Comparatively, we let $E_{{\rm B2}}={\rm Re}E_{j'}$ if $|\psi_{j'}\rangle$ is the last eigenstate of $\hat{H}$ whose ${\rm IPR}_{j'}$ becomes finite (${\rm IPR}_{\min}\simeq0\rightarrow>0$) and thus being localized with a complex energy $E_{j'}$. The winding number $w_{2}$ is then expected to mark the transition between the critical phase with a mixed spectrum and the localized phase with a purely complex spectrum following the increase of $\mu$ in our two-boson system. We would now have a vanishing $w_{1}$ in the extended phase and an integer quantized $w_{1}$ in both the critical and localized phases of $\hat{H}$. Instead, $w_{2}$ will become a nonzero integer only in the localized phase of $\hat{H}$. Combining the information offered by $(w_{1},w_{2})$ then allows us to establish a complete topological characterization of the spectral and localization transitions in our system.
	
	In Fig.~\ref{fig.WN.phase}, we present the topological phase diagram of $\hat{H}$ {[}Eq.~(\ref{eq.NHBHHM}){]} under the PBC with two bosons by evaluating the winding numbers $(w_{1},w_{2})$ at different values of system parameters $(\mu,U)$. Three distinct regions separated by two phase boundaries are clearly observed. 
	In the left region (in blue) of Fig.~\ref{fig.WN.phase}, we have $w_{1}=w_{2}=0$. The domain of this region is coincident with the PT-invariant phase in Figs.~\ref{fig.energy}(c)\textendash (d) and the extended phase in Fig.~\ref{fig.IPR}. In the right region (in yellow) of Fig.~\ref{fig.WN.phase}, we have $w_{1}>0$ and $w_{2}>0$. The domain of this region is consistent with the PT-broken phase with purely complex spectra in Figs.~\ref{fig.energy}(c)\textendash (d) and the localized phase in Fig.~\ref{fig.IPR}. In the middle region (in green) of Fig.~\ref{fig.WN.phase}, we have $(w_{1},w_{2})=(1,0)$. The domain of this region is identical to the PT-broken phase with mixed (real vs complex) spectra in Fig.~\ref{fig.energy}(d) and the critical phase in Fig.~\ref{fig.IPR}. All the phases and transitions induced by the interplay between correlated disorder and interaction in our Bose-Hubbard NHAAH model could thus be topologically characterized by the winding numbers introduced in Eq.~(\ref{eq.w12}).
	
	A summary of the main results obtained in this section is presented in Table \ref{table.1}. In Appendix \ref{app.consistency}, we further elaborate on the case with $U=0.8$ and provide some more details about the transitions characterized by different quantities in our system. We conclude that the presence of interactions indeed induces a critical phase sandwiched by two transitions in our two-body system. Moreover, by varying the strength of interaction, one can control the transition points and manipulate the range of mobility edges. Through a detailed characterization of the energy spectrum and states of the system\textquoteright s Hamiltonian $\hat{H}$ {[}Eq.~(\ref{eq.NHBHHM}){]}, we unveiled the boundaries between the extended, critical, and localized phases, as well as the phase transition processes. The winding numbers $(w_{1},w_{2})$ further serve as topological order parameters, providing a clear depiction of the topological phases and the transitions among them with consistent phase boundaries. The eigenstates and topological properties of our Bose-Hubbard NHAAH model collectively provide a comprehensive understanding of the PT-invariant extended phase, the PT-broken critical phase, and the PT-broken localized phase in two-body NHQCs. To make our observations more feasible in experiments, we will study and showcase the dynamical characteristics of our system in the next section.
	
	\section{Wavepacket and entanglement dynamics}\label{sec.dynamic}
	In this section, we explore the dynamics of two-body wavepackets and entanglement entropy (EE) in our system described by ${\hat H}$ in Eq.~(\ref{eq.NHBHHM}). This is partially motivated by the experimental accessibility to dynamical measurements \cite{Lin_2022, Lukin_2019}. We will trace the evolution of initial wavepackets and EE in the system over time and check whether our Bose-Hubbard NHAAH model could exhibit distinct behaviours in different phases and whether these behaviours could serve as indicators of these phases.
	
	\subsection{Wavepacket Dynamics}\label{subsec.wavepacket}
	
		\begin{figure*}
		\centering  
			\includegraphics[width=0.6\textwidth]{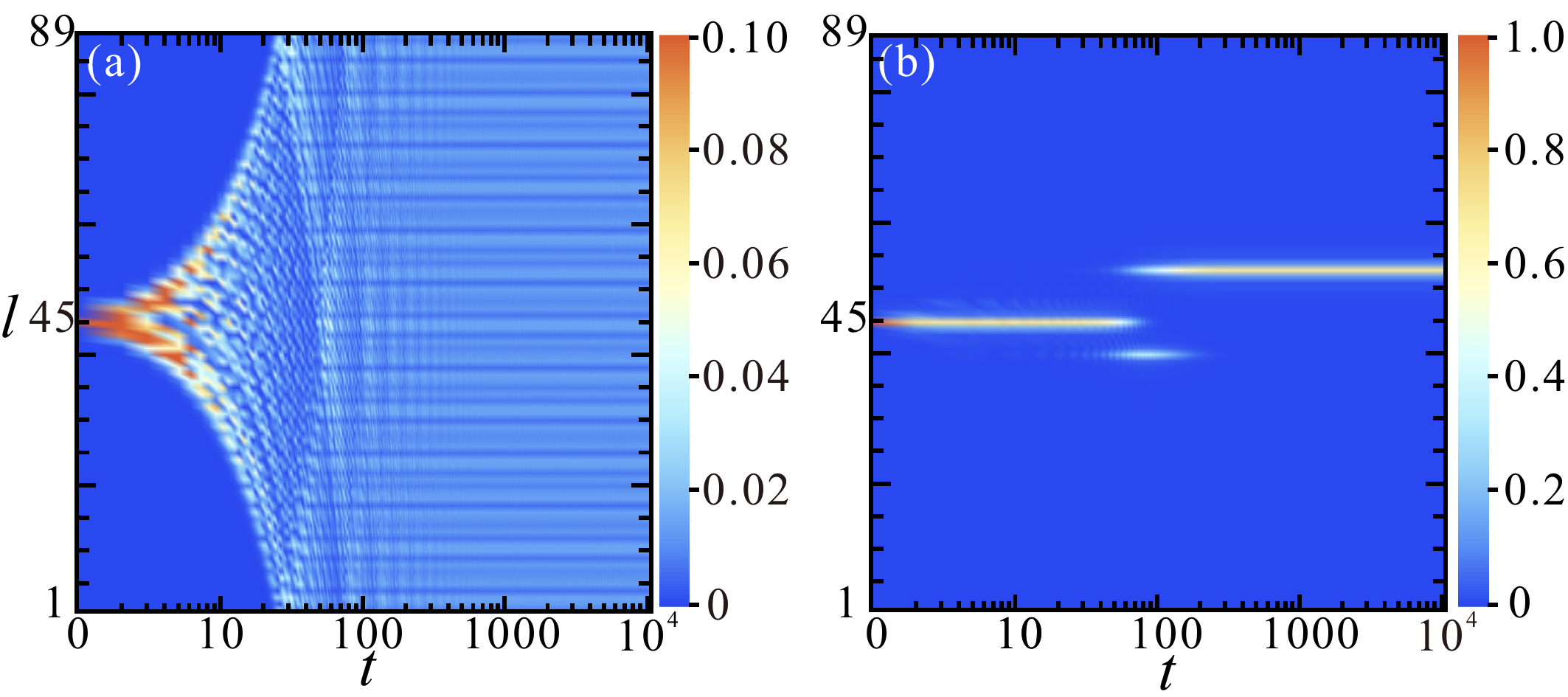}
			\label{fig.WP.U0.b}
		\caption{
		Dynamics of two free particles ($U=0$) in the Bose-Hubbard NHAAH model. The lattice size is $L=89$. The label $l$ along the vertical axis is the lattice site index. (a) and (b) show the spatial probability distributions of the same initial state $|\Psi_0\rangle$ evolved over time $t$, with system parameters set in the extended ($\mu=0.5$) and localized ($\mu=1.5$) phases, respectively.
		}\label{fig.WP.U0}
	\end{figure*}
	
		\begin{figure*}
		\centering 
			\includegraphics[width=0.9\textwidth]{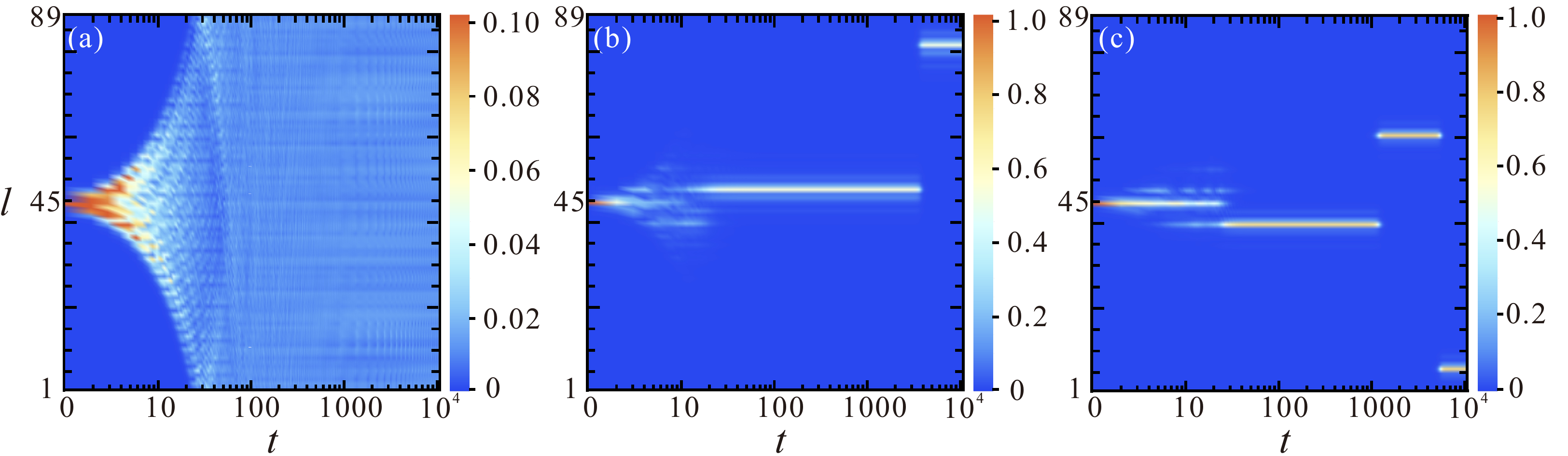}
		\caption{
		Dynamics of two-particle wavepacket for the Bose-Hubbard NHAAH model with the interaction strength $U=0.8$ and lattice size $L=89$. 
		The initial state is evolved under the same condition as in Fig.~\ref{fig.WP.U0}.
		(a)--(c) give the probability distributions of the same initial state over time $t$ for the extended, critical and localized phases with $\mu=0.5$, $1$ and $1.5$, respectively.
		}\label{fig.WP.U08}
	\end{figure*}
	
	We first consider the wavepacket dynamics in our two-body system. Without loss of generality, we let the two bosons to locate initially at the center of the lattice. At time $t=0$, the initial state of the system is thus $|\Psi_{0}\rangle=|2_{L/2}\rangle$. At a later time $t>0$, the initial wavepacket is evolved by $\hat{H}$ {[}Eq.~(\ref{eq.NHBHHM}){]} to the state (let $\hbar=1$) 
	\begin{equation}
		|\Psi'(t)\rangle=e^{-i\hat{H}t}|\Psi_{0}\rangle=\sum_{j}c_{j}e^{-iE_{j}t}|\psi_{j}\rangle,\label{eq.psi_t}
	\end{equation}
	where $|\psi_{j}\rangle$ is the $j$th eigenstate of $\hat{H}$ with energy $E_{j}$, and $c_{j}=\langle\psi_{j}|\Psi_{0}\rangle$ denotes the overlap between the initial state and $|\psi_{j}\rangle$. Since the $\hat{H}$ in Eq.~(\ref{eq.NHBHHM}) is not Hermitian, some of the eigenenergies $E_{j}$ may take complex values, yielding a nonunitary evolution and an unnormalized state $|\Psi'(t)\rangle$ in Eq.~(\ref{eq.psi_t}). Nevertheless, for dynamics of open quantum systems under postselection, we could arrive at a normalized state $|\Psi(t)\rangle$ given by 
	\begin{equation}
		|\Psi(t)\rangle=|\Psi'(t)\rangle/\sqrt{\langle\Psi'(t)|\Psi'(t)\rangle}.\label{eq.renorm}
	\end{equation}
	That is, at each time step, the state of the system is first evolved according to the Schr\"odinger equation with the $\hat{H}$ in Eq.~(\ref{eq.NHBHHM}). A normalization of the state then follows without going through any quantum jumps to generate $|\Psi(t)\rangle$. The spatial distribution of the two bosons at any time $t$ is finally obtained by projecting $|\Psi(t)\rangle$ onto the position eigenbasis {[}Eq.~(\ref{eq.Bas}){]} \cite{Longhi_2023c}.
	
	In Fig.~\ref{fig.WP.U0}, we illustrate representative dynamics of two-body wavepackets under the PBC in our Bose-Hubbard NHAAH model without interactions ($U=0$). We observe that when $|\mu|<|J|$, the localized initial state $|\Psi_{0}\rangle$ spreads ballistically and populates the whole lattice almost uniformly after a relatively short time window in Fig.~\ref{fig.WP.U0}(a). This is an expected behavior within a PT-invariant extended phase. Meanwhile, in Fig.~\ref{fig.WP.U0}(b), the initial wavepacket maintains its spatial localization over time up to certain non-Hermitian jumps when $|\mu|>|J|$, which is also expected in a PT-broken localized phase. The two distinct phases of free bosons in our system may thus be probed and distinguished via wavepacket dynamics.
	
	Typical time evolutions of two-body wavepackets under the PBC in our Bose-Hubbard NHAAH model with interactions ($U\neq0$) are shown in Fig.~\ref{fig.WP.U08}. When the system parameters are chosen inside the PT-invariant extended phase of $\hat{H}$ {[}$(\mu,U)=(0.5,0.8)$ in Fig.~\ref{fig.WP.U08}(a){]}, we again observe the fast expansion of initially localized wavepacket $|\Psi_{0}\rangle$ across the whole lattice within a relatively short time window. The interactions do not generate apparent effects on the spreading speed of wavepacket in this case. When the system parameters are set in the PT-broken localized phase of $\hat{H}$ {[}$(\mu,U)=(1.5,0.8)$ in Fig.~\ref{fig.WP.U08}(c){]}, the initial wavepacket $|\Psi_{0}\rangle$ is well localized over its entire time evolution around one particular lattice site. Meanwhile, two dynamically distinct regions can be observed in the wavepacket evolution when the system parameters are chosen in the PT-broken critical phase of $\hat{H}$ {[}$(\mu,U)=(1,0.8)$ in Fig.~\ref{fig.WP.U08}(b){]}. That is, the wavepacket $|\Psi_{0}\rangle$ is found to diffuse initially over a short time window ($t\sim10$), and then pinned strongly around a certain lattice site in its later dynamics. It is the non-Hermitian nature of the system that leads to prolonged lifetimes of particles in the critical and localized phases. The noticeable non-Hermitian jumps (co-tunneling) of two bosons in Figs.~\ref{fig.WP.U0}(b), \ref{fig.WP.U08}(b) and \ref{fig.WP.U08}(c) are further analyzed in the Appendix \ref{app.WP}.
	Note in passing that these jumps are not unique to interacting systems.

	We have checked the wavepacket dynamics for other parameter choices within each phase of our system and observed consistent results. In conclusion, notable changes in wavepacket dynamics across different phases (extended, critical and localized) are found in our Bose-Hubbard NHAAH model with two bosons, allowing us to distinguish these phases efficiently in cases without and with interactions.
	
	\subsection{Entanglement dynamics}\label{sec.dynamic.S}
	
	\begin{figure*}
		\centering
			\includegraphics[width=0.8\textwidth]{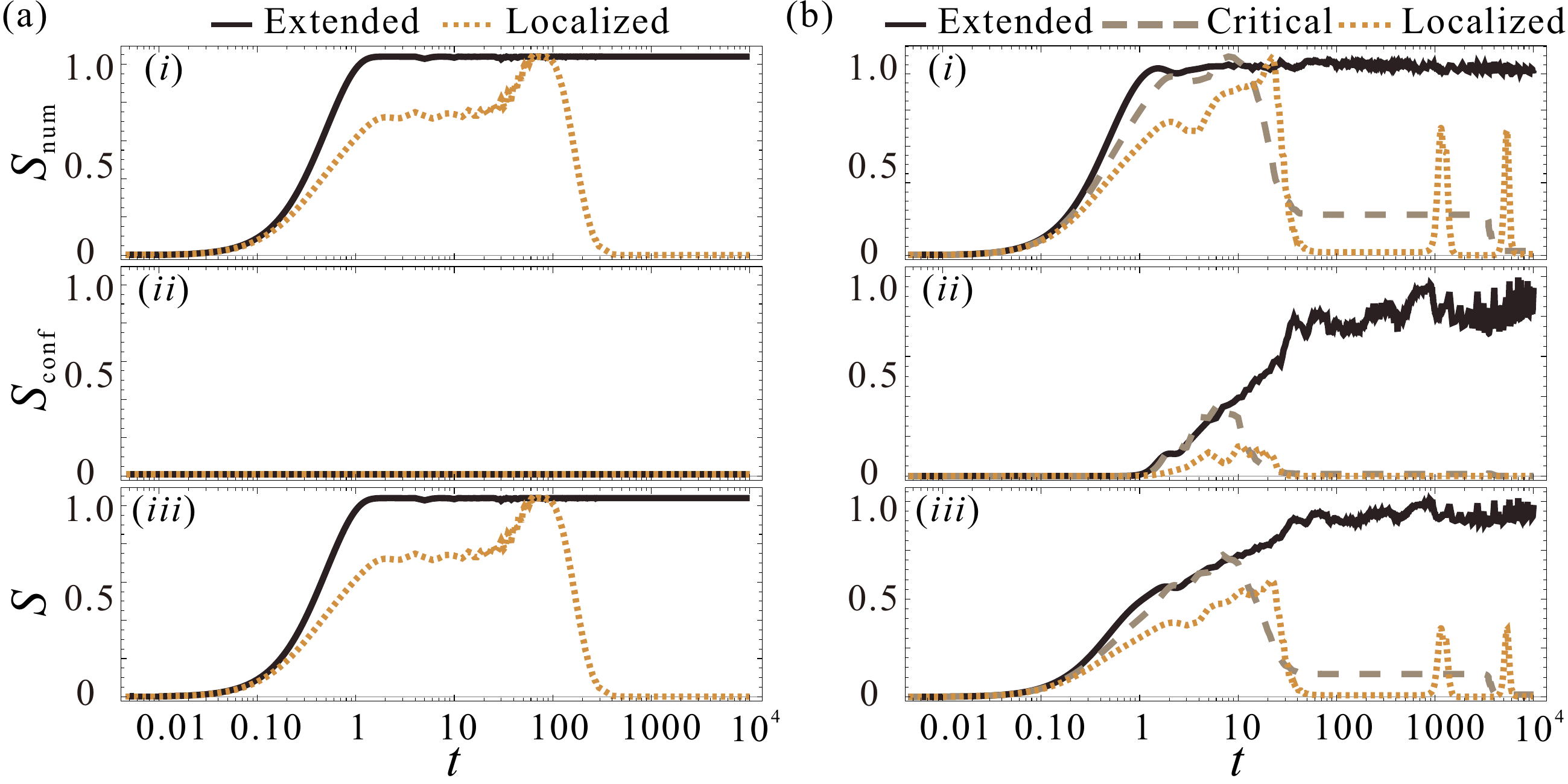}
		\caption{EE dynamics of two particles in the Bose-Hubbard NHAAH model. The lattice size is $L=89$ and the initial state is prepared to be the same as that used in the wavepacket dynamics. In (a) and (b), ($i$), ($ii$) and ($iii$) represent the time-dependence of $S_{\rm num}(t)$, $S_{\rm conf}(t)$ and the total $S(t)$ for the cases with $U=0$ and $0.8$, respectively. The solid, dashed and dotted lines in (b) denote the EE with $\mu=0.5$, $1$ and $1.5$.}\label{fig.entropy}
	\end{figure*}
	
	In this section, we focus on analyzing the dynamics of EE. Propagations of quasiparticles in a system and behaviours of entangled quantum states can be measured using the EE \cite{Altshuler_1997, Alba_2018, Agarwal_2015, Lukin_2019, Orito_2022}. The EE has also been identified as a valuable metric for characterizing many-body localizations \cite{Gornyi_2005, Basko_2006}. Here, we investigate the properties of extended, critical, and localized phases in our Bose-Hubbard NHAAH model with two bosons by calculating their related entanglement features in the time domain.
	
	We first make a distinction between two types of EE that may exist between two subsystems $A$ and $B$ of a composite system, which are referred to as number EE and configuration EE. The number EE describes the correlation between particle numbers in one subsystem and those in another. It arises through tunnelling effects between subsystems. The configuration EE depicts the correlation between particle configurations in one subsystem and those in another. It requires at least two particles to exist in the whole system. Tunnelling effects alone cannot generate configuration EE, as they only affect individual particles. Yet, interactions can entangle pairs of particles. Therefore, the combination of tunnelling effects and interactions can lead to the formation of configuration entanglement over longer distances \cite{Lukin_2019}. The number EE $S_{\rm num}$ and the configuration EE $S_{\rm conf}$ together constitute the total EE $S = S_{\rm num} + S_{\rm conf}$ \cite{Lukin_2019, Orito_2022}.
	
	The dynamics of total EE $S(t)$ can be decomposed as
	\begin{eqnarray}
		S(t) &=& S_{\rm num}(t)+S_{\rm conf}(t),\label{eq.St}\\
		S_{\rm num}(t)&=&-\sum_{N_A}p_{N_A} \log\ p_{N_A},\label{eq.Snt}	
		\\	S_{\rm conf}(t)&=&-\sum_{N_A}\sum_{i}
		p_{N_A}\tilde{\lambda}^{(N_A)}_{i}\log\ \tilde{\lambda}^{(N_A)}_{i}.\label{eq.Sct}
	\end{eqnarray}
	Here, $N_A$ is the number of particles in subsystem $A$, $p_{N_A}$ represents the sum of eigenvalues of all block diagonal matrices in subsystem $A$, and $\tilde{\lambda}^{(N_A)}_{i}$ denotes the normalized eigenvalue of the $i$th block diagonal matrix in the reduced density matrix of subsystem $A$. Detailed calculation strategies for $S_{\rm num}(t)$ and $S_{\rm conf}(t)$ are given in Appendix \ref{app.entropy.computation}.
	
	In Fig.~\ref{fig.entropy}(a), we show the dynamical evolution of the $(i)$ number EE $S_{\rm num}(t)$, $(ii)$ configuration EE $S_{\rm conf}(t)$, and $(iii)$ total EE $S(t)$ for our Bose-Hubbard NHAAH model with two bosons in the noninteracting limit [$U=0$ in Eq.~(\ref{eq.NHBHHM})]. The solid and dotted curves represent the cases when the initial state $|\Psi_{0}\rangle$ is prepared in the extended phase ($\mu=0.5$) and localized phase ($\mu=1.5$) of ${\hat H_0}$. Without interactions, we can focus on the dynamics of $S_{\rm num}(t)$. In the extended phase, $S_{\rm num}(t)$ initially increases rapidly, and then reaches a steady value at one. In the localized phase, $S_{\rm num}(t)$ also increases rapidly at first before reaching a metastable plateau. After that, it increases again up to one and finally decreases continuously until reaching zero. As $S_{\rm conf}(t)$ is always zero, the number EE $S_{\rm num}(t)$ serves as the total EE $S(t)$ in the noninteracting case. The observed behaviours of $S_{\rm num}(t)$ are typical for our system with two free bosons in its two distinct phases.
	
	The changes in $S_{\rm num}(t)$ are related to the wavepacket dynamics (Fig.~\ref{fig.WP.U0}). In the extended phase, the initial rapid growth of $S_{\rm num}(t)$ is caused by particle diffusion as observed in Fig.~\ref{fig.WP.U0}(a). As time goes by, particles diffuse throughout the system, leading to the equilibration and saturation of $S_{\rm num}(t)$. In the insulating phase, the diffusion is almost negligible and the wavepacket is localized at specific lattice sites. Since the initial particle distribution is centred at $L/2$, the boundary between subsystems $A$ and $B$, stable particle transport occurs between these subsystems, maintaining a metastable $S_{\rm num}(t)$ in the intermediate time domain. As discussed in Appendix \ref{app.WP}, with the increase of time, non-Hermiticity induced wavepacket jumps lead to the eventual collapse of the wavepacket onto the eigenstate of ${\hat H}_0$ with the largest ${\rm Im}E$, yielding a further increase of $S_{\rm num}(t)$ in the localized phase. Once the wavepacket jumps are complete and all particles are localized on one side of the system, there is no longer particle transport between subsystems $A$ and $B$, causing $S_{\rm num}(t)\rightarrow0$ in the end. Without interactions, nonlocal coherent dynamics between subsystems do not exist \cite{Lukin_2019}. Therefore, $S_{\rm conf}(t)$ remains to be zero regardless of whether the system is in the extended or the localized phase. In Fig.~\ref{fig.entropy}(a)($\romannumeral3$), the overall EE $S(t)$ is thus derived from the contribution of number EE only, leading to dynamical behaviors determined by $S_{\rm num}(t)$.
	
	Fig.~\ref{fig.entropy}(b) illustrates the dynamics of the $(i)$ number EE $S_{\rm num}(t)$, $(ii)$ configuration EE $S_{\rm conf}(t)$, and $(iii)$ total EE $S(t)$ for our Bose-Hubbard NHAAH model with two bosons and the interaction strength $U=0.8$. The initial state $|\Psi_{0}\rangle$ is the same as that used in the noninteracting case. The solid, dashed and dotted lines in each panel describe the EE for $\mu=0.5$, $1$ and $1.5$, corresponding to the extended, critical and localized phases, respectively. We focus on the growth of $S_{\rm num}(t)$ and $S_{\rm conf}(t)$ as well as their behaviors across different phases.
	
	In all the phases, $S_{\rm num}(t)$ increases rapidly in the initial stage. In the extended phase, $S_{\rm num}(t)$ saturates gradually and reaches a steady value after its initial growth. In the critical and localized phases, the $S_{\rm num}(t)$ grow first until reaching peak values around $t=10$ and then decrease. In the critical phase, $S_{\rm num}(t)$ drops to a finite value and reaches a metastable plateau. After a sufficiently long time, it decreases again towards zero. In the localized phase, $S_{\rm num}(t)$ drops to zero directly, then increases slightly twice in time before going back to zero.
	
    With interactions, the behavior of $S_{\rm num}(t)$ can be further understood by comparing the EE and wavepacket dynamics  [see Figs.~\ref{fig.entropy}(b) and \ref{fig.WP.U08}(c)]. In the extended phase, the initial increase of $S_{\rm num}(t)$ originates from the rapid spread of wavepacket over time, with the eventual saturation reflecting the uniform distribution of wavepacket across the entire system and the equilibration in the end. In the critical phase, the initial growth of $S_{\rm num}(t)$ results from imperfect localization of the system, leading to wavepacket diffusion in short-term. As time passes, wavepackets become localized around $L/2$	within the subsystem $B$, leading to a metastable and relatively small $S_{\rm num}(t)$ due to slight particle transport between subsystems $A$ and $B$. Non-Hermitian effects further allow the wavepacket to jump from one site of subsystem $B$ to another, suppressing the particle tunnelling between two subsystems and finally leading to $S_{\rm num}(t)$ goes to zero.
	
	In the localized phase, the initial increase of $S_{\rm num}(t)$ stems from the location of particles at the center of two subsystems [see Fig.~\ref{fig.WP.U08}(c)]. As time passes, we have $S_{\rm num}(t)\rightarrow0$ when wavepackets are fully localized within subsystem $A$. Over a longer evolution time, non-Hermitian effects cause wavepackets to undergo two jumps between subsystems $A$ and $B$, generating two increases in the number EE from $t\sim10^3-10^4$. Finally, the particles become pinned to the localized state with the largest ${\rm Im}E$, yielding $S_{\rm num}(t)=0$.
	Note that the appearance of two jumps in the late-time regime are not unique to interacting systems. The increases of $S_{\rm num}(t)$ in the localized phase are mainly caused by the non-Hermiticity of onsite potential. The presence of these jumps reflect that the state has not been evolved to its final location, i.e., the location of certain localized eigenstate of the system with the largest imaginary part of energy.
	The interaction further allows the bosons to form doublons that could jump together as a whole in space, which is a possible reason for the appearance of sharper peaks in the late-time regime of Fig.~\ref{fig.entropy}(b)(i) compared with the peak at $t\simeq100$ in Fig.~\ref{fig.entropy}(a)(i).
	Notably, interactions make the initially negligible $S_{\rm conf}(t)$ nontrivial. This is because configuration EE implies a correlation between particle configurations in subsystems $A$ and $B$, necessitating the presence of particles in each subsystem. Tunnelling alone cannot produce configuration EE, as it acts individually on each particle. Interactions thus allow particles to be configurationally entangled. The combined effects of tunnelling and interactions produce configuration entanglement over longer distances \cite{Lukin_2019}.
	
	In Fig.~\ref{fig.entropy}(b)($ii$), $S_{\rm conf}(t)$ experiences an initial growth, whose duration is nevertheless delayed significantly compared to the initial rapid increase of $S_{\rm num}(t)$. After $S_{\rm num}(t)$ saturates, $S_{\rm conf}(t)$ continues to grow, albeit at a slower rate, and gradually being stabilized. In the critical and localized phases, $S_{\rm conf}(t)$ grow initially and reach respective peaks before decaying. In the critical phase, $S_{\rm conf}(t)$ decreases to a finite value, while in the localized phase it tends to zero.
	
	\begin{figure}
		\centering
			\includegraphics[width=0.47\textwidth]{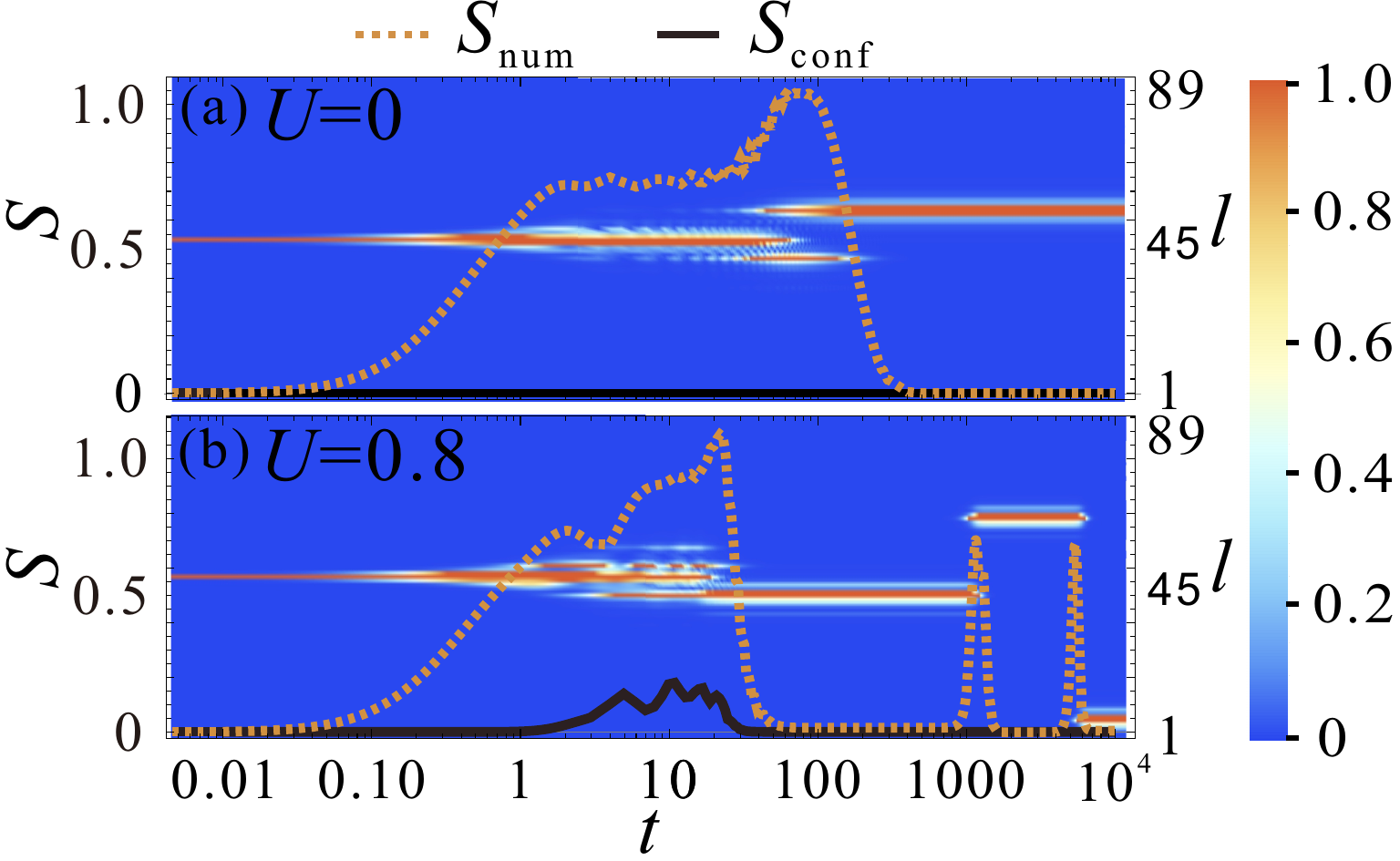}
		\caption{Wavepacket and EE dynamics of two particles in the Bose-Hubbard NHAAH model. The system parameters are $L=89$, $\mu=1.5$ and the initial state is prepared to be the same as that used in the wavepacket dynamics. In (a) and (b), solid and dash lines represent the time-dependence of $S_{\rm num}(t)$ and $S_{\rm conf}(t)$ for the cases with $U=0$ and $0.8$. 
		(a) and (b) also give probability distributions of the same initial state over time with $U=0$ and $0.8$, respectively.}\label{fig.WP.EE}
	\end{figure}

	In Fig.~\ref{fig.WP.EE}, the wavepacket and EE dynamics are shown in order to further demonstrate their correlations. 
	One can see that the $S_{\rm num}(t)$ increases in conjunction with the spreading of wavepacket at the beginning of the dynamics, regardless of the presence or absence of interactions. When the wavepacket is localized, the $S_{\rm num}(t)$ changes only when the position of the wavepacket is changed.
    The $S_{\rm conf}(t)$ exists only in systems with interactions. Its growth depends on the particle configuration of the subsystems. The $S_{\rm conf}(t)$ no longer changes and remains constant at zero once the particles in the system are localized at their final locations.
	
	We can draw the following conclusions in this section. First, the number EE $S_{\rm num}(t)$ signifies the correlation between particle numbers in different subsystems. The magnitude of $S_{\rm num}(t)$ reflects the strength of particle tunnelling between subsystems, providing insights into the behaviors of wavepacket dynamic. Second, the configuration EE $S_{\rm conf}(t)$ accounts for the correlation between particle configurations in different subsystems. Interaction is essential for its existence, as it facilitates entanglement between particle pairs. The combined efforts of tunnelling and interaction can thus lead to the formation of configuration entanglement, which persists over longer spatial distances \cite{Lukin_2019}. Third, in the extended, critical and localized phases, $S_{\rm num}(t)$ and $S_{\rm conf}(t)$ exhibit distinct properties. They reflect the degree of localization and allow us to distinguish between different phases based on the dynamics of EE. These insights deepened our understanding of EE in non-Hermitian systems, revealing diverse aspects of entanglement between particle numbers and configurations due to the interplay between disorder and interactions.
	
	In Appendix \ref{app.entropy.mean}, we present further results for time-averaged EE of our Bose-Hubbard NHAAH model with two bosons. The variations of EE suggest the presence of a possible entanglement phase transition across the PT and localization transition boundaries in our system.

	\section{Conclusion}\label{sec.sum}
	In this work, we unveiled PT transitions, localization transitions, topological transitions and mobility edges induced by the cooperation between correlated disorder and many-body interaction. Focusing on two bosons in a prototypical NHQC with onsite two-body interactions and quasiperiodically modulated gain and loss, we found that the interaction could trigger the PT and localization transition thresholds towards the regime of weaker disorder strengths. Moreover, the spectral, localization and topological triple phase transition point of the free bosonic system becomes unstable and expands into a whole critical phase with mobility edges in the presence of interactions. Phase diagrams of our system were established via systematic analyses of the spectrum, IPRs and topological winding numbers. Distinctive features of the extended, critical and localized two-body phases in our system were further revealed by investigating the dynamics of wavepackets and EE. Our findings thus laid the foundation for future studies of correlation-induced phenomena in NHQCs and in other non-Hermitian disordered systems. The PT symmetry breaking, localization transitions and mobility edges triggered by interactions are also expected to appear in NHQCs beyond the minimal model considered in this work.

	In experiments, the single-particle physics of NHQCs has been explored by implementing nonunitary photonic quantum walks, in which signatures of spectral, localization, topological triple phase transitions and mobility edges have been observed \cite{Weidemann_2022,Lin_2022}. Meanwhile, quantum walks of correlated photons have also been realized in various optical setups \cite{Peruzzo_2010,Esposito_2022,Jiao_2021,Poulios_2014}. The combination of these relevant techniques may allow us to realize our Bose-Hubbard NHAAH model and probe the associated phases and transitions in photonic systems. Ultracold atoms in optical lattices form another promising setup, in which the Bose-Hubbard model \cite{Greiner_2002} and quasiperiodic potentials \cite{Gommers_2006,Roux_2008,Viebahn_2019} have both been realized. Non-Hermitian effects can further be introduced by laser-induced atom losses \cite{Li_2019,Lapp_2019,Ren_2022,Xu_2017}. Therefore, our Bose-Hubbard NHAAH model may also be engineered in cold atoms and the interaction-induced phases and transitions might be detected in near-term experiments.

	In future work, it is interesting to explore correlation-induced phases and transitions in NHQCs with more than two particles and further in many-body regime. Other types of non-Hermitian terms, such as asymmetric hopping may induce delocalization transitions \cite{Yuki2023} and non-Hermitian skin effects \cite{Yao_2018}, whose interplay with disorder and interaction also deserves to be explored. 
	Our tentative calculations suggest that nonreciprocal hoppings could indeed transform bound states to scattering states and yield critical phases in interacting NHQCs.
	In non-Hermitian systems with quasiperiodic or quenched disorders, various types of entanglement phase transitions have been recently identified \cite{Li_2023b,Zhou_2023a,Li_2023c}. The impact of interactions \cite{Hamazaki_2019} on these intriguing entanglement transitions forms another interesting direction of future research.

	\textit{Note added}: During the finalization of this work, we noticed a paper that focused on the formation of doublons in a non-Hermitian Fermi-Hubbard model with two fermions \cite{Longhi_2023c}. It also demonstrated the emergence of mobility edges due to interactions, which is consistent with our findings in Bose-Hubbard NHQCs. Our work further generated a complete phase diagram in the parameter space of onsite Bose-Hubbard interaction and correlated disorder from a unified perspective of spectral, localization, topology, dynamics and entanglement.
	

	\begin{acknowledgments} 
		T.Q. acknowledges Xinhong Han, Wendong Li, Junjie Zeng and Javier Molina-Vilaplana for the assistance provided in numerical calculations. This work is supported by the National Natural Science Foundation of China (Grant Nos.~12275260 and 11905211), the Fundamental Research Funds for the Central Universities (Grant No.~202364008), and the Young Talents Project of Ocean University of China.
	\end{acknowledgments}		
	
	\appendix
	
	\begin{figure*}
	    \centering  
	    \includegraphics[width=0.9\textwidth]{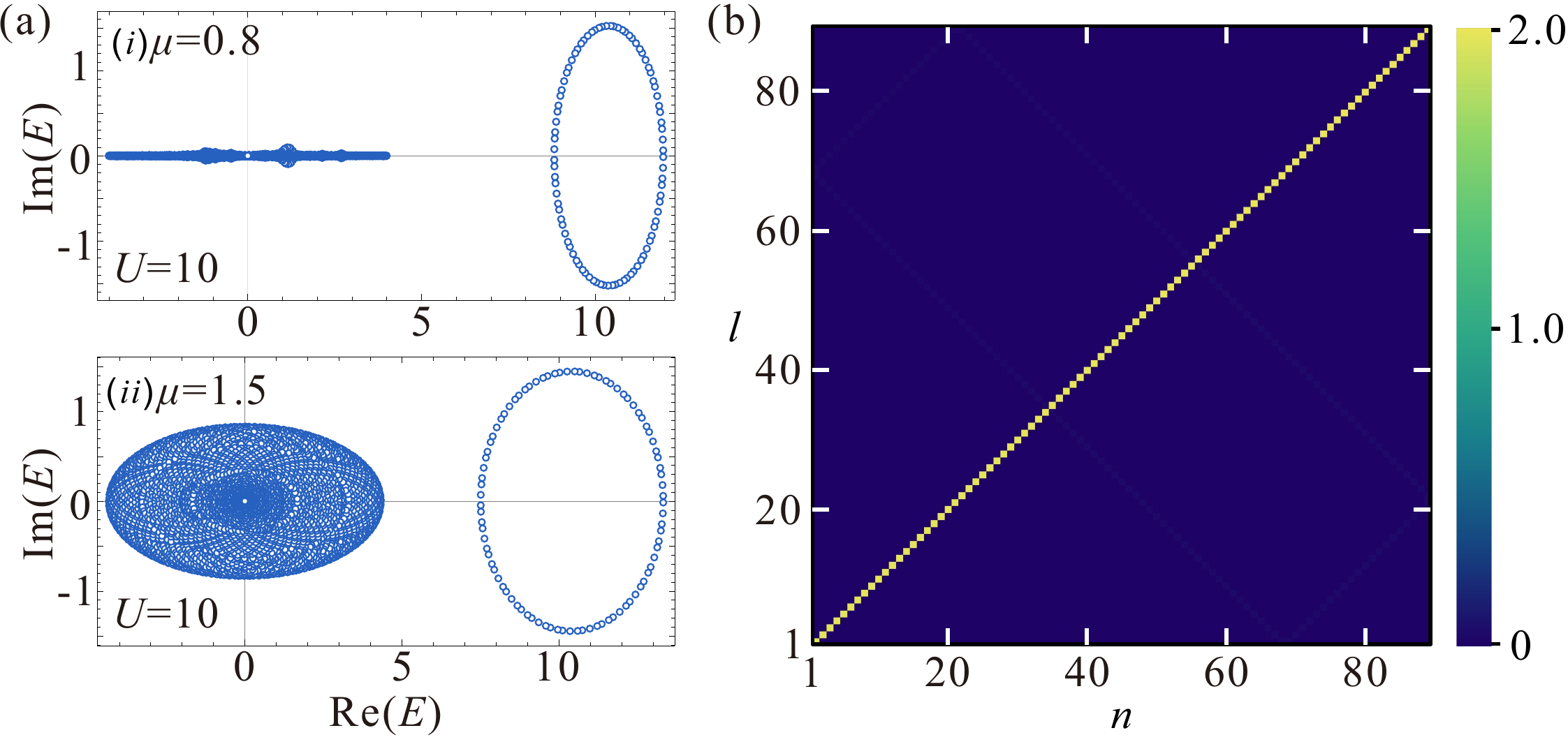}
	    \caption{
	    Energy spectrum and particle density distribution of the Bose-Hubbard NHAAH model under PBC and strong interactions, with $U=10$, $J=1$ and system size of $L=89$.
		(a) Energy spectrum $E$ of $\hat{H}$ for two typical values of $\mu$. 
		(b) Density distribution of bosons in the lattice for the isolated energy ring in (a)$(ii)$. The vertical axis, labeled by $l$ represents the lattice index. The horizontal axis, labeled by $n$ represents the eigenstate index.
		}\label{fig.doublon}
    \end{figure*}
	
	\section{Doublon states and effective Hamiltonian}\label{app.doublon}
	
	\subsection{Doublon states}
	In this appendix, we check whether the separated energy loop observed in Fig.~\ref{fig.energy}(b) of the main text is due to the formation of bounded two-boson states (doublons) induced by interactions. We will focus on the case of strong interactions.

	\begin{figure}
	    \centering  
	    \includegraphics[width=0.48\textwidth]{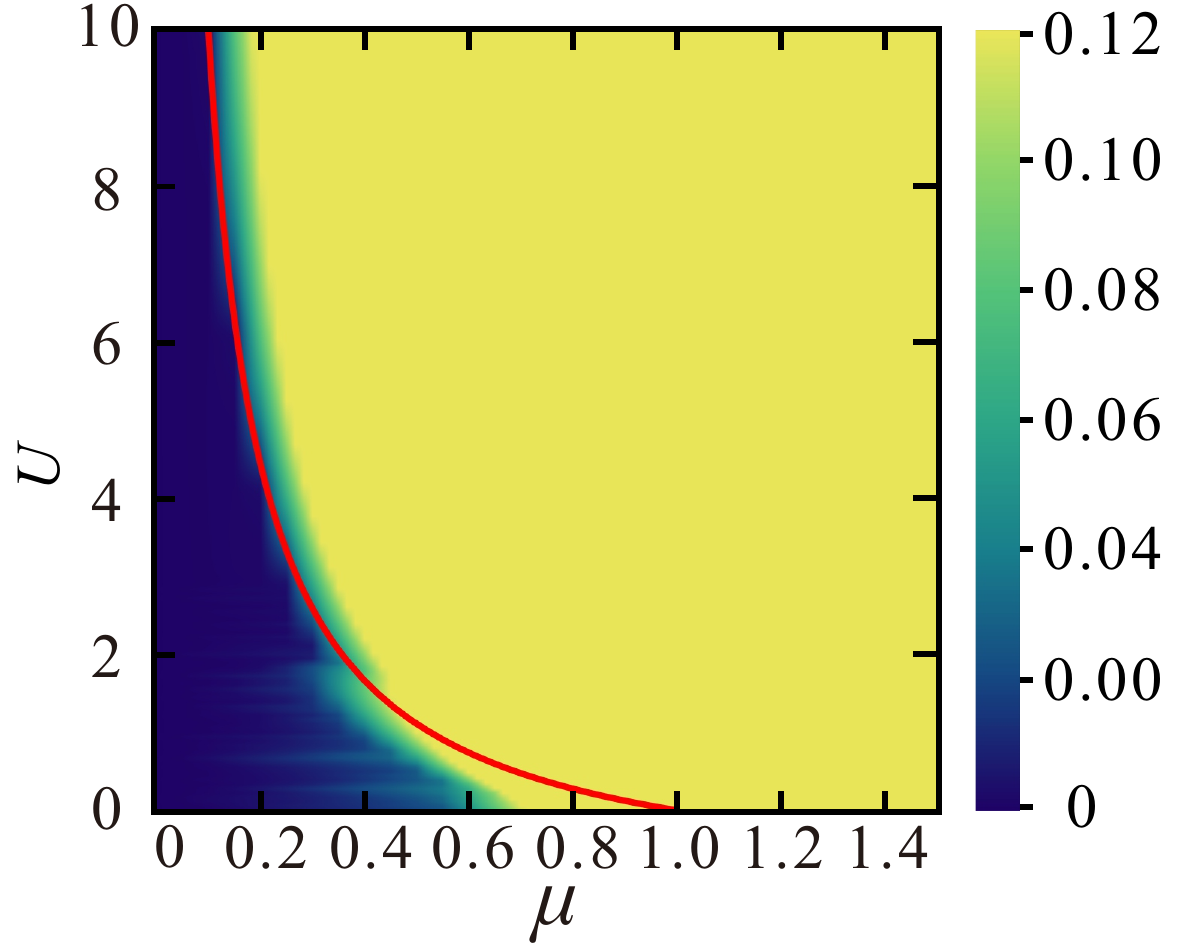}
	    \caption{
	    The PT phase diagram, given by the maximal imaginary parts of $E$, $|{\rm Im}(E)|_{\max}$ vs $U$ and $\mu$, computed with $J=1$ and the lattice size $L=89$. Each $|{\rm Im}(E)|_{\max}$ is rescaled by its maximum over the considered parameter space. The dark and light regions have $|{\rm Im}(E)|_{\max}\simeq0$ and $|{\rm Im}(E)|_{\max}>0$. The red curve represents 
	   $J^2=|U\mu|$. For large $U$, this curve coincides well with the PT phase boundary.
	   }\label{fig.doublon.boundry}
    \end{figure}

    In Fig.~\ref{fig.doublon}(a), we illustrate the energy spectrum of our system with the lattice size $L=89$ and interaction strength $U=10$ in both the critical ($\mu=0.8$) and localized ($\mu=1.5$) phases. The horizontal (vertical) axis represents the real (imaginary) part of energy. It is clear that in both the critical and localized phases, a separated energy loop is present on the right side of the spectrum. This isolated loop consists of $89$ eigenenergies, which are related to $89$ eigenstates of the system.

    To determine whether these energy loops describe doublons, we calculate spatial distributions of two-particle states associated with these loops. If both bosons are located in the same lattice site, it signifies a doublon state. In Fig.~\ref{fig.doublon}(b), we present the probability distributions of the $89$ eigenstates in space, with the same system parameters as of Fig.~\ref{fig.doublon}(a)($ii$). 
    These $89$ states correspond to the top $89$ eigenstates sorted in descending order by their real parts of energy. 
    We find that all two-particle eigenstates are located in individual lattice sites, forming doublons. This confirms that the two-particle states within the separated energy loop in Figs.~\ref{fig.energy}(b)$(ii)$ and Fig.~\ref{fig.doublon}(a) are formed by bounded boson pairs.
	
    An effective Hamiltonian for doublon states can be obtained by treating interactions as the dominant part and the remaining terms as perturbations. Applying degenerate perturbation theory up to second order \cite{Takahashi_1977,Ke_2017}, the form of effective Hamiltonian is found to be (see Appendix \ref{app.derivation} for derivation details)
	\begin{equation}\label{eq.H.eff}
	    \hat{H}_{\rm eff}=\frac{2J^2}{U}\sum_{l}(\hat{b}_l^{\dagger}\hat{b}_{l+1}+{\rm H.c.})-2\mu\sum_{l}e^{i2\pi\alpha l} \hat{n}_l+U+\frac{2J^2}{U}.
	\end{equation}
	We present the PT phase diagram of our model up to strong interactions in Fig.~\ref{fig.doublon.boundry}. The red curve in this figure is given by $J^2=|U\mu|$, as obtained from Eq.~(\ref{eq.H.eff}), which captures the PT phase boundary (and also the transition boundary between extended and critical phases) nicely for large $U$.
	Comparing the Hamiltonians in Eqs.~(\ref{eq.NHAAH}) and (\ref{eq.H.eff}), we could see that they share the same form up to some additional constants. Therefore, if the PT and localization transition points of ${\hat H}_0$ are at $|J|=|\mu|$ \cite{Longhi_2019a}, the PT and localization transition points of ${\hat H}_{\rm eff}$ must be at $|2J^2/U|=|2\mu|$. Note in passing that neither ${\hat H}_0$ nor ${\hat H}_{\rm eff}$ possesses the self-duality property of the Hermitian AAH model.
	Since we have set $J=1$, we also observe deflections of numerical results from the doublon phase boundary $J^2=|U\mu|$ when $U<2J^2=2$ for the domain of $\mu$ considered in our calculations. This is expected, as the perturbation theory should not work well in this parameter regime.
		
	\subsection{Derivation of $\hat{H}_{\rm eff}$}\label{app.derivation}
	To derive Eq.~(\ref{eq.H.eff}), we treat the noninteracting part of Hamiltonian ${\hat H}_{0}$ [Eq.~(\ref{eq.NHAAH})] as a perturbation to the interaction term ${\hat H}_{\rm int}$ in Eq.~(\ref{eq.NHBHHM}), where
	\begin{equation}
	    {\hat H}_{\rm int}=\frac{U}{2}\sum_{l}\hat{n}_{l}(\hat{n}_{l}-1).
	\end{equation}
	${\hat H}_{\rm int}$ has two eigenvalues $E_j = U$ with degenerate eigenstates $|2_j\rangle$ forming a subspace $\mathcal{U}$, and $E_{j,k} = 0$ with degenerate eigenstates $|1_j, 1_k\rangle$ forming the orthogonal complement subspace $\mathcal{V}$. We define projection operators upon $\mathcal{U}$ and $\mathcal{V}$ as
	\begin{equation}
	\begin{aligned}
	\hat{P}&=\sum_j|2_j\rangle\langle2_j|,\\
	\hat{S}&=\sum_{\substack{j,k\\ j\neq k}}\frac{1}{E_j-E_{j,k}}|1_j,1_k\rangle\langle1_j,1_k|\\
	&=\frac{1}{U}\sum_{\substack{j,k\\ j\neq k}}|1_j,1_k\rangle\langle1_j,1_k|.
	\end{aligned}
	\end{equation}
	Applying the degenerate perturbation theory up to the second order, the effective Hamiltonian for the subspace ${\cal U}$ is given by
	\begin{equation}
	\begin{aligned}
        {\hat H}_{\rm eff}=&E_j\hat{P}+\hat{P}{\hat H}_0\hat{P}+\hat{P}{\hat H}_0\hat{S}{\hat H}_0\hat{P}\\
        =&U\hat{P}+\sum_{j,k}|2_j\rangle\langle2_j|{\hat H}_0|2_k\rangle\langle2_k|\\
        +&\frac{1}{U}\sum_{j,k}\sum_{\substack{l,m\\ l\neq m}}|2_j\rangle\langle2_j|{\hat H}_0|1_l,1_m\rangle\langle1_l,1_m|{\hat H}_0|2_k\rangle\langle2_k|.
    \end{aligned}
    \end{equation}
	The matrix elements $\langle 2_j|{\hat H}_0|2_k\rangle$ do not vanish only for the onsite potential term, so
	\begin{equation}
	\begin{aligned}
        \hat{P}{\hat H}_0\hat{P}=& \sum_{j,k}|2_{j}\rangle \langle2_{j}|{\hat H}_{0}|2_{k}\rangle \langle2_{k}|\\
        =&-\mu\sum_{l}e^{i2\pi \alpha l}\sum_{j,k}|2_{j}\rangle \langle2_{j}|\hat{n}_{l}|2_{k}\rangle \langle2_{k}|  \\
        =&-\mu\sum_{l}e^{i2\pi\alpha l}\sum_{k}|2_{k}\rangle \langle2_k|\hat{n}_{l}|2_{k}\rangle \langle2_k| \\
        =&-2\mu\sum_{l}e^{i2\pi\alpha l}|2_l\rangle \langle2_l|.
    \end{aligned}
    \end{equation}
    Further calculations yield that
    
    \begin{equation}
    \begin{aligned}
    	& \hat{P}\hat{H_0}\hat{S}{\hat H}_0\hat{P}\\
   =&\frac{1}{U}\sum_{k,l}\sum_{m,n}|2_k\rangle \langle2_k|{\hat H}_{0}|1_{l},1_m\rangle \langle 1_l,1_m| {\hat H}_{0}|2_n\rangle \langle 2_n|\\
    =&\frac{1}{U}\sum_{k,l}\sum_{m,n}|2_k\rangle \langle2_k|\sum_{j=1}^{D}-J({\hat b}_{j}^{\dagger}{\hat b}_{j+1}+{\rm H.c.})|1_k,1_m\rangle\\
    \times& \langle 1_l,1_m|\sum_{r=1}^{D} -J({\hat b}_{r}^{\dagger}{\hat b}_{r+1}+{\rm H.c.})|2_n\rangle \langle 2_n|\\
	=&\frac{J^2}{U}\sum_{j,k}\sum_{\substack{l,m\\ l\neq m}}\sum_{j=1}^{D}|2_k\rangle\langle2_k|(\hat{b}_{j}^\dagger\hat{b}_{j+1}+{\rm H.c.})|1_l,1_m\rangle  \\
		\times& \sum_{r=1}^{D}\langle1_l,1_m|(\hat{b}_{r+1}^\dagger\hat{b}_r+{\rm H.c.})|2_n\rangle\langle2_n|.
    \end{aligned}
    \end{equation}
	These summations will produce four terms
	\begin{equation}
    \begin{aligned}\label{eq.heff.1}
    &\sum_{k,l}\sum_{\substack{m,n\\ m\neq n}}\sum_{j=1}^{D}\sum_{r=1}^{D}|2_{k}\rangle \langle2_{k}|{\hat b}_{j}^{\dagger}{\hat b}_{j+1}|1_{l},1_{m}\rangle\\
    \times&\langle 1_{l},1_{m}|{\hat b}_{r}^{\dagger}{\hat b}_{r+1}|2_n\rangle \langle 2_n|
    =2\sum_{m}|2_m\rangle \langle2_{m-1}|,
    \end{aligned}
    \end{equation}
	\begin{equation}
	\begin{aligned}\label{eq.heff.2}
    &\sum_{k,l}\sum_{\substack{m,n\\ m\neq n}}\sum_{j=1}^{D}\sum_{r=1}^{D}|2_k\rangle \langle 2_k|{\hat b}_{j+1}^{\dagger}{\hat b}_j|1_l,1_m\rangle\\
    \times& \langle 1_{l} 1_m |{\hat b}_{r}^{\dagger}{\hat b}_{r+1}|2_n\rangle \langle 2_n|
    =2\sum_{m}|2_{m-1}\rangle \langle 2_m|,
    \end{aligned}
    \end{equation}
    \begin{equation}
	\begin{aligned}\label{eq.heff.3}
    &\sum_{k,l}\sum_{\substack{m,n\\ m\neq n}}\sum_{j=1}^{D}\sum_{r=1}^{D} |2_k\rangle \langle 2_k|{\hat b}_{j}^{\dagger}{\hat b}_{j+1}|1_l,1_m\rangle\\
    \times& \langle 1_l,1_m|{\hat b}_{r+1}^{\dagger}{\hat b}_r|2_n\rangle \langle 2_n|
    =2\sum_{m}|2_m\rangle \langle 2_m|,
    \end{aligned}
    \end{equation}
    \begin{equation}
	\begin{aligned}\label{eq.heff.4}
    &\sum_{k,l}\sum_{\substack{m,n\\ m\neq n}}\sum_{j=1}^{D}\sum_{r=1}^{D} |2_k\rangle \langle 2_k|{\hat b}_{j+1}^{\dagger}{\hat b}_{j}|1_l,1_m\rangle\\
    \times& \langle 1_l,1_m|{\hat b}_{r}^{\dagger}{\hat b}_{r+1}|2_n\rangle \langle 2_n|=2\sum_{m}|2_m\rangle \langle 2_m|.
    \end{aligned}
    \end{equation}
    Putting Eq.~(\ref{eq.heff.1})--Eq.~(\ref{eq.heff.4}) together, we find
    \begin{equation}
    \begin{aligned}
    {\hat H}_{\rm eff}=&(E_{0}+2J^2/U)\hat{P}-2\mu\sum_{l}e^{i2\pi\alpha l}|2_l\rangle \langle 2_l|\\
    +&2\frac{J^2}{U}\sum_{m}(|2_m\rangle\langle 2_{m+1}|+{\rm H.c.}).
    \end{aligned}
    \end{equation}
    Defining $|2_l\rangle={\hat b}_l^{\dagger}|0\rangle$, we finally obtain the expected expression of ${\hat H}_{\rm eff}$.    	

	\section{Critical phase and mobility edge}\label{app.mobility.edge}
	
	\begin{figure*}
	    \centering  
	    \includegraphics[width=0.8\textwidth]{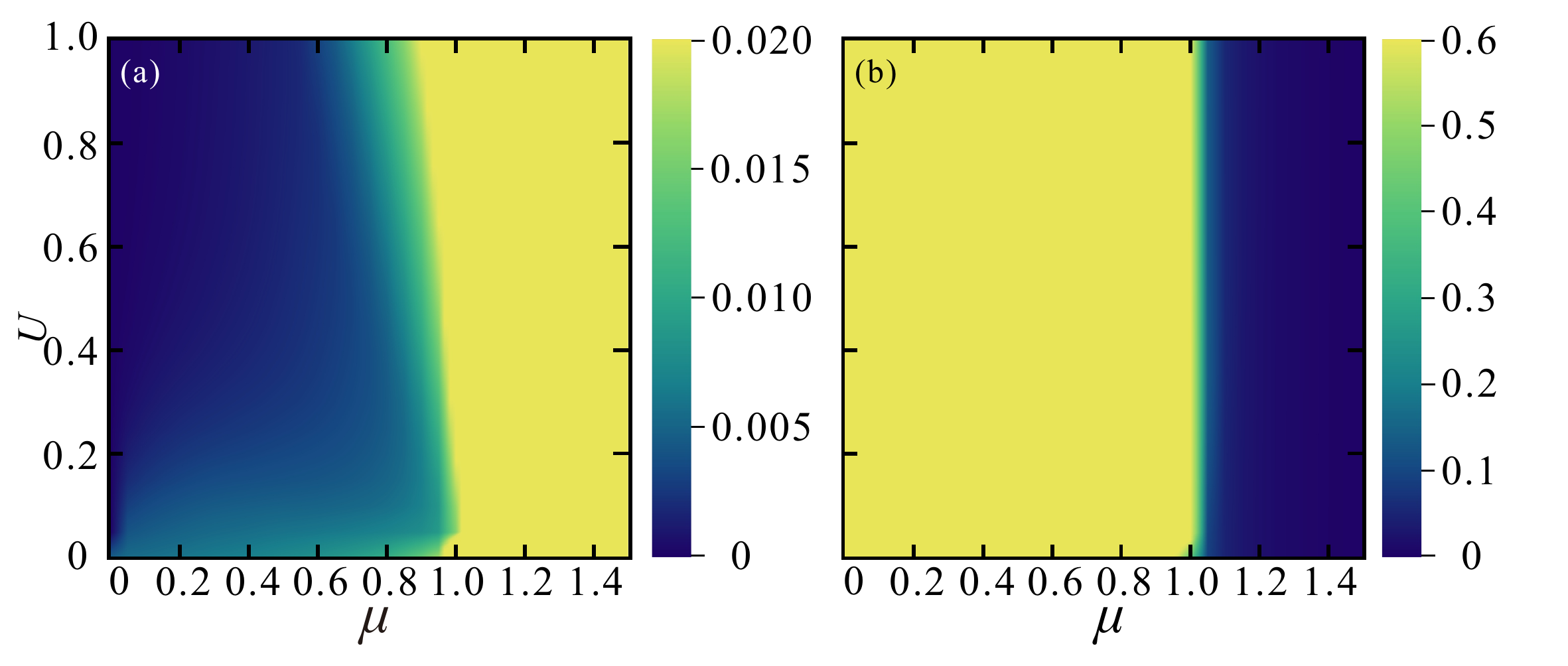}
	    \caption{
	    State properties of the Bose-Hubbard NHAAH model under PBC, computed with $J = 1$ and the lattice size $L = 144$. (a) and (b) show the ${\rm IPR_{ave}}$ [Eq.~(\ref{eq.IPRave})] and ${\rm NPR_{ave}}$ [Eq.~(\ref{eq.NPRave})].
	   }\label{fig.ave.IPRNPR}
    \end{figure*}
    
	\subsection{Critical phase characterized by ${\rm IPR_{ave}}$ and ${\rm NPR_{ave}}$ }
	In section~\ref{sec.states}, we defined the regions of extended and localized phases using IPR and NPR, while simultaneously utilizing $\zeta$ to define the region of critical phase. We refer to the domain in which both ${\rm IPR_{ave}}$ and ${\rm NPR_{ave}}$ are finite as the critical phase \cite{Li_2020a}. Fig.~\ref{fig.ave.IPRNPR} presents separate results of ${\rm IPR_{ave}}$ and ${\rm NPR_{ave}}$ over the $\mu-U$ parameter space, showcasing the presence of a critical phase.
	
	In Fig.~\ref{fig.ave.IPRNPR}(a), we can also observe a  boundary separating an extended phase (${\rm IPR_{ave}}\simeq0$) from a region with localized eigenstates (${\rm IPR_{ave}}>0$), which is the borderline of localization transitions in our system. These transitions happen at weaker disorder strengths $\mu$ with the increase of $U$. The Fig.~\ref{fig.ave.IPRNPR}(a) is similar to the Fig.~\ref{fig.IPR}(a) which demonstrates the extended phase with ${\rm IPR}_{\max}$. The latter is more sensitive to the onset of the first localized eigenstate in the system for a large system size $L$.
    Fig.~\ref{fig.ave.IPRNPR}(b) displays the region characterized by $\rm NPR_{ave}$, which aligns with the results shown in Fig.~\ref{fig.IPR}(b) with ${\rm IPR}_{\min}$.

	\subsection{Mobility edge}
	
	\begin{figure*}
	    \centering  
	    \includegraphics[width=0.8\textwidth]{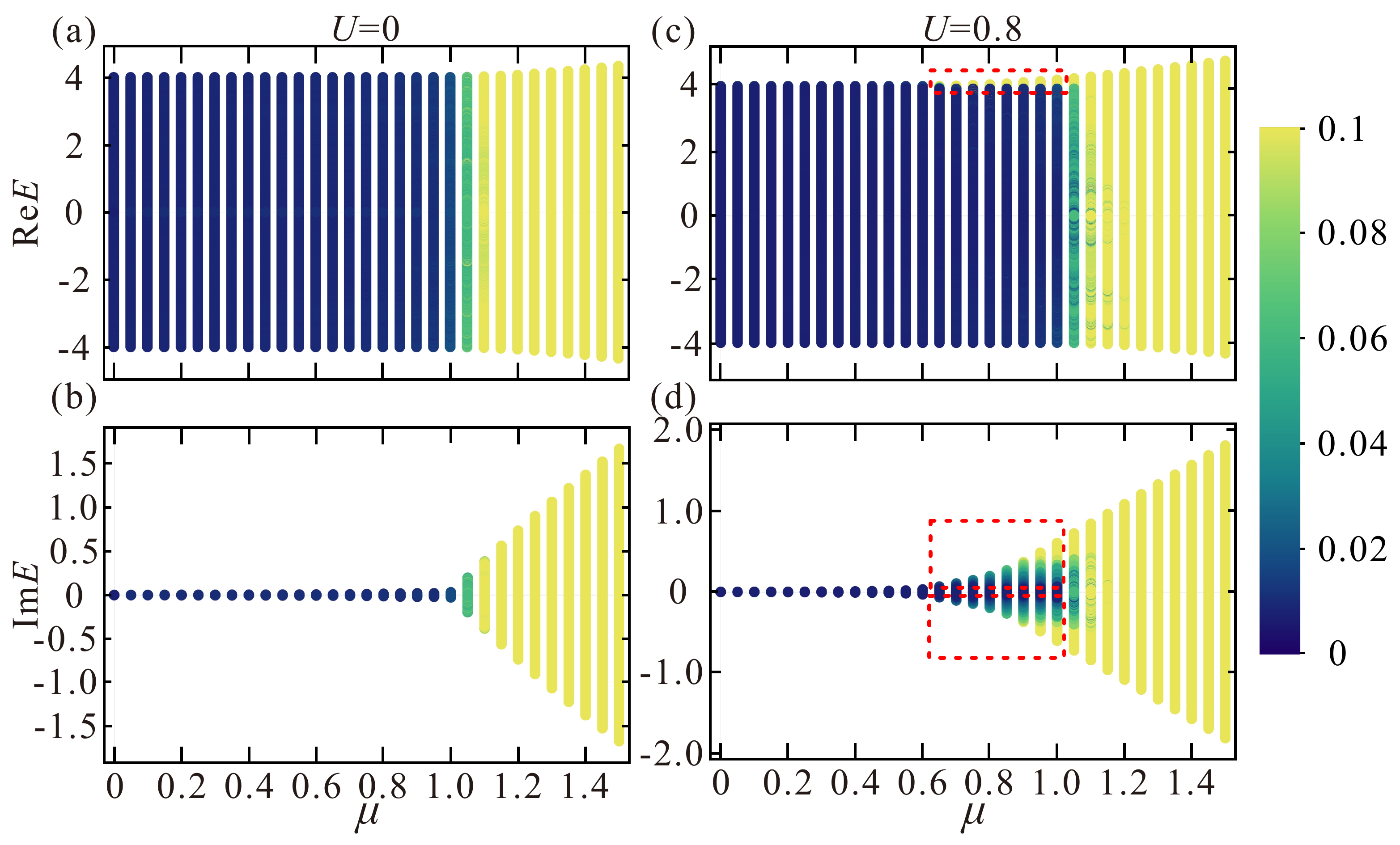}
	    \caption{
	    IPRs of all the eigenstates of Bose-Hubbard NHAAH model under PBC with $J = 1$ and the lattice size $L = 144$ against $\mu$ and their eigenenergies.
	    (a) and (b) show the the real and imaginary parts of energy spectrum vs the disorder strength $\mu$ with $U=0$.
	    (c) and (d) show the the real and imaginary parts of energy spectrum vs $\mu$ with $U=0.8$.
	    The red dashed boxes guide the view of mobility edges at $\mu\in(0.65,1.0)$ for $U=0.8$. (a)--(d) share the same color bar.\label{fig.mobility.edge}}
    \end{figure*}
    
	To further illustrate the presence of interaction-induced mobility edge in our critical phase, we present the real and imaginary parts of the energy spectrum of our system ($L=144$) vs the disorder strength $\mu$ for $U=0$ and $U=0.8$, as shown in Fig.~\ref{fig.mobility.edge}. The color map of each data point is further given by the IPR of the associated eigenstate. Clear signatures of mobility edge are observed in the critical phase when $U=0.8$. Therefore, the existence of mobility edges in the complex spectra of our system within the critical phase is justified. 

	\section{Spectral and topological properties of different phases}\label{app.consistency}
	In the Sec.~\ref{sec.static} of the main text, we find that the interplay among interactions, non-Hermiticity, and quasiperiodic potential gives rise to various phases and transitions. They are elucidated with both spectral characteristics [Fig.~\ref{fig.IPR}(c)] and topological properties (Fig.~\ref{fig.WN.phase}). It is worth noting that the phases described by spectrum and topological numbers are consistent. To further validate this point, we discuss in this appendix the changes of various quantities in the presence (or absence) of interactions.
	
\begin{figure}
		\centering  
			\includegraphics[width=0.485\textwidth]{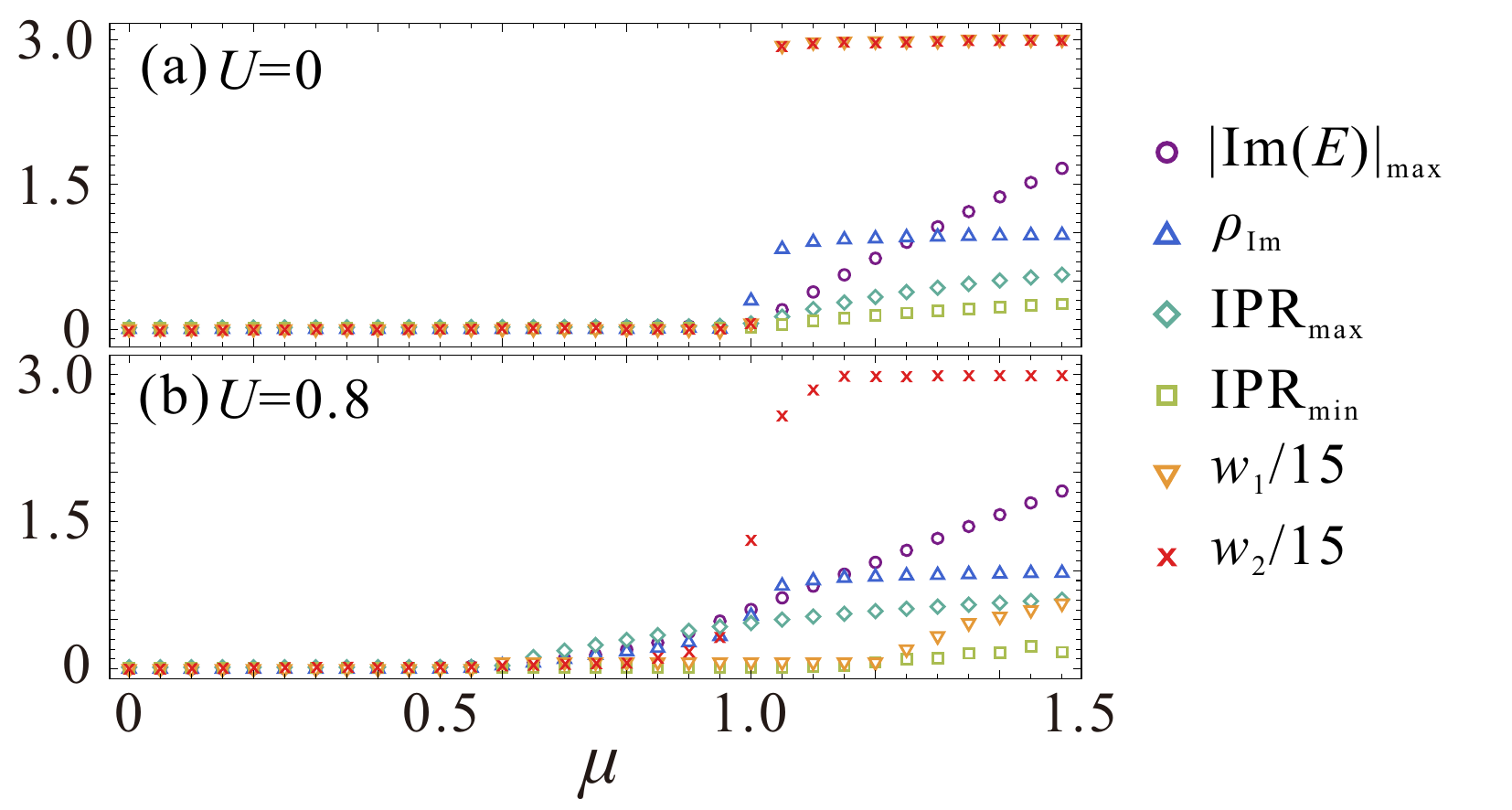}
		\caption{Collection of different quantities vs $\mu$ for the Bose-Hubbard NHAAH model with $J=1$ and $L=89$. (a) and (b) show the $|{\rm Im}(E)|_{\max}$, $\rho_{\rm Im}$, ${\rm IPR}_{\max}$, $\rm IPR_{\min}$, $w_1$ and $w_2$ with $U=0$ and $U=0.8$, respectively. The values of $w_{1,2}$ are rescaled by a factor $1/15$ to show them together with other quantities more compactly in the same figures.}\label{fig.merge}
	\end{figure}

	Fig.~\ref{fig.merge} illustrates two phase boundaries collectively determined by $|{\rm Im}(E)|_{\max}$, $\rho_{\rm Im}$, ${\rm IPR}_{\max}$, ${\rm IPR}_{\min}$, $w_1$, and $w_2$. Without interactions ($U=0$), the PT-symmetric extended phase has $(w_{1},w_{2})=(0,0)$ and also $|{\rm Im}(E)|_{\max} \simeq \rho_{\rm Im} \simeq {\rm IPR}_{\max} \simeq {\rm IPR}_{\min}\simeq0$. This persists until $\mu=1$, where a PT/localization transition occurs as both $(w_{1},w_{2})$ become finite and $|{\rm Im}(E)|_{\max}$, $\rho_{\rm Im}$, $\rm IPR_{\max}$, $\rm IPR_{\min}$ all deviate from zero. The system then enters a localized phase.
	
	With interactions [Fig.~\ref{fig.merge}(b)], we find $(w_{1},w_{2})=(0,0)$ in the PT-invariant extended phase, where $|{\rm Im}(E)|_{\max} \simeq \rho_{\rm Im} \simeq {\rm IPR}_{\max} \simeq {\rm IPR}_{\min}\simeq0$.
	Upon crossing the first transition point at $\mu_{c_1}\approx 0.55$, $w_{1}\rightarrow1$ and $|{\rm Im}(E)|_{\max}$, $\rho_{\rm Im}$, $\rm IPR_{\max}$ all deviate from zero, leading to a critical phase with winding numbers $(w_{1},w_{2})=(1,0)$.
    Such a phase does not exist when $U=0$, revealing the role of  interactions in creating new phases and transitions in NHQCs.
	As $\mu$ surpasses the second critical point $\mu_{c_2}\approx1.1$, the winding number $w_{1}$ jumps again, while $w_{2}$ goes from zero to a finite integer. The ${\rm IPR}_{\min}$ also deviates from zero and $\rho_{\rm Im}\simeq1$.
	The system thus enters a phase where all the eigenstates are localized, characterized by the winding numbers $(w_{1},w_{2})\neq(0,0)$. $(w_{1},w_{2})$ could thus distinguish the three possible phases with distinct spectrum and localization properties and describe the transitions among them.
	
	\section{Jump analysis in wavepacket dynamics}\label{app.WP}
	
	In Figs.~\ref{fig.WP.U0} and \ref{fig.WP.U08} of the main text, we observe certain anomalous jumps of wavepackets in the lattice.
    In this appendix, we analyze the reason behind these jumps and determine when they will reach their finales.

    In the critical phase, the wavepacket first diffuses from the initial position [$l_0=45$ in Fig.~\ref{fig.WP.U08}(b)], and then localizes near $l=48$. However, at $t=3577$, the wavepacket jumps from $l=48$ to the vicinity of $l=82$, where it remains localized thereafter.
    In the localized phase [Fig.~\ref{fig.WP.U08}(c)], the wavepacket center first starts from the site $l_0=45$ and localizes near $l=40$. However, at $t=1138$, it jumps to $l=61$. With the passage of time till around $t=5941$, the wavepacket once again jumps to the vicinity of $l=6$, and then localized thereafter.
	
	Our analysis indicates that this behavior stems from the longer lifetimes of excitations in non-Hermitian systems. The eigenenergies, denoted as $E_{j}$, consist of real part ${\rm Re}E_{j}$ and imaginary part ${\rm Im}E_{j}$. Eq.~(\ref{eq.psi_t}) can then be represented as
	\begin{eqnarray}
		|\Psi'(t) \rangle	&=&\sum_{j} c_{j} e^{-i E_{j} t} |\psi_j\rangle\\
		&=&\sum_{j} \langle \psi_j|\Psi_{0} \rangle e^{-i t {\rm Re}E_{j}}  e^{t {\rm Im}E_{j}}  |\psi_j\rangle.
		\label{eq.WP.E}
	\end{eqnarray}
	$|\Psi'(t) \rangle$ comprises three components $e^{-i t {\rm Re}E_{j}}$, $e^{t {\rm Im}E_{j}}$, and $\langle \psi_j|\Psi_{0} \rangle$.
	Due to the PT symmetry breaking, we have ${\rm Im}(E_{j})\neq0$ in the localized phase for most eigenstates. Consequently, $|\Psi'(t) \rangle$ is predominantly determined by the competition between $e^{t {\rm Im}E_{j}}$ and $\langle \psi_j|\Psi_{0} \rangle$.
	As time progresses, $e^{t {\rm Im}E_{j}}$ exhibits exponential growth if $E_j>0$. Therefore, in the long-time domain, the ultimate location of the wavepacket will be controlled by $e^{t {\rm Im}(E_{j})}$, especially the eigenstate with the largest imaginary part $({\rm Im}E_{j})_{\max}$. However, before the wavepacket reaches its final position, due to the interplay between $e^{t {\rm Im}E_{j}}$ and $\langle \psi_j|\Psi_{0} \rangle$, jump phenomena of wavepackets could occur in the critical and localized phases as depicted in Fig.~\ref{fig.WP.U08}.
	
	We found that within the system, the jumps in wavepacket dynamics occur only in the critical and localized phases. The difference between these phases compared with the extended phase is the PT symmetry, namely whether or not there are complex eigenenergies. The existence of complex eigenenergies due to non-Hermitian effects is thus the key reason behind the observed jumping of wavepacket locations.

	
	
	In Table \ref{table.WP}, we present four eigenstates $|\psi_j\rangle$ with $j=1,2,3,4$ of the system under the conditions $L=89$, $N=2$, $U=0.8$ and $\mu=1.5$. These states are the top four eigenstates obtained by sorting the imaginary parts of the system's eigenenergy ${\rm Im}E_j$ in descending order. $l_j$ denotes the site index with the maximum probability in the distribution of $|\psi_j\rangle$ and the absolute overlap of $|\psi_j\rangle$ with the initial state $|\Psi_0\rangle$.

    Based on the parameters in Table \ref{table.WP}, calculations reveal that each jump in Fig.~\ref{fig.WP.U08}(c) arises from the competition between $c_j$ and ${\rm Im}E_j$ as a function of time $t$. For example, according to the parameters in the table, at $t\approx1138$, the term $c_2  e^{t {\rm Im}E_{2}}$ corresponding to $|\psi_2\rangle$ surpasses the term $c_3  e^{t {\rm Im}E_{3}}$ corresponding to $|\psi_3\rangle$. This precisely corresponds to the wave packet jump occurring at $t\approx1100$ in Fig.~\ref{fig.WP.U08}(c).
    
	\begin{table}
		\begin{center}
		\caption{
			Parameters for the jump analysis in the Bose-Hubbard NHAAH model at $L=89, U=0.8,\mu =1.5$ [same as in Fig.~\ref{fig.WP.U08}(c)]. $|\psi_j\rangle$ with $j=1,2,3,4$ represent the top four eigenstates corresponding to the descending order of the imaginary parts of the eigenenergy ${\rm Im}E_j$. $l_j$ denotes the site index corresponding to the maximum probability in the probability distribution of $|\psi_j\rangle$. ${\rm Im}E_j$ is the value of the imaginary part of the eigenenergy after sorting in descending order. $|c_j|$ is the absolute overlap between $|\psi_j\rangle$ and the initial state $|\Psi_0\rangle$.
				\label{table.WP}}
			\begin{tabular}{cccc}
				\toprule[1pt]
				\qquad ${|\psi_{j}\rangle}$   &\qquad $l_j$  & \qquad${\rm Im}E_j$  & \qquad $|c_j|$  \\
				\midrule
				\qquad${| \psi_1\rangle}$& \qquad6& \qquad1.81162& \qquad$5.2534*10^{-17}$\\
				\qquad${| \psi_2\rangle}$& \qquad61& \qquad1.80848& \qquad$6.6423*10^{-9}$\\
				\qquad${| \psi_3\rangle}$& \qquad40&\qquad 1.79762&\qquad $1.54865*10^{-3}$\\
				\qquad${| \psi_4\rangle}$& \qquad27&\qquad 1.78668&\qquad $4.163*10^{-10}$\\
				\bottomrule[1pt]
			\end{tabular}
		\end{center}
	\end{table}

    %
    %
    
    			\begin{figure*}
    	\centering
    	\includegraphics[width=0.99\textwidth]{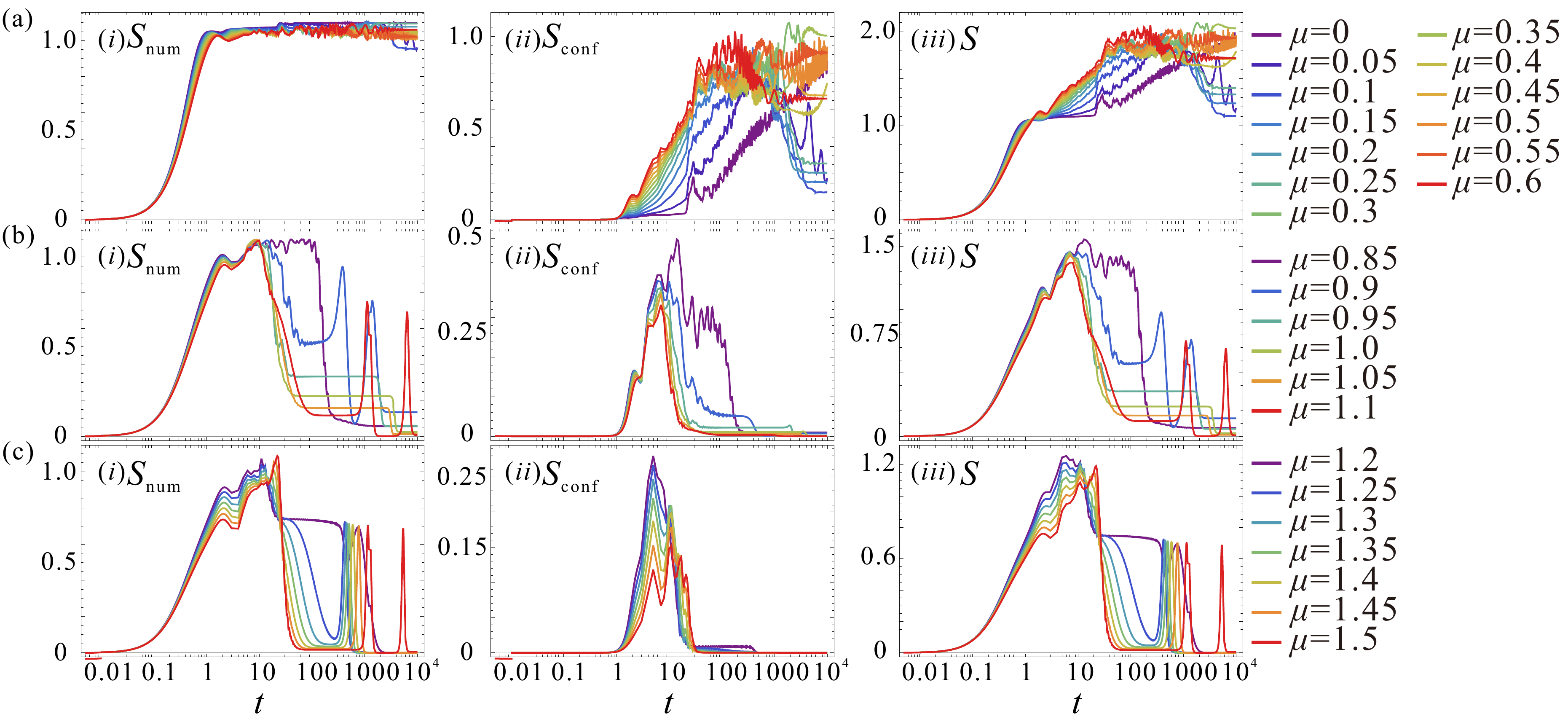}
    	\caption{EE dynamics of the Bose-Hubbard NHAAH model, with $U=0.8$ and the lattice size $L=89$. The values of $\mu$ for different solid lines are given on the right side. (a)($i$)--(a)($iii$), (b)($i$)--(b)($iii$) and (c)($i$)--(c)($iii$) show results in the extended, critical and localized phases, respectively. $(i)-(iii)$ in each row represent $S_{\rm num}(t)$, $S_{\rm conf}(t)$ and $S(t)$.}\label{fig.S.all.U08}
    \end{figure*}
    \begin{figure*}
    	\centering
    	\includegraphics[width=0.99\textwidth]{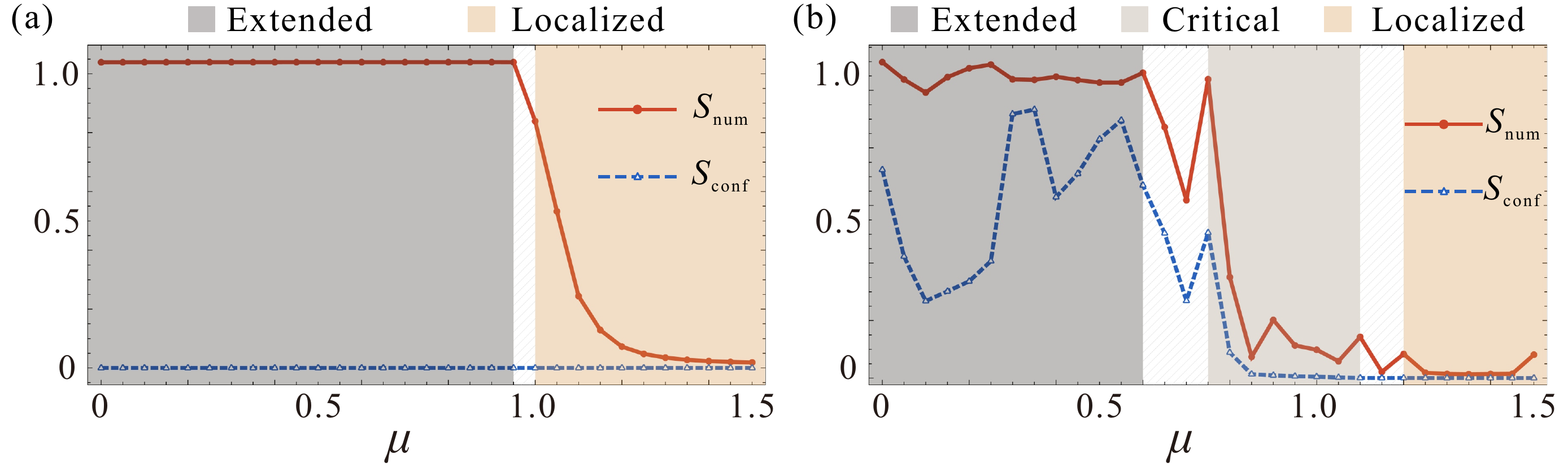}
    	\caption{Averaged EE $\overline{S}_{\rm num}$ (red solid lines) and $\overline{S}_{\rm conf}$ (blue dashed lines) vs $\mu$ for the Bose-Hubbard NHAAH model. The lattice size is $L=89$. (a) and (b) depict the cases with $U=0$ and $U=0.8$. The three colored regions denote extended, critical, and localized phases. Shaded regions in (b) highlight oscillations around phase boundaries.}\label{fig.entropy.mean}
    \end{figure*}
	
	\section{Numerical methods of calculating EE}\label{app.entropy.computation}
	In this appendix, we provide a detailed exposition of the method we used to compute the components $S_{\rm num}(t)$ and $S_{\rm conf}(t)$ of EE $S(t)$ following Refs.~\cite{Lukin_2019, Orito_2022}.
	The bipartite EE can be calculated by taking the partial trace over half of the system. This involves dividing the system of length $L$ into two subsystems of length $L/2$, denoted as $A$ and $B$. The positions $j$ within subsystem $A$ satisfy $j \in j_A = \{1, 2, \dots, L/2\}$, while positions $j$ within subsystem $B$ satisfy $j \in j_B = \{L/2 + 1, L/2 + 2, \dots, L\}$. By tracing out degrees of freedom belonging to subsystem $B$, we find the reduced density matrix $\Omega_A (t)$ of subsystem $A$.
	
	We begin by utilizing the density matrix
	\begin{equation}
		\Omega (t) = |\psi (t)\rangle \langle\psi (t)|,
		\label{eq.DM}
	\end{equation}
	and taking the partial trace as
	\begin{equation}
		\Omega_A (t)={\rm Tr}_{B}\Omega(t),
		\label{eq.RDM}
	\end{equation}
	In Eq.~(\ref{eq.DM}),
	$|\psi (t)\rangle$ represents the state evolved from the initial state $|\psi_0\rangle$ over a time $t$. Subsequently, the EE is found as
	\begin{equation}
		S_A(t) = - {\rm Tr}\ \Omega_A (t)\log\Omega_A (t),
	\end{equation}
	which is related to the eigenvalues $\lambda_{i}(t)$ of $\Omega_A (t)$ by
	\begin{equation}
		S_A(t) = - \sum_i \lambda_i(t) \log \lambda_i(t).
		\label{eq.S_A}
	\end{equation}
	
	If the particle number within subsystem $A$, denoted as
	\begin{equation}
		N_A=\sum_{j\in j_A} n_j,
	\end{equation}
	constitutes a conserved quantity, i.e.,
	\begin{equation}
		[\Omega_A,N_A]=0,
		\label{eq.commu}
	\end{equation}
	then the EE $S(t)=S_A(t)$ can be divided into two components, the number EE and configuration EE, i.e.,
	\begin{equation}
		S(t)=S_{\rm num}(t)+S_{\rm conf}(t),
	\end{equation}
	\begin{eqnarray}
		S_{\rm num}&=&-\sum_{N_A}p_{N_A} \log\ p_{N_A},
		\label{eq.S_num}	
		\\	S_{\rm conf}&=&-\sum_{N_A}\sum_{i}
		p_{N_A}\tilde{\lambda}^{(N_A)}_{i}\log\ \tilde{\lambda}^{(N_A)}_{i}.
		\label{eq.S_conf}
	\end{eqnarray}
	To obtain Eqs.~(\ref{eq.S_num}) and (\ref{eq.S_conf}),
	we first note that the $\Omega_A$
	is block-diagonal
	owing to the simultaneous diagonalizability of $N_A$ and $\Omega_A$ (see Eq.~\ref{eq.commu}), i.e.,
	\begin{equation}
		\Omega_A=\Omega_{N_1}\oplus\Omega_{N_2}\oplus\Omega_{N_3}\oplus\cdots,
		\label{RDM_bd}
	\end{equation}
	where
	$\Omega_{N_A}$ is a block-diagonal matrix of $\Omega_A$ restricted to the particle number $N_A$ within subsystem $A$.
	Consequently, the eigenvalues $\lambda_i$ of $\Omega_A$ are divided among the eigenvalues of $\Omega_{N_A}$. Let us denote one such eigenvalue as $\lambda^{(N_A)}_{i}$.
	Utilizing $\lambda^{(N_A)}_{i}$, we define $p_{N_A}$ and rescale the eigenvalues within this subspace to obtain normalized eigenvalues $\tilde{\lambda}^{(N_A)}_{i}$ as
	\begin{eqnarray}
		p_{N_A}&=&\sum_{i} \lambda^{(N_A)}_{i},
		\label{eq.pn}
		\\
		\tilde{\lambda}^{(N_A)}_{i}&=&\lambda^{(N_A)}_{i}/p_{N_A}.
		\label{eq.lambdan}
	\end{eqnarray}
	In contrast to Eq.~(\ref{eq.S_A}), in Eqs.~(\ref{eq.S_conf}), (\ref{eq.pn}), and (\ref{eq.lambdan}), the summation over $i$ is not extended across all eigenstates of the total reduced density matrix $\Omega_A$, but only within the subspace of fixed $N_A$. With these steps, we arrive at the expressions for both $S_{\rm num}(t)$ and $S_{\rm conf}(t)$.
	
	\section{Further details of EE dynamics}\label{app.entropy.detail}
	
	In Sec.~\ref{sec.dynamic.S} of the main text, we showed distinct dynamical behaviors of EE across the extended, critical, and localized phases. There, we present the results of EE dynamics for $\mu=0.5, 1, 1.5$ at $L=89$, $J=1$, and $U=0.8$. In this appendix, we showcase all results of EE for $\mu=0\sim0.6$, $0.85\sim1.1$ and $1.2\sim1.5$ in Fig.~\ref{fig.S.all.U08}. This comprehensive illustration captures the full details of EE dynamics in the aforementioned parameter regions.
	
	\section{Averaged EE and possible entanglement transitions}\label{app.entropy.mean}
	In order to provide a clear characterization of the behaviors of $S_{\rm num}$ and $S_{\rm conf}$ in the extended, critical and localized phases, we present the changes of mean number EE $\overline{S}_{\rm num}$ (red solid lines) and configuration-averaged EE $\overline{S}_{\rm conf}$ (blue dashed lines) for our system after reaching a steady state. They are illustrated for different $\mu$ in Fig.~\ref{fig.entropy.mean}.
	In the extended phase (deep grey), $\overline{S}_{\rm num}$ remains saturated, while $\overline{S}_{\rm conf}$ exhibits pronounced oscillations. As $\mu$ reaches the boundary between extended and critical phases, both $\overline{S}_{\rm num}$ and $\overline{S}_{\rm conf}$ display similar declined oscillatory behaviors. After entering the critical phase (light grey), $\overline{S}_{\rm num}$ and $\overline{S}_{\rm conf}$ both take finite values, showing a decreasing trend with respect to $\mu$. As the $\mu$ increases further so that the system enters into the localized phase (orange), we find $\overline{S}_{\rm num} = \overline{S}_{\rm conf} = 0$.
	$\overline{S}_{\rm num}$ and $\overline{S}_{\rm conf}$ thus exhibit different behaviors in the three phases, showing distinct entanglement properties within each phase. This observation is valuable for quantifying phase transitions in interacting non-Hermitian systems.

	\bibliography{nhbhm.bib}

\begin{thebibliography}{104}%
\makeatletter
\providecommand \@ifxundefined [1]{%
 \@ifx{#1\undefined}
}%
\providecommand \@ifnum [1]{%
 \ifnum #1\expandafter \@firstoftwo
 \else \expandafter \@secondoftwo
 \fi
}%
\providecommand \@ifx [1]{%
 \ifx #1\expandafter \@firstoftwo
 \else \expandafter \@secondoftwo
 \fi
}%
\providecommand \natexlab [1]{#1}%
\providecommand \enquote  [1]{``#1''}%
\providecommand \bibnamefont  [1]{#1}%
\providecommand \bibfnamefont [1]{#1}%
\providecommand \citenamefont [1]{#1}%
\providecommand \href@noop [0]{\@secondoftwo}%
\providecommand \href [0]{\begingroup \@sanitize@url \@href}%
\providecommand \@href[1]{\@@startlink{#1}\@@href}%
\providecommand \@@href[1]{\endgroup#1\@@endlink}%
\providecommand \@sanitize@url [0]{\catcode `\\12\catcode `\$12\catcode
  `\&12\catcode `\#12\catcode `\^12\catcode `\_12\catcode `\%12\relax}%
\providecommand \@@startlink[1]{}%
\providecommand \@@endlink[0]{}%
\providecommand \url  [0]{\begingroup\@sanitize@url \@url }%
\providecommand \@url [1]{\endgroup\@href {#1}{\urlprefix }}%
\providecommand \urlprefix  [0]{URL }%
\providecommand \Eprint [0]{\href }%
\providecommand \doibase [0]{https://doi.org/}%
\providecommand \selectlanguage [0]{\@gobble}%
\providecommand \bibinfo  [0]{\@secondoftwo}%
\providecommand \bibfield  [0]{\@secondoftwo}%
\providecommand \translation [1]{[#1]}%
\providecommand \BibitemOpen [0]{}%
\providecommand \bibitemStop [0]{}%
\providecommand \bibitemNoStop [0]{.\EOS\space}%
\providecommand \EOS [0]{\spacefactor3000\relax}%
\providecommand \BibitemShut  [1]{\csname bibitem#1\endcsname}%
\let\auto@bib@innerbib\@empty
\bibitem [{\citenamefont {Ashida}\ \emph {et~al.}(2020)\citenamefont {Ashida},
  \citenamefont {Gong},\ and\ \citenamefont {Ueda}}]{Ashida_2020}%
  \BibitemOpen
  \bibfield  {author} {\bibinfo {author} {\bibfnamefont {Y.}~\bibnamefont
  {Ashida}}, \bibinfo {author} {\bibfnamefont {Z.}~\bibnamefont {Gong}},\ and\
  \bibinfo {author} {\bibfnamefont {M.}~\bibnamefont {Ueda}},\ }\bibfield
  {title} {\bibinfo {title} {Non-{{Hermitian}} physics},\ }\href
  {https://doi.org/10.1080/00018732.2021.1876991} {\bibfield  {journal}
  {\bibinfo  {journal} {Adv. Phys.}\ }\textbf {\bibinfo {volume} {69}},\
  \bibinfo {pages} {249} (\bibinfo {year} {2020})}\BibitemShut {NoStop}%
\bibitem [{\citenamefont {{El-Ganainy}}\ \emph {et~al.}(2018)\citenamefont
  {{El-Ganainy}}, \citenamefont {Makris}, \citenamefont {Khajavikhan},
  \citenamefont {Musslimani}, \citenamefont {Rotter},\ and\ \citenamefont
  {Christodoulides}}]{El-Ganainy_2018}%
  \BibitemOpen
  \bibfield  {author} {\bibinfo {author} {\bibfnamefont {R.}~\bibnamefont
  {{El-Ganainy}}}, \bibinfo {author} {\bibfnamefont {K.}~\bibnamefont
  {Makris}}, \bibinfo {author} {\bibfnamefont {M.}~\bibnamefont {Khajavikhan}},
  \bibinfo {author} {\bibfnamefont {Z.}~\bibnamefont {Musslimani}}, \bibinfo
  {author} {\bibfnamefont {S.}~\bibnamefont {Rotter}},\ and\ \bibinfo {author}
  {\bibfnamefont {D.}~\bibnamefont {Christodoulides}},\ }\bibfield  {title}
  {\bibinfo {title} {Non-{{Hermitian}} physics and {{PT}} symmetry},\ }\href
  {https://doi.org/10.1038/nphys4323} {\bibfield  {journal} {\bibinfo
  {journal} {Nat. Phys.}\ }\textbf {\bibinfo {volume} {14}},\ \bibinfo {pages}
  {11} (\bibinfo {year} {2018})}\BibitemShut {NoStop}%
\bibitem [{\citenamefont {Gong}\ \emph {et~al.}(2018)\citenamefont {Gong},
  \citenamefont {Ashida}, \citenamefont {Kawabata}, \citenamefont {Takasan},
  \citenamefont {Higashikawa},\ and\ \citenamefont {Ueda}}]{Gong_2018}%
  \BibitemOpen
  \bibfield  {author} {\bibinfo {author} {\bibfnamefont {Z.}~\bibnamefont
  {Gong}}, \bibinfo {author} {\bibfnamefont {Y.}~\bibnamefont {Ashida}},
  \bibinfo {author} {\bibfnamefont {K.}~\bibnamefont {Kawabata}}, \bibinfo
  {author} {\bibfnamefont {K.}~\bibnamefont {Takasan}}, \bibinfo {author}
  {\bibfnamefont {S.}~\bibnamefont {Higashikawa}},\ and\ \bibinfo {author}
  {\bibfnamefont {M.}~\bibnamefont {Ueda}},\ }\bibfield  {title} {\bibinfo
  {title} {Topological {{Phases}} of {{Non-Hermitian Systems}}},\ }\href
  {https://doi.org/10.1103/PhysRevX.8.031079} {\bibfield  {journal} {\bibinfo
  {journal} {Phys. Rev. X}\ }\textbf {\bibinfo {volume} {8}},\ \bibinfo {pages}
  {031079} (\bibinfo {year} {2018})}\BibitemShut {NoStop}%
\bibitem [{\citenamefont {Bergholtz}\ \emph {et~al.}(2021)\citenamefont
  {Bergholtz}, \citenamefont {Budich},\ and\ \citenamefont
  {Kunst}}]{Bergholtz_2021}%
  \BibitemOpen
  \bibfield  {author} {\bibinfo {author} {\bibfnamefont {E.~J.}\ \bibnamefont
  {Bergholtz}}, \bibinfo {author} {\bibfnamefont {J.~C.}\ \bibnamefont
  {Budich}},\ and\ \bibinfo {author} {\bibfnamefont {F.~K.}\ \bibnamefont
  {Kunst}},\ }\bibfield  {title} {\bibinfo {title} {Exceptional topology of
  non-hermitian systems},\ }\href
  {https://doi.org/10.1103/RevModPhys.93.015005} {\bibfield  {journal}
  {\bibinfo  {journal} {Rev. Mod. Phys.}\ }\textbf {\bibinfo {volume} {93}},\
  \bibinfo {pages} {015005} (\bibinfo {year} {2021})}\BibitemShut {NoStop}%
\bibitem [{\citenamefont {Martinez~Alvarez}\ \emph
  {et~al.}(2018{\natexlab{a}})\citenamefont {Martinez~Alvarez}, \citenamefont
  {Barrios~Vargas}, \citenamefont {Berdakin},\ and\ \citenamefont
  {Foa~Torres}}]{MartinezAlvarez_2018}%
  \BibitemOpen
  \bibfield  {author} {\bibinfo {author} {\bibfnamefont {V.~M.}\ \bibnamefont
  {Martinez~Alvarez}}, \bibinfo {author} {\bibfnamefont {J.~E.}\ \bibnamefont
  {Barrios~Vargas}}, \bibinfo {author} {\bibfnamefont {M.}~\bibnamefont
  {Berdakin}},\ and\ \bibinfo {author} {\bibfnamefont {L.~E.~F.}\ \bibnamefont
  {Foa~Torres}},\ }\bibfield  {title} {\bibinfo {title} {Topological states of
  non-{{Hermitian}} systems},\ }\href
  {https://doi.org/10.1140/epjst/e2018-800091-5} {\bibfield  {journal}
  {\bibinfo  {journal} {Eur. Phys. J. Special Topics}\ }\textbf {\bibinfo
  {volume} {227}},\ \bibinfo {pages} {1295} (\bibinfo {year}
  {2018}{\natexlab{a}})}\BibitemShut {NoStop}%
\bibitem [{\citenamefont {Ghatak}\ and\ \citenamefont
  {Das}(2019)}]{Ghatak_2019}%
  \BibitemOpen
  \bibfield  {author} {\bibinfo {author} {\bibfnamefont {A.}~\bibnamefont
  {Ghatak}}\ and\ \bibinfo {author} {\bibfnamefont {T.}~\bibnamefont {Das}},\
  }\bibfield  {title} {\bibinfo {title} {New topological invariants in
  non-{{Hermitian}} systems},\ }\href
  {https://doi.org/10.1088/1361-648X/ab11b3} {\bibfield  {journal} {\bibinfo
  {journal} {J. Phys.: Condens. Matter}\ }\textbf {\bibinfo {volume} {31}},\
  \bibinfo {pages} {263001} (\bibinfo {year} {2019})}\BibitemShut {NoStop}%
\bibitem [{\citenamefont {Bender}\ and\ \citenamefont
  {Boettcher}(1998)}]{Bender_1998}%
  \BibitemOpen
  \bibfield  {author} {\bibinfo {author} {\bibfnamefont {C.~M.}\ \bibnamefont
  {Bender}}\ and\ \bibinfo {author} {\bibfnamefont {S.}~\bibnamefont
  {Boettcher}},\ }\bibfield  {title} {\bibinfo {title} {Real {{Spectra}} in
  {{Non-Hermitian Hamiltonians Having}} $\mathcal{PT}$ {{Symmetry}}},\ }\href
  {https://doi.org/10.1103/PhysRevLett.80.5243} {\bibfield  {journal} {\bibinfo
   {journal} {Phys. Rev. Lett.}\ }\textbf {\bibinfo {volume} {80}},\ \bibinfo
  {pages} {5243} (\bibinfo {year} {1998})}\BibitemShut {NoStop}%
\bibitem [{\citenamefont {Berry}(2004)}]{Berry_2004}%
  \BibitemOpen
  \bibfield  {author} {\bibinfo {author} {\bibfnamefont {M.~V.}\ \bibnamefont
  {Berry}},\ }\bibfield  {title} {\bibinfo {title} {Physics of {{Nonhermitian
  Degeneracies}}},\ }\href {https://doi.org/10.1023/B:CJOP.0000044002.05657.04}
  {\bibfield  {journal} {\bibinfo  {journal} {Czech. J. Phys.}\ }\textbf
  {\bibinfo {volume} {54}},\ \bibinfo {pages} {1039} (\bibinfo {year}
  {2004})}\BibitemShut {NoStop}%
\bibitem [{\citenamefont {Heiss}(2012)}]{Heiss_2012}%
  \BibitemOpen
  \bibfield  {author} {\bibinfo {author} {\bibfnamefont {W.~D.}\ \bibnamefont
  {Heiss}},\ }\bibfield  {title} {\bibinfo {title} {The physics of exceptional
  points},\ }\href {https://doi.org/10.1088/1751-8113/45/44/444016} {\bibfield
  {journal} {\bibinfo  {journal} {J. Phys. A: Math. Theor.}\ }\textbf {\bibinfo
  {volume} {45}},\ \bibinfo {pages} {444016} (\bibinfo {year}
  {2012})}\BibitemShut {NoStop}%
\bibitem [{\citenamefont {Miri}\ and\ \citenamefont
  {Al{\`u}}(2019)}]{Miri_2019}%
  \BibitemOpen
  \bibfield  {author} {\bibinfo {author} {\bibfnamefont {M.-A.}\ \bibnamefont
  {Miri}}\ and\ \bibinfo {author} {\bibfnamefont {A.}~\bibnamefont {Al{\`u}}},\
  }\bibfield  {title} {\bibinfo {title} {Exceptional points in optics and
  photonics},\ }\href {https://doi.org/10.1126/science.aar7709} {\bibfield
  {journal} {\bibinfo  {journal} {Science}\ }\textbf {\bibinfo {volume}
  {363}},\ \bibinfo {pages} {eaar7709} (\bibinfo {year} {2019})}\BibitemShut
  {NoStop}%
\bibitem [{\citenamefont {Yao}\ and\ \citenamefont {Wang}(2018)}]{Yao_2018}%
  \BibitemOpen
  \bibfield  {author} {\bibinfo {author} {\bibfnamefont {S.}~\bibnamefont
  {Yao}}\ and\ \bibinfo {author} {\bibfnamefont {Z.}~\bibnamefont {Wang}},\
  }\bibfield  {title} {\bibinfo {title} {Edge {{States}} and {{Topological
  Invariants}} of {{Non-Hermitian Systems}}},\ }\href
  {https://doi.org/10.1103/PhysRevLett.121.086803} {\bibfield  {journal}
  {\bibinfo  {journal} {Phys. Rev. Lett.}\ }\textbf {\bibinfo {volume} {121}},\
  \bibinfo {pages} {086803} (\bibinfo {year} {2018})}\BibitemShut {NoStop}%
\bibitem [{\citenamefont {Kunst}\ \emph {et~al.}(2018)\citenamefont {Kunst},
  \citenamefont {Edvardsson}, \citenamefont {Budich},\ and\ \citenamefont
  {Bergholtz}}]{Kunst_2018}%
  \BibitemOpen
  \bibfield  {author} {\bibinfo {author} {\bibfnamefont {F.~K.}\ \bibnamefont
  {Kunst}}, \bibinfo {author} {\bibfnamefont {E.}~\bibnamefont {Edvardsson}},
  \bibinfo {author} {\bibfnamefont {J.~C.}\ \bibnamefont {Budich}},\ and\
  \bibinfo {author} {\bibfnamefont {E.~J.}\ \bibnamefont {Bergholtz}},\
  }\bibfield  {title} {\bibinfo {title} {Biorthogonal {{Bulk-Boundary
  Correspondence}} in {{Non-Hermitian Systems}}},\ }\href
  {https://doi.org/10.1103/PhysRevLett.121.026808} {\bibfield  {journal}
  {\bibinfo  {journal} {Phys. Rev. Lett.}\ }\textbf {\bibinfo {volume} {121}},\
  \bibinfo {pages} {026808} (\bibinfo {year} {2018})}\BibitemShut {NoStop}%
\bibitem [{\citenamefont {Martinez~Alvarez}\ \emph
  {et~al.}(2018{\natexlab{b}})\citenamefont {Martinez~Alvarez}, \citenamefont
  {Barrios~Vargas},\ and\ \citenamefont {Foa~Torres}}]{MartinezAlvarez_2018a}%
  \BibitemOpen
  \bibfield  {author} {\bibinfo {author} {\bibfnamefont {V.~M.}\ \bibnamefont
  {Martinez~Alvarez}}, \bibinfo {author} {\bibfnamefont {J.~E.}\ \bibnamefont
  {Barrios~Vargas}},\ and\ \bibinfo {author} {\bibfnamefont {L.~E.~F.}\
  \bibnamefont {Foa~Torres}},\ }\bibfield  {title} {\bibinfo {title}
  {Non-{{Hermitian}} robust edge states in one dimension: {{Anomalous}}
  localization and eigenspace condensation at exceptional points},\ }\href
  {https://doi.org/10.1103/PhysRevB.97.121401} {\bibfield  {journal} {\bibinfo
  {journal} {Phys. Rev. B}\ }\textbf {\bibinfo {volume} {97}},\ \bibinfo
  {pages} {121401} (\bibinfo {year} {2018}{\natexlab{b}})}\BibitemShut
  {NoStop}%
\bibitem [{\citenamefont {Lee}\ and\ \citenamefont {Thomale}(2019)}]{Lee_2019}%
  \BibitemOpen
  \bibfield  {author} {\bibinfo {author} {\bibfnamefont {C.~H.}\ \bibnamefont
  {Lee}}\ and\ \bibinfo {author} {\bibfnamefont {R.}~\bibnamefont {Thomale}},\
  }\bibfield  {title} {\bibinfo {title} {Anatomy of skin modes and topology in
  non-{{Hermitian}} systems},\ }\href
  {https://doi.org/10.1103/PhysRevB.99.201103} {\bibfield  {journal} {\bibinfo
  {journal} {Phys. Rev. B}\ }\textbf {\bibinfo {volume} {99}},\ \bibinfo
  {pages} {201103} (\bibinfo {year} {2019})}\BibitemShut {NoStop}%
\bibitem [{\citenamefont {Hatano}\ and\ \citenamefont
  {Nelson}(1996)}]{Hatano_1996}%
  \BibitemOpen
  \bibfield  {author} {\bibinfo {author} {\bibfnamefont {N.}~\bibnamefont
  {Hatano}}\ and\ \bibinfo {author} {\bibfnamefont {D.~R.}\ \bibnamefont
  {Nelson}},\ }\bibfield  {title} {\bibinfo {title} {Localization
  {{Transitions}} in {{Non-Hermitian Quantum Mechanics}}},\ }\href
  {https://doi.org/10.1103/PhysRevLett.77.570} {\bibfield  {journal} {\bibinfo
  {journal} {Phys. Rev. Lett.}\ }\textbf {\bibinfo {volume} {77}},\ \bibinfo
  {pages} {570} (\bibinfo {year} {1996})}\BibitemShut {NoStop}%
\bibitem [{\citenamefont {Feinberg}\ and\ \citenamefont
  {Zee}(1999{\natexlab{a}})}]{Feinberg_1999}%
  \BibitemOpen
  \bibfield  {author} {\bibinfo {author} {\bibfnamefont {J.}~\bibnamefont
  {Feinberg}}\ and\ \bibinfo {author} {\bibfnamefont {A.}~\bibnamefont {Zee}},\
  }\bibfield  {title} {\bibinfo {title} {Non-{{Hermitian}} localization and
  delocalization},\ }\href {https://doi.org/10.1103/PhysRevE.59.6433}
  {\bibfield  {journal} {\bibinfo  {journal} {Phys. Rev. E}\ }\textbf {\bibinfo
  {volume} {59}},\ \bibinfo {pages} {6433} (\bibinfo {year}
  {1999}{\natexlab{a}})}\BibitemShut {NoStop}%
\bibitem [{\citenamefont {Feinberg}\ and\ \citenamefont
  {Zee}(1999{\natexlab{b}})}]{Feinberg_1999a}%
  \BibitemOpen
  \bibfield  {author} {\bibinfo {author} {\bibfnamefont {J.}~\bibnamefont
  {Feinberg}}\ and\ \bibinfo {author} {\bibfnamefont {A.}~\bibnamefont {Zee}},\
  }\bibfield  {title} {\bibinfo {title} {Spectral curves of non-hermitian
  hamiltonians},\ }\href {https://doi.org/10.1016/S0550-3213(99)00246-1}
  {\bibfield  {journal} {\bibinfo  {journal} {Nucl. Phys. B}\ }\textbf
  {\bibinfo {volume} {552}},\ \bibinfo {pages} {599} (\bibinfo {year}
  {1999}{\natexlab{b}})}\BibitemShut {NoStop}%
\bibitem [{\citenamefont {Hatano}\ and\ \citenamefont
  {Feinberg}(2016)}]{Hatano_2016}%
  \BibitemOpen
  \bibfield  {author} {\bibinfo {author} {\bibfnamefont {N.}~\bibnamefont
  {Hatano}}\ and\ \bibinfo {author} {\bibfnamefont {J.}~\bibnamefont
  {Feinberg}},\ }\bibfield  {title} {\bibinfo {title} {Chebyshev-polynomial
  expansion of the localization length of {{Hermitian}} and non-{{Hermitian}}
  random chains},\ }\href {https://doi.org/10.1103/PhysRevE.94.063305}
  {\bibfield  {journal} {\bibinfo  {journal} {Phys. Rev. E}\ }\textbf {\bibinfo
  {volume} {94}},\ \bibinfo {pages} {063305} (\bibinfo {year}
  {2016})}\BibitemShut {NoStop}%
\bibitem [{\citenamefont {Kawabata}\ \emph {et~al.}(2019)\citenamefont
  {Kawabata}, \citenamefont {Shiozaki}, \citenamefont {Ueda},\ and\
  \citenamefont {Sato}}]{Kawabata_2019b}%
  \BibitemOpen
  \bibfield  {author} {\bibinfo {author} {\bibfnamefont {K.}~\bibnamefont
  {Kawabata}}, \bibinfo {author} {\bibfnamefont {K.}~\bibnamefont {Shiozaki}},
  \bibinfo {author} {\bibfnamefont {M.}~\bibnamefont {Ueda}},\ and\ \bibinfo
  {author} {\bibfnamefont {M.}~\bibnamefont {Sato}},\ }\bibfield  {title}
  {\bibinfo {title} {Symmetry and {{Topology}} in {{Non-Hermitian Physics}}},\
  }\href {https://doi.org/10.1103/PhysRevX.9.041015} {\bibfield  {journal}
  {\bibinfo  {journal} {Phys. Rev. X}\ }\textbf {\bibinfo {volume} {9}},\
  \bibinfo {pages} {041015} (\bibinfo {year} {2019})}\BibitemShut {NoStop}%
\bibitem [{\citenamefont {Zhou}\ and\ \citenamefont {Lee}(2019)}]{Zhou_2019}%
  \BibitemOpen
  \bibfield  {author} {\bibinfo {author} {\bibfnamefont {H.}~\bibnamefont
  {Zhou}}\ and\ \bibinfo {author} {\bibfnamefont {J.~Y.}\ \bibnamefont {Lee}},\
  }\bibfield  {title} {\bibinfo {title} {Periodic table for topological bands
  with non-{{Hermitian}} symmetries},\ }\href
  {https://doi.org/10.1103/PhysRevB.99.235112} {\bibfield  {journal} {\bibinfo
  {journal} {Phys. Rev. B}\ }\textbf {\bibinfo {volume} {99}},\ \bibinfo
  {pages} {235112} (\bibinfo {year} {2019})}\BibitemShut {NoStop}%
\bibitem [{\citenamefont {Wojcik}\ \emph {et~al.}(2020)\citenamefont {Wojcik},
  \citenamefont {Sun}, \citenamefont {Bzdu{\v s}ek},\ and\ \citenamefont
  {Fan}}]{Wojcik_2020}%
  \BibitemOpen
  \bibfield  {author} {\bibinfo {author} {\bibfnamefont {C.~C.}\ \bibnamefont
  {Wojcik}}, \bibinfo {author} {\bibfnamefont {X.-Q.}\ \bibnamefont {Sun}},
  \bibinfo {author} {\bibfnamefont {T.}~\bibnamefont {Bzdu{\v s}ek}},\ and\
  \bibinfo {author} {\bibfnamefont {S.}~\bibnamefont {Fan}},\ }\bibfield
  {title} {\bibinfo {title} {Homotopy characterization of non-{{Hermitian
  Hamiltonians}}},\ }\href {https://doi.org/10.1103/PhysRevB.101.205417}
  {\bibfield  {journal} {\bibinfo  {journal} {Phys. Rev. B}\ }\textbf {\bibinfo
  {volume} {101}},\ \bibinfo {pages} {205417} (\bibinfo {year}
  {2020})}\BibitemShut {NoStop}%
\bibitem [{\citenamefont {Shiozaki}\ and\ \citenamefont
  {Ono}(2021)}]{Shiozaki_2021}%
  \BibitemOpen
  \bibfield  {author} {\bibinfo {author} {\bibfnamefont {K.}~\bibnamefont
  {Shiozaki}}\ and\ \bibinfo {author} {\bibfnamefont {S.}~\bibnamefont {Ono}},\
  }\bibfield  {title} {\bibinfo {title} {Symmetry indicator in
  non-{{Hermitian}} systems},\ }\href
  {https://doi.org/10.1103/PhysRevB.104.035424} {\bibfield  {journal} {\bibinfo
   {journal} {Phys. Rev. B}\ }\textbf {\bibinfo {volume} {104}},\ \bibinfo
  {pages} {035424} (\bibinfo {year} {2021})}\BibitemShut {NoStop}%
\bibitem [{\citenamefont {Gou}\ \emph {et~al.}(2020)\citenamefont {Gou},
  \citenamefont {Chen}, \citenamefont {Xie}, \citenamefont {Xiao},
  \citenamefont {Deng}, \citenamefont {Gadway}, \citenamefont {Yi},\ and\
  \citenamefont {Yan}}]{Gou_2020}%
  \BibitemOpen
  \bibfield  {author} {\bibinfo {author} {\bibfnamefont {W.}~\bibnamefont
  {Gou}}, \bibinfo {author} {\bibfnamefont {T.}~\bibnamefont {Chen}}, \bibinfo
  {author} {\bibfnamefont {D.}~\bibnamefont {Xie}}, \bibinfo {author}
  {\bibfnamefont {T.}~\bibnamefont {Xiao}}, \bibinfo {author} {\bibfnamefont
  {T.-S.}\ \bibnamefont {Deng}}, \bibinfo {author} {\bibfnamefont
  {B.}~\bibnamefont {Gadway}}, \bibinfo {author} {\bibfnamefont
  {W.}~\bibnamefont {Yi}},\ and\ \bibinfo {author} {\bibfnamefont
  {B.}~\bibnamefont {Yan}},\ }\bibfield  {title} {\bibinfo {title} {Tunable
  {{Nonreciprocal Quantum Transport}} through a {{Dissipative Aharonov-Bohm
  Ring}} in {{Ultracold Atoms}}},\ }\href
  {https://doi.org/10.1103/PhysRevLett.124.070402} {\bibfield  {journal}
  {\bibinfo  {journal} {Phys. Rev. Lett.}\ }\textbf {\bibinfo {volume} {124}},\
  \bibinfo {pages} {070402} (\bibinfo {year} {2020})}\BibitemShut {NoStop}%
\bibitem [{\citenamefont {Li}\ \emph {et~al.}(2019)\citenamefont {Li},
  \citenamefont {Harter}, \citenamefont {Liu}, \citenamefont {{de Melo}},
  \citenamefont {Joglekar},\ and\ \citenamefont {Luo}}]{Li_2019}%
  \BibitemOpen
  \bibfield  {author} {\bibinfo {author} {\bibfnamefont {J.}~\bibnamefont
  {Li}}, \bibinfo {author} {\bibfnamefont {A.~K.}\ \bibnamefont {Harter}},
  \bibinfo {author} {\bibfnamefont {J.}~\bibnamefont {Liu}}, \bibinfo {author}
  {\bibfnamefont {L.}~\bibnamefont {{de Melo}}}, \bibinfo {author}
  {\bibfnamefont {Y.~N.}\ \bibnamefont {Joglekar}},\ and\ \bibinfo {author}
  {\bibfnamefont {L.}~\bibnamefont {Luo}},\ }\bibfield  {title} {\bibinfo
  {title} {Observation of parity-time symmetry breaking transitions in a
  dissipative {{Floquet}} system of ultracold atoms},\ }\href
  {https://doi.org/10.1038/s41467-019-08596-1} {\bibfield  {journal} {\bibinfo
  {journal} {Nat. Commun.}\ }\textbf {\bibinfo {volume} {10}},\ \bibinfo
  {pages} {855} (\bibinfo {year} {2019})}\BibitemShut {NoStop}%
\bibitem [{\citenamefont {Xu}\ \emph {et~al.}(2017)\citenamefont {Xu},
  \citenamefont {Wang},\ and\ \citenamefont {Duan}}]{Xu_2017}%
  \BibitemOpen
  \bibfield  {author} {\bibinfo {author} {\bibfnamefont {Y.}~\bibnamefont
  {Xu}}, \bibinfo {author} {\bibfnamefont {S.-T.}\ \bibnamefont {Wang}},\ and\
  \bibinfo {author} {\bibfnamefont {L.-M.}\ \bibnamefont {Duan}},\ }\bibfield
  {title} {\bibinfo {title} {Weyl {{Exceptional Rings}} in a
  {{Three-Dimensional Dissipative Cold Atomic Gas}}},\ }\href
  {https://doi.org/10.1103/PhysRevLett.118.045701} {\bibfield  {journal}
  {\bibinfo  {journal} {Phys. Rev. Lett.}\ }\textbf {\bibinfo {volume} {118}},\
  \bibinfo {pages} {045701} (\bibinfo {year} {2017})}\BibitemShut {NoStop}%
\bibitem [{\citenamefont {Lapp}\ \emph {et~al.}(2019)\citenamefont {Lapp},
  \citenamefont {Ang'ong'a}, \citenamefont {An},\ and\ \citenamefont
  {Gadway}}]{Lapp_2019}%
  \BibitemOpen
  \bibfield  {author} {\bibinfo {author} {\bibfnamefont {S.}~\bibnamefont
  {Lapp}}, \bibinfo {author} {\bibfnamefont {J.}~\bibnamefont {Ang'ong'a}},
  \bibinfo {author} {\bibfnamefont {F.~A.}\ \bibnamefont {An}},\ and\ \bibinfo
  {author} {\bibfnamefont {B.}~\bibnamefont {Gadway}},\ }\bibfield  {title}
  {\bibinfo {title} {Engineering tunable local loss in a synthetic lattice of
  momentum states},\ }\href {https://doi.org/10.1088/1367-2630/ab1147}
  {\bibfield  {journal} {\bibinfo  {journal} {New J. Phys.}\ }\textbf {\bibinfo
  {volume} {21}},\ \bibinfo {pages} {045006} (\bibinfo {year}
  {2019})}\BibitemShut {NoStop}%
\bibitem [{\citenamefont {Ren}\ \emph {et~al.}(2022)\citenamefont {Ren},
  \citenamefont {Liu}, \citenamefont {Zhao}, \citenamefont {He}, \citenamefont
  {Pak}, \citenamefont {Li},\ and\ \citenamefont {Jo}}]{Ren_2022}%
  \BibitemOpen
  \bibfield  {author} {\bibinfo {author} {\bibfnamefont {Z.}~\bibnamefont
  {Ren}}, \bibinfo {author} {\bibfnamefont {D.}~\bibnamefont {Liu}}, \bibinfo
  {author} {\bibfnamefont {E.}~\bibnamefont {Zhao}}, \bibinfo {author}
  {\bibfnamefont {C.}~\bibnamefont {He}}, \bibinfo {author} {\bibfnamefont
  {K.~K.}\ \bibnamefont {Pak}}, \bibinfo {author} {\bibfnamefont
  {J.}~\bibnamefont {Li}},\ and\ \bibinfo {author} {\bibfnamefont {G.-B.}\
  \bibnamefont {Jo}},\ }\bibfield  {title} {\bibinfo {title} {Chiral control of
  quantum states in non-{{Hermitian}} spin\textendash orbit-coupled fermions},\
  }\href {https://doi.org/10.1038/s41567-021-01491-x} {\bibfield  {journal}
  {\bibinfo  {journal} {Nat. Phys.}\ }\textbf {\bibinfo {volume} {18}},\
  \bibinfo {pages} {385} (\bibinfo {year} {2022})}\BibitemShut {NoStop}%
\bibitem [{\citenamefont {Zeuner}\ \emph {et~al.}(2015)\citenamefont {Zeuner},
  \citenamefont {Rechtsman}, \citenamefont {Plotnik}, \citenamefont {Lumer},
  \citenamefont {Nolte}, \citenamefont {Rudner}, \citenamefont {Segev},\ and\
  \citenamefont {Szameit}}]{Zeuner_2015}%
  \BibitemOpen
  \bibfield  {author} {\bibinfo {author} {\bibfnamefont {J.~M.}\ \bibnamefont
  {Zeuner}}, \bibinfo {author} {\bibfnamefont {M.~C.}\ \bibnamefont
  {Rechtsman}}, \bibinfo {author} {\bibfnamefont {Y.}~\bibnamefont {Plotnik}},
  \bibinfo {author} {\bibfnamefont {Y.}~\bibnamefont {Lumer}}, \bibinfo
  {author} {\bibfnamefont {S.}~\bibnamefont {Nolte}}, \bibinfo {author}
  {\bibfnamefont {M.~S.}\ \bibnamefont {Rudner}}, \bibinfo {author}
  {\bibfnamefont {M.}~\bibnamefont {Segev}},\ and\ \bibinfo {author}
  {\bibfnamefont {A.}~\bibnamefont {Szameit}},\ }\bibfield  {title} {\bibinfo
  {title} {Observation of a {{Topological Transition}} in the {{Bulk}} of a
  {{Non-Hermitian System}}},\ }\href
  {https://doi.org/10.1103/PhysRevLett.115.040402} {\bibfield  {journal}
  {\bibinfo  {journal} {Phys. Rev. Lett.}\ }\textbf {\bibinfo {volume} {115}},\
  \bibinfo {pages} {040402} (\bibinfo {year} {2015})}\BibitemShut {NoStop}%
\bibitem [{\citenamefont {Weimann}\ \emph {et~al.}(2017)\citenamefont
  {Weimann}, \citenamefont {Kremer}, \citenamefont {Plotnik}, \citenamefont
  {Lumer}, \citenamefont {Nolte}, \citenamefont {Makris}, \citenamefont
  {Segev}, \citenamefont {Rechtsman},\ and\ \citenamefont
  {Szameit}}]{Weimann_2017}%
  \BibitemOpen
  \bibfield  {author} {\bibinfo {author} {\bibfnamefont {S.}~\bibnamefont
  {Weimann}}, \bibinfo {author} {\bibfnamefont {M.}~\bibnamefont {Kremer}},
  \bibinfo {author} {\bibfnamefont {Y.}~\bibnamefont {Plotnik}}, \bibinfo
  {author} {\bibfnamefont {Y.}~\bibnamefont {Lumer}}, \bibinfo {author}
  {\bibfnamefont {S.}~\bibnamefont {Nolte}}, \bibinfo {author} {\bibfnamefont
  {K.~G.}\ \bibnamefont {Makris}}, \bibinfo {author} {\bibfnamefont
  {M.}~\bibnamefont {Segev}}, \bibinfo {author} {\bibfnamefont {M.~C.}\
  \bibnamefont {Rechtsman}},\ and\ \bibinfo {author} {\bibfnamefont
  {A.}~\bibnamefont {Szameit}},\ }\bibfield  {title} {\bibinfo {title}
  {Topologically protected bound states in photonic parity\textendash
  time-symmetric crystals},\ }\href {https://doi.org/10.1038/nmat4811}
  {\bibfield  {journal} {\bibinfo  {journal} {Nature Mater.}\ }\textbf
  {\bibinfo {volume} {16}},\ \bibinfo {pages} {433} (\bibinfo {year}
  {2017})}\BibitemShut {NoStop}%
\bibitem [{\citenamefont {Wang}\ \emph {et~al.}(2019)\citenamefont {Wang},
  \citenamefont {Qiu}, \citenamefont {Xiao}, \citenamefont {Zhan},
  \citenamefont {Bian}, \citenamefont {Sanders}, \citenamefont {Yi},\ and\
  \citenamefont {Xue}}]{Wang_2019}%
  \BibitemOpen
  \bibfield  {author} {\bibinfo {author} {\bibfnamefont {K.}~\bibnamefont
  {Wang}}, \bibinfo {author} {\bibfnamefont {X.}~\bibnamefont {Qiu}}, \bibinfo
  {author} {\bibfnamefont {L.}~\bibnamefont {Xiao}}, \bibinfo {author}
  {\bibfnamefont {X.}~\bibnamefont {Zhan}}, \bibinfo {author} {\bibfnamefont
  {Z.}~\bibnamefont {Bian}}, \bibinfo {author} {\bibfnamefont {B.~C.}\
  \bibnamefont {Sanders}}, \bibinfo {author} {\bibfnamefont {W.}~\bibnamefont
  {Yi}},\ and\ \bibinfo {author} {\bibfnamefont {P.}~\bibnamefont {Xue}},\
  }\bibfield  {title} {\bibinfo {title} {Observation of emergent
  momentum\textendash time skyrmions in parity\textendash time-symmetric
  non-unitary quench dynamics},\ }\href
  {https://doi.org/10.1038/s41467-019-10252-7} {\bibfield  {journal} {\bibinfo
  {journal} {Nat. Commun.}\ }\textbf {\bibinfo {volume} {10}},\ \bibinfo
  {pages} {2293} (\bibinfo {year} {2019})}\BibitemShut {NoStop}%
\bibitem [{\citenamefont {Xiao}\ \emph {et~al.}(2020)\citenamefont {Xiao},
  \citenamefont {Deng}, \citenamefont {Wang}, \citenamefont {Zhu},
  \citenamefont {Wang}, \citenamefont {Yi},\ and\ \citenamefont
  {Xue}}]{Xiao_2020}%
  \BibitemOpen
  \bibfield  {author} {\bibinfo {author} {\bibfnamefont {L.}~\bibnamefont
  {Xiao}}, \bibinfo {author} {\bibfnamefont {T.}~\bibnamefont {Deng}}, \bibinfo
  {author} {\bibfnamefont {K.}~\bibnamefont {Wang}}, \bibinfo {author}
  {\bibfnamefont {G.}~\bibnamefont {Zhu}}, \bibinfo {author} {\bibfnamefont
  {Z.}~\bibnamefont {Wang}}, \bibinfo {author} {\bibfnamefont {W.}~\bibnamefont
  {Yi}},\ and\ \bibinfo {author} {\bibfnamefont {P.}~\bibnamefont {Xue}},\
  }\bibfield  {title} {\bibinfo {title} {Non-{{Hermitian}} bulk\textendash
  boundary correspondence in quantum dynamics},\ }\href
  {https://doi.org/10.1038/s41567-020-0836-6} {\bibfield  {journal} {\bibinfo
  {journal} {Nat. Phys.}\ }\textbf {\bibinfo {volume} {16}},\ \bibinfo {pages}
  {761} (\bibinfo {year} {2020})}\BibitemShut {NoStop}%
\bibitem [{\citenamefont {Lukin}\ \emph {et~al.}(2019)\citenamefont {Lukin},
  \citenamefont {Rispoli}, \citenamefont {Schittko}, \citenamefont {Tai},
  \citenamefont {Kaufman}, \citenamefont {Choi}, \citenamefont {Khemani},
  \citenamefont {L{\'e}onard},\ and\ \citenamefont {Greiner}}]{Lukin_2019}%
  \BibitemOpen
  \bibfield  {author} {\bibinfo {author} {\bibfnamefont {A.}~\bibnamefont
  {Lukin}}, \bibinfo {author} {\bibfnamefont {M.}~\bibnamefont {Rispoli}},
  \bibinfo {author} {\bibfnamefont {R.}~\bibnamefont {Schittko}}, \bibinfo
  {author} {\bibfnamefont {M.~E.}\ \bibnamefont {Tai}}, \bibinfo {author}
  {\bibfnamefont {A.~M.}\ \bibnamefont {Kaufman}}, \bibinfo {author}
  {\bibfnamefont {S.}~\bibnamefont {Choi}}, \bibinfo {author} {\bibfnamefont
  {V.}~\bibnamefont {Khemani}}, \bibinfo {author} {\bibfnamefont
  {J.}~\bibnamefont {L{\'e}onard}},\ and\ \bibinfo {author} {\bibfnamefont
  {M.}~\bibnamefont {Greiner}},\ }\bibfield  {title} {\bibinfo {title} {Probing
  entanglement in a many-body\textendash localized system},\ }\bibfield
  {journal} {\bibinfo  {journal} {Science}\ }\href
  {https://doi.org/10.1126/science.aau0818} {10.1126/science.aau0818} (\bibinfo
  {year} {2019})\BibitemShut {NoStop}%
\bibitem [{\citenamefont {Lin}\ \emph {et~al.}(2022)\citenamefont {Lin},
  \citenamefont {Li}, \citenamefont {Xiao}, \citenamefont {Wang}, \citenamefont
  {Yi},\ and\ \citenamefont {Xue}}]{Lin_2022}%
  \BibitemOpen
  \bibfield  {author} {\bibinfo {author} {\bibfnamefont {Q.}~\bibnamefont
  {Lin}}, \bibinfo {author} {\bibfnamefont {T.}~\bibnamefont {Li}}, \bibinfo
  {author} {\bibfnamefont {L.}~\bibnamefont {Xiao}}, \bibinfo {author}
  {\bibfnamefont {K.}~\bibnamefont {Wang}}, \bibinfo {author} {\bibfnamefont
  {W.}~\bibnamefont {Yi}},\ and\ \bibinfo {author} {\bibfnamefont
  {P.}~\bibnamefont {Xue}},\ }\bibfield  {title} {\bibinfo {title} {Topological
  {{Phase Transitions}} and {{Mobility Edges}} in {{Non-Hermitian
  Quasicrystals}}},\ }\href {https://doi.org/10.1103/PhysRevLett.129.113601}
  {\bibfield  {journal} {\bibinfo  {journal} {Phys. Rev. Lett.}\ }\textbf
  {\bibinfo {volume} {129}},\ \bibinfo {pages} {113601} (\bibinfo {year}
  {2022})}\BibitemShut {NoStop}%
\bibitem [{\citenamefont {Weidemann}\ \emph {et~al.}(2022)\citenamefont
  {Weidemann}, \citenamefont {Kremer}, \citenamefont {Longhi},\ and\
  \citenamefont {Szameit}}]{Weidemann_2022}%
  \BibitemOpen
  \bibfield  {author} {\bibinfo {author} {\bibfnamefont {S.}~\bibnamefont
  {Weidemann}}, \bibinfo {author} {\bibfnamefont {M.}~\bibnamefont {Kremer}},
  \bibinfo {author} {\bibfnamefont {S.}~\bibnamefont {Longhi}},\ and\ \bibinfo
  {author} {\bibfnamefont {A.}~\bibnamefont {Szameit}},\ }\bibfield  {title}
  {\bibinfo {title} {Topological triple phase transition in non-{{Hermitian
  Floquet}} quasicrystals},\ }\href
  {https://doi.org/10.1038/s41586-021-04253-0} {\bibfield  {journal} {\bibinfo
  {journal} {Nature}\ }\textbf {\bibinfo {volume} {601}},\ \bibinfo {pages}
  {354} (\bibinfo {year} {2022})}\BibitemShut {NoStop}%
\bibitem [{\citenamefont {Zhu}\ \emph {et~al.}(2018)\citenamefont {Zhu},
  \citenamefont {Fang}, \citenamefont {Li}, \citenamefont {Sun}, \citenamefont
  {Li}, \citenamefont {Jing},\ and\ \citenamefont {Chen}}]{Zhu_2018}%
  \BibitemOpen
  \bibfield  {author} {\bibinfo {author} {\bibfnamefont {W.}~\bibnamefont
  {Zhu}}, \bibinfo {author} {\bibfnamefont {X.}~\bibnamefont {Fang}}, \bibinfo
  {author} {\bibfnamefont {D.}~\bibnamefont {Li}}, \bibinfo {author}
  {\bibfnamefont {Y.}~\bibnamefont {Sun}}, \bibinfo {author} {\bibfnamefont
  {Y.}~\bibnamefont {Li}}, \bibinfo {author} {\bibfnamefont {Y.}~\bibnamefont
  {Jing}},\ and\ \bibinfo {author} {\bibfnamefont {H.}~\bibnamefont {Chen}},\
  }\bibfield  {title} {\bibinfo {title} {Simultaneous {{Observation}} of a
  {{Topological Edge State}} and {{Exceptional Point}} in an {{Open}} and
  {{Non-Hermitian Acoustic System}}},\ }\href
  {https://doi.org/10.1103/PhysRevLett.121.124501} {\bibfield  {journal}
  {\bibinfo  {journal} {Phys. Rev. Lett.}\ }\textbf {\bibinfo {volume} {121}},\
  \bibinfo {pages} {124501} (\bibinfo {year} {2018})}\BibitemShut {NoStop}%
\bibitem [{\citenamefont {Shen}\ \emph {et~al.}(2018)\citenamefont {Shen},
  \citenamefont {Zhen},\ and\ \citenamefont {Fu}}]{Shen_2018}%
  \BibitemOpen
  \bibfield  {author} {\bibinfo {author} {\bibfnamefont {H.}~\bibnamefont
  {Shen}}, \bibinfo {author} {\bibfnamefont {B.}~\bibnamefont {Zhen}},\ and\
  \bibinfo {author} {\bibfnamefont {L.}~\bibnamefont {Fu}},\ }\bibfield
  {title} {\bibinfo {title} {Topological {{Band Theory}} for {{Non-Hermitian
  Hamiltonians}}},\ }\href {https://doi.org/10.1103/PhysRevLett.120.146402}
  {\bibfield  {journal} {\bibinfo  {journal} {Phys. Rev. Lett.}\ }\textbf
  {\bibinfo {volume} {120}},\ \bibinfo {pages} {146402} (\bibinfo {year}
  {2018})}\BibitemShut {NoStop}%
\bibitem [{\citenamefont {Gao}\ \emph {et~al.}(2020)\citenamefont {Gao},
  \citenamefont {Xue}, \citenamefont {Wang}, \citenamefont {Gu}, \citenamefont
  {Liu}, \citenamefont {Zhu},\ and\ \citenamefont {Zhang}}]{Gao_2020}%
  \BibitemOpen
  \bibfield  {author} {\bibinfo {author} {\bibfnamefont {H.}~\bibnamefont
  {Gao}}, \bibinfo {author} {\bibfnamefont {H.}~\bibnamefont {Xue}}, \bibinfo
  {author} {\bibfnamefont {Q.}~\bibnamefont {Wang}}, \bibinfo {author}
  {\bibfnamefont {Z.}~\bibnamefont {Gu}}, \bibinfo {author} {\bibfnamefont
  {T.}~\bibnamefont {Liu}}, \bibinfo {author} {\bibfnamefont {J.}~\bibnamefont
  {Zhu}},\ and\ \bibinfo {author} {\bibfnamefont {B.}~\bibnamefont {Zhang}},\
  }\bibfield  {title} {\bibinfo {title} {Observation of topological edge states
  induced solely by non-{{Hermiticity}} in an acoustic crystal},\ }\href
  {https://doi.org/10.1103/PhysRevB.101.180303} {\bibfield  {journal} {\bibinfo
   {journal} {Phys. Rev. B}\ }\textbf {\bibinfo {volume} {101}},\ \bibinfo
  {pages} {180303} (\bibinfo {year} {2020})}\BibitemShut {NoStop}%
\bibitem [{\citenamefont {Hofmann}\ \emph {et~al.}(2020)\citenamefont
  {Hofmann}, \citenamefont {Helbig}, \citenamefont {Schindler}, \citenamefont
  {Salgo}, \citenamefont {Brzezi{\'n}ska}, \citenamefont {Greiter},
  \citenamefont {Kiessling}, \citenamefont {Wolf}, \citenamefont {Vollhardt},
  \citenamefont {Kaba{\v s}i}, \citenamefont {Lee}, \citenamefont {Bilu{\v
  s}i{\'c}}, \citenamefont {Thomale},\ and\ \citenamefont
  {Neupert}}]{Hofmann_2020}%
  \BibitemOpen
  \bibfield  {author} {\bibinfo {author} {\bibfnamefont {T.}~\bibnamefont
  {Hofmann}}, \bibinfo {author} {\bibfnamefont {T.}~\bibnamefont {Helbig}},
  \bibinfo {author} {\bibfnamefont {F.}~\bibnamefont {Schindler}}, \bibinfo
  {author} {\bibfnamefont {N.}~\bibnamefont {Salgo}}, \bibinfo {author}
  {\bibfnamefont {M.}~\bibnamefont {Brzezi{\'n}ska}}, \bibinfo {author}
  {\bibfnamefont {M.}~\bibnamefont {Greiter}}, \bibinfo {author} {\bibfnamefont
  {T.}~\bibnamefont {Kiessling}}, \bibinfo {author} {\bibfnamefont
  {D.}~\bibnamefont {Wolf}}, \bibinfo {author} {\bibfnamefont {A.}~\bibnamefont
  {Vollhardt}}, \bibinfo {author} {\bibfnamefont {A.}~\bibnamefont {Kaba{\v
  s}i}}, \bibinfo {author} {\bibfnamefont {C.~H.}\ \bibnamefont {Lee}},
  \bibinfo {author} {\bibfnamefont {A.}~\bibnamefont {Bilu{\v s}i{\'c}}},
  \bibinfo {author} {\bibfnamefont {R.}~\bibnamefont {Thomale}},\ and\ \bibinfo
  {author} {\bibfnamefont {T.}~\bibnamefont {Neupert}},\ }\bibfield  {title}
  {\bibinfo {title} {Reciprocal skin effect and its realization in a
  topolectrical circuit},\ }\href
  {https://doi.org/10.1103/PhysRevResearch.2.023265} {\bibfield  {journal}
  {\bibinfo  {journal} {Phys. Rev. Res.}\ }\textbf {\bibinfo {volume} {2}},\
  \bibinfo {pages} {023265} (\bibinfo {year} {2020})}\BibitemShut {NoStop}%
\bibitem [{\citenamefont {Helbig}\ \emph {et~al.}(2020)\citenamefont {Helbig},
  \citenamefont {Hofmann}, \citenamefont {Imhof}, \citenamefont {Abdelghany},
  \citenamefont {Kiessling}, \citenamefont {Molenkamp}, \citenamefont {Lee},
  \citenamefont {Szameit}, \citenamefont {Greiter},\ and\ \citenamefont
  {Thomale}}]{Helbig_2020}%
  \BibitemOpen
  \bibfield  {author} {\bibinfo {author} {\bibfnamefont {T.}~\bibnamefont
  {Helbig}}, \bibinfo {author} {\bibfnamefont {T.}~\bibnamefont {Hofmann}},
  \bibinfo {author} {\bibfnamefont {S.}~\bibnamefont {Imhof}}, \bibinfo
  {author} {\bibfnamefont {M.}~\bibnamefont {Abdelghany}}, \bibinfo {author}
  {\bibfnamefont {T.}~\bibnamefont {Kiessling}}, \bibinfo {author}
  {\bibfnamefont {L.~W.}\ \bibnamefont {Molenkamp}}, \bibinfo {author}
  {\bibfnamefont {C.~H.}\ \bibnamefont {Lee}}, \bibinfo {author} {\bibfnamefont
  {A.}~\bibnamefont {Szameit}}, \bibinfo {author} {\bibfnamefont
  {M.}~\bibnamefont {Greiter}},\ and\ \bibinfo {author} {\bibfnamefont
  {R.}~\bibnamefont {Thomale}},\ }\bibfield  {title} {\bibinfo {title}
  {Generalized bulk\textendash boundary correspondence in non-{{Hermitian}}
  topolectrical circuits},\ }\href {https://doi.org/10.1038/s41567-020-0922-9}
  {\bibfield  {journal} {\bibinfo  {journal} {Nat. Phys.}\ }\textbf {\bibinfo
  {volume} {16}},\ \bibinfo {pages} {747} (\bibinfo {year} {2020})}\BibitemShut
  {NoStop}%
\bibitem [{\citenamefont {Liu}\ \emph {et~al.}(2020{\natexlab{a}})\citenamefont
  {Liu}, \citenamefont {Ma}, \citenamefont {Yang}, \citenamefont {Zhang},
  \citenamefont {Gao}, \citenamefont {Xiang}, \citenamefont {Cui},\ and\
  \citenamefont {Zhang}}]{Liu_2020b}%
  \BibitemOpen
  \bibfield  {author} {\bibinfo {author} {\bibfnamefont {S.}~\bibnamefont
  {Liu}}, \bibinfo {author} {\bibfnamefont {S.}~\bibnamefont {Ma}}, \bibinfo
  {author} {\bibfnamefont {C.}~\bibnamefont {Yang}}, \bibinfo {author}
  {\bibfnamefont {L.}~\bibnamefont {Zhang}}, \bibinfo {author} {\bibfnamefont
  {W.}~\bibnamefont {Gao}}, \bibinfo {author} {\bibfnamefont {Y.~J.}\
  \bibnamefont {Xiang}}, \bibinfo {author} {\bibfnamefont {T.~J.}\ \bibnamefont
  {Cui}},\ and\ \bibinfo {author} {\bibfnamefont {S.}~\bibnamefont {Zhang}},\
  }\bibfield  {title} {\bibinfo {title} {Gain- and {{Loss-Induced Topological
  Insulating Phase}} in a {{Non-Hermitian Electrical Circuit}}},\ }\href
  {https://doi.org/10.1103/PhysRevApplied.13.014047} {\bibfield  {journal}
  {\bibinfo  {journal} {Phys. Rev. Applied}\ }\textbf {\bibinfo {volume}
  {13}},\ \bibinfo {pages} {014047} (\bibinfo {year}
  {2020}{\natexlab{a}})}\BibitemShut {NoStop}%
\bibitem [{\citenamefont {Wu}\ \emph {et~al.}(2019)\citenamefont {Wu},
  \citenamefont {Liu}, \citenamefont {Geng}, \citenamefont {Song},
  \citenamefont {Ye}, \citenamefont {Duan}, \citenamefont {Rong},\ and\
  \citenamefont {Du}}]{Wu_2019}%
  \BibitemOpen
  \bibfield  {author} {\bibinfo {author} {\bibfnamefont {Y.}~\bibnamefont
  {Wu}}, \bibinfo {author} {\bibfnamefont {W.}~\bibnamefont {Liu}}, \bibinfo
  {author} {\bibfnamefont {J.}~\bibnamefont {Geng}}, \bibinfo {author}
  {\bibfnamefont {X.}~\bibnamefont {Song}}, \bibinfo {author} {\bibfnamefont
  {X.}~\bibnamefont {Ye}}, \bibinfo {author} {\bibfnamefont {C.-K.}\
  \bibnamefont {Duan}}, \bibinfo {author} {\bibfnamefont {X.}~\bibnamefont
  {Rong}},\ and\ \bibinfo {author} {\bibfnamefont {J.}~\bibnamefont {Du}},\
  }\bibfield  {title} {\bibinfo {title} {Observation of parity-time symmetry
  breaking in a single-spin system},\ }\href
  {https://doi.org/10.1126/science.aaw8205} {\bibfield  {journal} {\bibinfo
  {journal} {Science}\ }\textbf {\bibinfo {volume} {364}},\ \bibinfo {pages}
  {878} (\bibinfo {year} {2019})}\BibitemShut {NoStop}%
\bibitem [{\citenamefont {Sarnak}(1982)}]{Sarnak_1982}%
  \BibitemOpen
  \bibfield  {author} {\bibinfo {author} {\bibfnamefont {P.}~\bibnamefont
  {Sarnak}},\ }\bibfield  {title} {\bibinfo {title} {Spectral behavior of quasi
  periodic potentials},\ }\href {https://doi.org/10.1007/BF01208483} {\bibfield
   {journal} {\bibinfo  {journal} {Commun. Math. Phys.}\ }\textbf {\bibinfo
  {volume} {84}},\ \bibinfo {pages} {377} (\bibinfo {year} {1982})}\BibitemShut
  {NoStop}%
\bibitem [{\citenamefont {Jazaeri}\ and\ \citenamefont
  {Satija}(2001)}]{Jazaeri_2001}%
  \BibitemOpen
  \bibfield  {author} {\bibinfo {author} {\bibfnamefont {A.}~\bibnamefont
  {Jazaeri}}\ and\ \bibinfo {author} {\bibfnamefont {I.~I.}\ \bibnamefont
  {Satija}},\ }\bibfield  {title} {\bibinfo {title} {Localization transition in
  incommensurate non-{{Hermitian}} systems},\ }\href
  {https://doi.org/10.1103/PhysRevE.63.036222} {\bibfield  {journal} {\bibinfo
  {journal} {Phys. Rev. E}\ }\textbf {\bibinfo {volume} {63}},\ \bibinfo
  {pages} {036222} (\bibinfo {year} {2001})}\BibitemShut {NoStop}%
\bibitem [{\citenamefont {Zeng}\ \emph {et~al.}(2017)\citenamefont {Zeng},
  \citenamefont {Chen},\ and\ \citenamefont {L{\"u}}}]{Zeng_2017}%
  \BibitemOpen
  \bibfield  {author} {\bibinfo {author} {\bibfnamefont {Q.-B.}\ \bibnamefont
  {Zeng}}, \bibinfo {author} {\bibfnamefont {S.}~\bibnamefont {Chen}},\ and\
  \bibinfo {author} {\bibfnamefont {R.}~\bibnamefont {L{\"u}}},\ }\bibfield
  {title} {\bibinfo {title} {Anderson localization in the non-{{Hermitian
  Aubry-Andr}}{\'e}-{{Harper}} model with physical gain and loss},\ }\href
  {https://doi.org/10.1103/PhysRevA.95.062118} {\bibfield  {journal} {\bibinfo
  {journal} {Phys. Rev. A}\ }\textbf {\bibinfo {volume} {95}},\ \bibinfo
  {pages} {062118} (\bibinfo {year} {2017})}\BibitemShut {NoStop}%
\bibitem [{\citenamefont {Jiang}\ \emph {et~al.}(2019)\citenamefont {Jiang},
  \citenamefont {Lang}, \citenamefont {Yang}, \citenamefont {Zhu},\ and\
  \citenamefont {Chen}}]{Jiang_2019}%
  \BibitemOpen
  \bibfield  {author} {\bibinfo {author} {\bibfnamefont {H.}~\bibnamefont
  {Jiang}}, \bibinfo {author} {\bibfnamefont {L.-J.}\ \bibnamefont {Lang}},
  \bibinfo {author} {\bibfnamefont {C.}~\bibnamefont {Yang}}, \bibinfo {author}
  {\bibfnamefont {S.-L.}\ \bibnamefont {Zhu}},\ and\ \bibinfo {author}
  {\bibfnamefont {S.}~\bibnamefont {Chen}},\ }\bibfield  {title} {\bibinfo
  {title} {Interplay of non-{{Hermitian}} skin effects and {{Anderson}}
  localization in nonreciprocal quasiperiodic lattices},\ }\href
  {https://doi.org/10.1103/PhysRevB.100.054301} {\bibfield  {journal} {\bibinfo
   {journal} {Phys. Rev. B}\ }\textbf {\bibinfo {volume} {100}},\ \bibinfo
  {pages} {054301} (\bibinfo {year} {2019})}\BibitemShut {NoStop}%
\bibitem [{\citenamefont {Longhi}(2019{\natexlab{a}})}]{Longhi_2019}%
  \BibitemOpen
  \bibfield  {author} {\bibinfo {author} {\bibfnamefont {S.}~\bibnamefont
  {Longhi}},\ }\bibfield  {title} {\bibinfo {title} {Topological {{Phase
  Transition}} in non-{{Hermitian Quasicrystals}}},\ }\href
  {https://doi.org/10.1103/PhysRevLett.122.237601} {\bibfield  {journal}
  {\bibinfo  {journal} {Phys. Rev. Lett.}\ }\textbf {\bibinfo {volume} {122}},\
  \bibinfo {pages} {237601} (\bibinfo {year} {2019}{\natexlab{a}})}\BibitemShut
  {NoStop}%
\bibitem [{\citenamefont {Longhi}(2019{\natexlab{b}})}]{Longhi_2019a}%
  \BibitemOpen
  \bibfield  {author} {\bibinfo {author} {\bibfnamefont {S.}~\bibnamefont
  {Longhi}},\ }\bibfield  {title} {\bibinfo {title} {Metal-insulator phase
  transition in a non-{{Hermitian Aubry-Andr}}{\'e}-{{Harper}} model},\ }\href
  {https://doi.org/10.1103/PhysRevB.100.125157} {\bibfield  {journal} {\bibinfo
   {journal} {Phys. Rev. B}\ }\textbf {\bibinfo {volume} {100}},\ \bibinfo
  {pages} {125157} (\bibinfo {year} {2019}{\natexlab{b}})}\BibitemShut
  {NoStop}%
\bibitem [{\citenamefont {Liu}\ \emph {et~al.}(2020{\natexlab{b}})\citenamefont
  {Liu}, \citenamefont {Guo}, \citenamefont {Pu},\ and\ \citenamefont
  {Longhi}}]{Liu_2020a}%
  \BibitemOpen
  \bibfield  {author} {\bibinfo {author} {\bibfnamefont {T.}~\bibnamefont
  {Liu}}, \bibinfo {author} {\bibfnamefont {H.}~\bibnamefont {Guo}}, \bibinfo
  {author} {\bibfnamefont {Y.}~\bibnamefont {Pu}},\ and\ \bibinfo {author}
  {\bibfnamefont {S.}~\bibnamefont {Longhi}},\ }\bibfield  {title} {\bibinfo
  {title} {Generalized {{Aubry-Andr}}{\'e} self-duality and mobility edges in
  non-{{Hermitian}} quasiperiodic lattices},\ }\href
  {https://doi.org/10.1103/PhysRevB.102.024205} {\bibfield  {journal} {\bibinfo
   {journal} {Phys. Rev. B}\ }\textbf {\bibinfo {volume} {102}},\ \bibinfo
  {pages} {024205} (\bibinfo {year} {2020}{\natexlab{b}})}\BibitemShut
  {NoStop}%
\bibitem [{\citenamefont {Liu}\ \emph {et~al.}(2020{\natexlab{c}})\citenamefont
  {Liu}, \citenamefont {Jiang}, \citenamefont {Cao},\ and\ \citenamefont
  {Chen}}]{Liu_2020}%
  \BibitemOpen
  \bibfield  {author} {\bibinfo {author} {\bibfnamefont {Y.}~\bibnamefont
  {Liu}}, \bibinfo {author} {\bibfnamefont {X.-P.}\ \bibnamefont {Jiang}},
  \bibinfo {author} {\bibfnamefont {J.}~\bibnamefont {Cao}},\ and\ \bibinfo
  {author} {\bibfnamefont {S.}~\bibnamefont {Chen}},\ }\bibfield  {title}
  {\bibinfo {title} {Non-{{Hermitian}} mobility edges in one-dimensional
  quasicrystals with parity-time symmetry},\ }\href
  {https://doi.org/10.1103/PhysRevB.101.174205} {\bibfield  {journal} {\bibinfo
   {journal} {Phys. Rev. B}\ }\textbf {\bibinfo {volume} {101}},\ \bibinfo
  {pages} {174205} (\bibinfo {year} {2020}{\natexlab{c}})}\BibitemShut
  {NoStop}%
\bibitem [{\citenamefont {Zeng}\ \emph
  {et~al.}(2020{\natexlab{a}})\citenamefont {Zeng}, \citenamefont {Yang},\ and\
  \citenamefont {Xu}}]{Zeng_2020}%
  \BibitemOpen
  \bibfield  {author} {\bibinfo {author} {\bibfnamefont {Q.-B.}\ \bibnamefont
  {Zeng}}, \bibinfo {author} {\bibfnamefont {Y.-B.}\ \bibnamefont {Yang}},\
  and\ \bibinfo {author} {\bibfnamefont {Y.}~\bibnamefont {Xu}},\ }\bibfield
  {title} {\bibinfo {title} {Topological phases in non-{{Hermitian
  Aubry-Andr}}{\'e}-{{Harper}} models},\ }\href
  {https://doi.org/10.1103/PhysRevB.101.020201} {\bibfield  {journal} {\bibinfo
   {journal} {Phys. Rev. B}\ }\textbf {\bibinfo {volume} {101}},\ \bibinfo
  {pages} {020201} (\bibinfo {year} {2020}{\natexlab{a}})}\BibitemShut
  {NoStop}%
\bibitem [{\citenamefont {Zeng}\ \emph
  {et~al.}(2020{\natexlab{b}})\citenamefont {Zeng}, \citenamefont {Yang},\ and\
  \citenamefont {L{\"u}}}]{Zeng_2020b}%
  \BibitemOpen
  \bibfield  {author} {\bibinfo {author} {\bibfnamefont {Q.-B.}\ \bibnamefont
  {Zeng}}, \bibinfo {author} {\bibfnamefont {Y.-B.}\ \bibnamefont {Yang}},\
  and\ \bibinfo {author} {\bibfnamefont {R.}~\bibnamefont {L{\"u}}},\
  }\bibfield  {title} {\bibinfo {title} {Topological phases in one-dimensional
  nonreciprocal superlattices},\ }\href
  {https://doi.org/10.1103/PhysRevB.101.125418} {\bibfield  {journal} {\bibinfo
   {journal} {Phys. Rev. B}\ }\textbf {\bibinfo {volume} {101}},\ \bibinfo
  {pages} {125418} (\bibinfo {year} {2020}{\natexlab{b}})}\BibitemShut
  {NoStop}%
\bibitem [{\citenamefont {Zhai}\ \emph {et~al.}(2020)\citenamefont {Zhai},
  \citenamefont {Yin},\ and\ \citenamefont {Huang}}]{Zhai_2020}%
  \BibitemOpen
  \bibfield  {author} {\bibinfo {author} {\bibfnamefont {L.-J.}\ \bibnamefont
  {Zhai}}, \bibinfo {author} {\bibfnamefont {S.}~\bibnamefont {Yin}},\ and\
  \bibinfo {author} {\bibfnamefont {G.-Y.}\ \bibnamefont {Huang}},\ }\bibfield
  {title} {\bibinfo {title} {Many-body localization in a non-{{Hermitian}}
  quasiperiodic system},\ }\href {https://doi.org/10.1103/PhysRevB.102.064206}
  {\bibfield  {journal} {\bibinfo  {journal} {Phys. Rev. B}\ }\textbf {\bibinfo
  {volume} {102}},\ \bibinfo {pages} {064206} (\bibinfo {year}
  {2020})}\BibitemShut {NoStop}%
\bibitem [{\citenamefont {Liu}\ \emph {et~al.}(2021{\natexlab{a}})\citenamefont
  {Liu}, \citenamefont {Wang}, \citenamefont {Liu}, \citenamefont {Zhou},\ and\
  \citenamefont {Chen}}]{Liu_2021a}%
  \BibitemOpen
  \bibfield  {author} {\bibinfo {author} {\bibfnamefont {Y.}~\bibnamefont
  {Liu}}, \bibinfo {author} {\bibfnamefont {Y.}~\bibnamefont {Wang}}, \bibinfo
  {author} {\bibfnamefont {X.-J.}\ \bibnamefont {Liu}}, \bibinfo {author}
  {\bibfnamefont {Q.}~\bibnamefont {Zhou}},\ and\ \bibinfo {author}
  {\bibfnamefont {S.}~\bibnamefont {Chen}},\ }\bibfield  {title} {\bibinfo
  {title} {Exact mobility edges, $\mathcal{PT}$-symmetry breaking, and skin
  effect in one-dimensional non-{{Hermitian}} quasicrystals},\ }\href
  {https://doi.org/10.1103/PhysRevB.103.014203} {\bibfield  {journal} {\bibinfo
   {journal} {Phys. Rev. B}\ }\textbf {\bibinfo {volume} {103}},\ \bibinfo
  {pages} {014203} (\bibinfo {year} {2021}{\natexlab{a}})}\BibitemShut
  {NoStop}%
\bibitem [{\citenamefont {Liu}\ \emph {et~al.}(2021{\natexlab{b}})\citenamefont
  {Liu}, \citenamefont {Wang}, \citenamefont {Zheng},\ and\ \citenamefont
  {Chen}}]{Liu_2021b}%
  \BibitemOpen
  \bibfield  {author} {\bibinfo {author} {\bibfnamefont {Y.}~\bibnamefont
  {Liu}}, \bibinfo {author} {\bibfnamefont {Y.}~\bibnamefont {Wang}}, \bibinfo
  {author} {\bibfnamefont {Z.}~\bibnamefont {Zheng}},\ and\ \bibinfo {author}
  {\bibfnamefont {S.}~\bibnamefont {Chen}},\ }\bibfield  {title} {\bibinfo
  {title} {Exact non-{{Hermitian}} mobility edges in one-dimensional
  quasicrystal lattice with exponentially decaying hopping and its dual
  lattice},\ }\href {https://doi.org/10.1103/PhysRevB.103.134208} {\bibfield
  {journal} {\bibinfo  {journal} {Phys. Rev. B}\ }\textbf {\bibinfo {volume}
  {103}},\ \bibinfo {pages} {134208} (\bibinfo {year}
  {2021}{\natexlab{b}})}\BibitemShut {NoStop}%
\bibitem [{\citenamefont {Xu}\ and\ \citenamefont {Chen}(2021)}]{Xu_2021}%
  \BibitemOpen
  \bibfield  {author} {\bibinfo {author} {\bibfnamefont {Z.}~\bibnamefont
  {Xu}}\ and\ \bibinfo {author} {\bibfnamefont {S.}~\bibnamefont {Chen}},\
  }\bibfield  {title} {\bibinfo {title} {Dynamical evolution in a
  one-dimensional incommensurate lattice with $\mathcal{PT}$ symmetry},\ }\href
  {https://doi.org/10.1103/PhysRevA.103.043325} {\bibfield  {journal} {\bibinfo
   {journal} {Phys. Rev. A}\ }\textbf {\bibinfo {volume} {103}},\ \bibinfo
  {pages} {043325} (\bibinfo {year} {2021})}\BibitemShut {NoStop}%
\bibitem [{\citenamefont {Cai}(2021{\natexlab{a}})}]{Cai_2021}%
  \BibitemOpen
  \bibfield  {author} {\bibinfo {author} {\bibfnamefont {X.}~\bibnamefont
  {Cai}},\ }\bibfield  {title} {\bibinfo {title} {Boundary-dependent
  self-dualities, winding numbers, and asymmetrical localization in
  non-{{Hermitian}} aperiodic one-dimensional models},\ }\href
  {https://doi.org/10.1103/PhysRevB.103.014201} {\bibfield  {journal} {\bibinfo
   {journal} {Phys. Rev. B}\ }\textbf {\bibinfo {volume} {103}},\ \bibinfo
  {pages} {014201} (\bibinfo {year} {2021}{\natexlab{a}})}\BibitemShut
  {NoStop}%
\bibitem [{\citenamefont {Tang}\ \emph {et~al.}(2021)\citenamefont {Tang},
  \citenamefont {Zhang}, \citenamefont {Zhang},\ and\ \citenamefont
  {Zhang}}]{Tang_2021}%
  \BibitemOpen
  \bibfield  {author} {\bibinfo {author} {\bibfnamefont {L.-Z.}\ \bibnamefont
  {Tang}}, \bibinfo {author} {\bibfnamefont {G.-Q.}\ \bibnamefont {Zhang}},
  \bibinfo {author} {\bibfnamefont {L.-F.}\ \bibnamefont {Zhang}},\ and\
  \bibinfo {author} {\bibfnamefont {D.-W.}\ \bibnamefont {Zhang}},\ }\bibfield
  {title} {\bibinfo {title} {Localization and topological transitions in
  non-{{Hermitian}} quasiperiodic lattices},\ }\href
  {https://doi.org/10.1103/PhysRevA.103.033325} {\bibfield  {journal} {\bibinfo
   {journal} {Phys. Rev. A}\ }\textbf {\bibinfo {volume} {103}},\ \bibinfo
  {pages} {033325} (\bibinfo {year} {2021})}\BibitemShut {NoStop}%
\bibitem [{\citenamefont {Liu}\ \emph {et~al.}(2021{\natexlab{c}})\citenamefont
  {Liu}, \citenamefont {Cheng}, \citenamefont {Guo},\ and\ \citenamefont
  {Xianlong}}]{Liu_2021c}%
  \BibitemOpen
  \bibfield  {author} {\bibinfo {author} {\bibfnamefont {T.}~\bibnamefont
  {Liu}}, \bibinfo {author} {\bibfnamefont {S.}~\bibnamefont {Cheng}}, \bibinfo
  {author} {\bibfnamefont {H.}~\bibnamefont {Guo}},\ and\ \bibinfo {author}
  {\bibfnamefont {G.}~\bibnamefont {Xianlong}},\ }\bibfield  {title} {\bibinfo
  {title} {Fate of {{Majorana}} zero modes, exact location of critical states,
  and unconventional real-complex transition in non-{{Hermitian}} quasiperiodic
  lattices},\ }\href {https://doi.org/10.1103/PhysRevB.103.104203} {\bibfield
  {journal} {\bibinfo  {journal} {Phys. Rev. B}\ }\textbf {\bibinfo {volume}
  {103}},\ \bibinfo {pages} {104203} (\bibinfo {year}
  {2021}{\natexlab{c}})}\BibitemShut {NoStop}%
\bibitem [{\citenamefont {Zhai}\ \emph {et~al.}(2021)\citenamefont {Zhai},
  \citenamefont {Huang},\ and\ \citenamefont {Yin}}]{Zhai_2021}%
  \BibitemOpen
  \bibfield  {author} {\bibinfo {author} {\bibfnamefont {L.-J.}\ \bibnamefont
  {Zhai}}, \bibinfo {author} {\bibfnamefont {G.-Y.}\ \bibnamefont {Huang}},\
  and\ \bibinfo {author} {\bibfnamefont {S.}~\bibnamefont {Yin}},\ }\bibfield
  {title} {\bibinfo {title} {Cascade of the delocalization transition in a
  non-{{Hermitian}} interpolating {{Aubry-Andr}}{\'e}-{{Fibonacci}} chain},\
  }\href {https://doi.org/10.1103/PhysRevB.104.014202} {\bibfield  {journal}
  {\bibinfo  {journal} {Phys. Rev. B}\ }\textbf {\bibinfo {volume} {104}},\
  \bibinfo {pages} {014202} (\bibinfo {year} {2021})}\BibitemShut {NoStop}%
\bibitem [{\citenamefont {Zhou}\ and\ \citenamefont {Han}(2021)}]{Zhou_2021b}%
  \BibitemOpen
  \bibfield  {author} {\bibinfo {author} {\bibfnamefont {L.}~\bibnamefont
  {Zhou}}\ and\ \bibinfo {author} {\bibfnamefont {W.}~\bibnamefont {Han}},\
  }\bibfield  {title} {\bibinfo {title} {Non-hermitian quasicrystal in
  dimerized lattices},\ }\href {https://doi.org/10.1088/1674-1056/ac1efc}
  {\bibfield  {journal} {\bibinfo  {journal} {Chin. Phys. B}\ }\textbf
  {\bibinfo {volume} {30}},\ \bibinfo {pages} {100308} (\bibinfo {year}
  {2021})}\BibitemShut {NoStop}%
\bibitem [{\citenamefont {Wang}\ \emph {et~al.}(2021)\citenamefont {Wang},
  \citenamefont {Xu}, \citenamefont {Li}, \citenamefont {Xu},\ and\
  \citenamefont {Wang}}]{Wang_2021c}%
  \BibitemOpen
  \bibfield  {author} {\bibinfo {author} {\bibfnamefont {Z.-H.}\ \bibnamefont
  {Wang}}, \bibinfo {author} {\bibfnamefont {F.}~\bibnamefont {Xu}}, \bibinfo
  {author} {\bibfnamefont {L.}~\bibnamefont {Li}}, \bibinfo {author}
  {\bibfnamefont {D.-H.}\ \bibnamefont {Xu}},\ and\ \bibinfo {author}
  {\bibfnamefont {B.}~\bibnamefont {Wang}},\ }\bibfield  {title} {\bibinfo
  {title} {Unconventional real-complex spectral transition and majorana zero
  modes in nonreciprocal quasicrystals},\ }\href
  {https://doi.org/10.1103/PhysRevB.104.174501} {\bibfield  {journal} {\bibinfo
   {journal} {Phys. Rev. B}\ }\textbf {\bibinfo {volume} {104}},\ \bibinfo
  {pages} {174501} (\bibinfo {year} {2021})}\BibitemShut {NoStop}%
\bibitem [{\citenamefont {Liu}\ \emph {et~al.}(2021{\natexlab{d}})\citenamefont
  {Liu}, \citenamefont {Zhou},\ and\ \citenamefont {Chen}}]{Liu_2021}%
  \BibitemOpen
  \bibfield  {author} {\bibinfo {author} {\bibfnamefont {Y.}~\bibnamefont
  {Liu}}, \bibinfo {author} {\bibfnamefont {Q.}~\bibnamefont {Zhou}},\ and\
  \bibinfo {author} {\bibfnamefont {S.}~\bibnamefont {Chen}},\ }\bibfield
  {title} {\bibinfo {title} {Localization transition, spectrum structure, and
  winding numbers for one-dimensional non-{{Hermitian}} quasicrystals},\ }\href
  {https://doi.org/10.1103/PhysRevB.104.024201} {\bibfield  {journal} {\bibinfo
   {journal} {Phys. Rev. B}\ }\textbf {\bibinfo {volume} {104}},\ \bibinfo
  {pages} {024201} (\bibinfo {year} {2021}{\natexlab{d}})}\BibitemShut
  {NoStop}%
\bibitem [{\citenamefont {Cai}(2021{\natexlab{b}})}]{Cai_2021a}%
  \BibitemOpen
  \bibfield  {author} {\bibinfo {author} {\bibfnamefont {X.}~\bibnamefont
  {Cai}},\ }\bibfield  {title} {\bibinfo {title} {Localization and topological
  phase transitions in non-{{Hermitian Aubry-Andr}}{\'e}-{{Harper}} models with
  $p$-wave pairing},\ }\href {https://doi.org/10.1103/PhysRevB.103.214202}
  {\bibfield  {journal} {\bibinfo  {journal} {Phys. Rev. B}\ }\textbf {\bibinfo
  {volume} {103}},\ \bibinfo {pages} {214202} (\bibinfo {year}
  {2021}{\natexlab{b}})}\BibitemShut {NoStop}%
\bibitem [{\citenamefont {Longhi}(2021{\natexlab{a}})}]{Longhi_2021}%
  \BibitemOpen
  \bibfield  {author} {\bibinfo {author} {\bibfnamefont {S.}~\bibnamefont
  {Longhi}},\ }\bibfield  {title} {\bibinfo {title} {Non-{{Hermitian Maryland}}
  model},\ }\href {https://doi.org/10.1103/PhysRevB.103.224206} {\bibfield
  {journal} {\bibinfo  {journal} {Phys. Rev. B}\ }\textbf {\bibinfo {volume}
  {103}},\ \bibinfo {pages} {224206} (\bibinfo {year}
  {2021}{\natexlab{a}})}\BibitemShut {NoStop}%
\bibitem [{\citenamefont {Longhi}(2021{\natexlab{b}})}]{Longhi_2021a}%
  \BibitemOpen
  \bibfield  {author} {\bibinfo {author} {\bibfnamefont {S.}~\bibnamefont
  {Longhi}},\ }\bibfield  {title} {\bibinfo {title} {Phase transitions in a
  non-{{Hermitian Aubry-Andr}}{\'e}-{{Harper}} model},\ }\href
  {https://doi.org/10.1103/PhysRevB.103.054203} {\bibfield  {journal} {\bibinfo
   {journal} {Phys. Rev. B}\ }\textbf {\bibinfo {volume} {103}},\ \bibinfo
  {pages} {054203} (\bibinfo {year} {2021}{\natexlab{b}})}\BibitemShut
  {NoStop}%
\bibitem [{\citenamefont {Zhou}(2021)}]{Zhou_2021a}%
  \BibitemOpen
  \bibfield  {author} {\bibinfo {author} {\bibfnamefont {L.}~\bibnamefont
  {Zhou}},\ }\bibfield  {title} {\bibinfo {title} {Floquet engineering of
  topological localization transitions and mobility edges in one-dimensional
  non-{{Hermitian}} quasicrystals},\ }\href
  {https://doi.org/10.1103/PhysRevResearch.3.033184} {\bibfield  {journal}
  {\bibinfo  {journal} {Phys. Rev. Res.}\ }\textbf {\bibinfo {volume} {3}},\
  \bibinfo {pages} {033184} (\bibinfo {year} {2021})}\BibitemShut {NoStop}%
\bibitem [{\citenamefont {Acharya}\ \emph {et~al.}(2022)\citenamefont
  {Acharya}, \citenamefont {Chakrabarty}, \citenamefont {Sahu},\ and\
  \citenamefont {Datta}}]{Acharya_2022}%
  \BibitemOpen
  \bibfield  {author} {\bibinfo {author} {\bibfnamefont {A.~P.}\ \bibnamefont
  {Acharya}}, \bibinfo {author} {\bibfnamefont {A.}~\bibnamefont
  {Chakrabarty}}, \bibinfo {author} {\bibfnamefont {D.~K.}\ \bibnamefont
  {Sahu}},\ and\ \bibinfo {author} {\bibfnamefont {S.}~\bibnamefont {Datta}},\
  }\bibfield  {title} {\bibinfo {title} {Localization, $\mathcal{PT}$ symmetry
  breaking, and topological transitions in non-{{Hermitian}} quasicrystals},\
  }\href {https://doi.org/10.1103/PhysRevB.105.014202} {\bibfield  {journal}
  {\bibinfo  {journal} {Phys. Rev. B}\ }\textbf {\bibinfo {volume} {105}},\
  \bibinfo {pages} {014202} (\bibinfo {year} {2022})}\BibitemShut {NoStop}%
\bibitem [{\citenamefont {Zhou}\ and\ \citenamefont {Gu}(2022)}]{Zhou_2022}%
  \BibitemOpen
  \bibfield  {author} {\bibinfo {author} {\bibfnamefont {L.}~\bibnamefont
  {Zhou}}\ and\ \bibinfo {author} {\bibfnamefont {Y.}~\bibnamefont {Gu}},\
  }\bibfield  {title} {\bibinfo {title} {Topological delocalization transitions
  and mobility edges in the nonreciprocal {{Maryland}} model},\ }\href
  {https://doi.org/10.1088/1361-648X/ac4530} {\bibfield  {journal} {\bibinfo
  {journal} {J. Phys.: Condens. Matter}\ }\textbf {\bibinfo {volume} {34}},\
  \bibinfo {pages} {115402} (\bibinfo {year} {2022})}\BibitemShut {NoStop}%
\bibitem [{\citenamefont {Xia}\ \emph {et~al.}(2022)\citenamefont {Xia},
  \citenamefont {Huang}, \citenamefont {Wang},\ and\ \citenamefont
  {Li}}]{Xia_2022}%
  \BibitemOpen
  \bibfield  {author} {\bibinfo {author} {\bibfnamefont {X.}~\bibnamefont
  {Xia}}, \bibinfo {author} {\bibfnamefont {K.}~\bibnamefont {Huang}}, \bibinfo
  {author} {\bibfnamefont {S.}~\bibnamefont {Wang}},\ and\ \bibinfo {author}
  {\bibfnamefont {X.}~\bibnamefont {Li}},\ }\bibfield  {title} {\bibinfo
  {title} {Exact mobility edges in the non-{{Hermitian}} $t_{1}-{t}_{2}$ model:
  {{Theory}} and possible experimental realizations},\ }\href
  {https://doi.org/10.1103/PhysRevB.105.014207} {\bibfield  {journal} {\bibinfo
   {journal} {Phys. Rev. B}\ }\textbf {\bibinfo {volume} {105}},\ \bibinfo
  {pages} {014207} (\bibinfo {year} {2022})}\BibitemShut {NoStop}%
\bibitem [{\citenamefont {Zeng}\ and\ \citenamefont {Xu}(2020)}]{Zeng_2020a}%
  \BibitemOpen
  \bibfield  {author} {\bibinfo {author} {\bibfnamefont {Q.-B.}\ \bibnamefont
  {Zeng}}\ and\ \bibinfo {author} {\bibfnamefont {Y.}~\bibnamefont {Xu}},\
  }\bibfield  {title} {\bibinfo {title} {Winding numbers and generalized
  mobility edges in non-{{Hermitian}} systems},\ }\href
  {https://doi.org/10.1103/PhysRevResearch.2.033052} {\bibfield  {journal}
  {\bibinfo  {journal} {Phys. Rev. Res.}\ }\textbf {\bibinfo {volume} {2}},\
  \bibinfo {pages} {033052} (\bibinfo {year} {2020})}\BibitemShut {NoStop}%
\bibitem [{\citenamefont {Cai}(2022)}]{Cai_2022}%
  \BibitemOpen
  \bibfield  {author} {\bibinfo {author} {\bibfnamefont {X.}~\bibnamefont
  {Cai}},\ }\bibfield  {title} {\bibinfo {title} {Localization transitions and
  winding numbers for non-{{Hermitian Aubry-Andr}}{\'e}-{{Harper}} models with
  off-diagonal modulations},\ }\href
  {https://doi.org/10.1103/PhysRevB.106.214207} {\bibfield  {journal} {\bibinfo
   {journal} {Phys. Rev. B}\ }\textbf {\bibinfo {volume} {106}},\ \bibinfo
  {pages} {214207} (\bibinfo {year} {2022})}\BibitemShut {NoStop}%
\bibitem [{\citenamefont {Zhai}\ \emph {et~al.}(2022)\citenamefont {Zhai},
  \citenamefont {Huang},\ and\ \citenamefont {Yin}}]{Zhai_2022}%
  \BibitemOpen
  \bibfield  {author} {\bibinfo {author} {\bibfnamefont {L.-J.}\ \bibnamefont
  {Zhai}}, \bibinfo {author} {\bibfnamefont {G.-Y.}\ \bibnamefont {Huang}},\
  and\ \bibinfo {author} {\bibfnamefont {S.}~\bibnamefont {Yin}},\ }\bibfield
  {title} {\bibinfo {title} {Nonequilibrium dynamics of the
  localization-delocalization transition in the non-{{Hermitian
  Aubry-Andr}}{\'e} model},\ }\href
  {https://doi.org/10.1103/PhysRevB.106.014204} {\bibfield  {journal} {\bibinfo
   {journal} {Phys. Rev. B}\ }\textbf {\bibinfo {volume} {106}},\ \bibinfo
  {pages} {014204} (\bibinfo {year} {2022})}\BibitemShut {NoStop}%
\bibitem [{\citenamefont {Chakrabarty}\ and\ \citenamefont
  {Datta}(2023)}]{Chakrabarty_2023}%
  \BibitemOpen
  \bibfield  {author} {\bibinfo {author} {\bibfnamefont {A.}~\bibnamefont
  {Chakrabarty}}\ and\ \bibinfo {author} {\bibfnamefont {S.}~\bibnamefont
  {Datta}},\ }\bibfield  {title} {\bibinfo {title} {Skin effect and dynamical
  delocalization in non-{{Hermitian}} quasicrystals with spin-orbit
  interaction},\ }\href {https://doi.org/10.1103/PhysRevB.107.064305}
  {\bibfield  {journal} {\bibinfo  {journal} {Phys. Rev. B}\ }\textbf {\bibinfo
  {volume} {107}},\ \bibinfo {pages} {064305} (\bibinfo {year}
  {2023})}\BibitemShut {NoStop}%
\bibitem [{\citenamefont {Zhou}(2023{\natexlab{a}})}]{Zhou_2023}%
  \BibitemOpen
  \bibfield  {author} {\bibinfo {author} {\bibfnamefont {L.}~\bibnamefont
  {Zhou}},\ }\bibfield  {title} {\bibinfo {title} {Non-abelian generalization
  of non-hermitian quasicrystals: Pt-symmetry breaking, localization,
  entanglement, and topological transitions},\ }\href
  {https://doi.org/10.1103/PhysRevB.108.014202} {\bibfield  {journal} {\bibinfo
   {journal} {Phys. Rev. B}\ }\textbf {\bibinfo {volume} {108}},\ \bibinfo
  {pages} {014202} (\bibinfo {year} {2023}{\natexlab{a}})}\BibitemShut
  {NoStop}%
\bibitem [{\citenamefont {Zhou}\ and\ \citenamefont {Han}(2022)}]{Zhou_2022a}%
  \BibitemOpen
  \bibfield  {author} {\bibinfo {author} {\bibfnamefont {L.}~\bibnamefont
  {Zhou}}\ and\ \bibinfo {author} {\bibfnamefont {W.}~\bibnamefont {Han}},\
  }\bibfield  {title} {\bibinfo {title} {Driving-induced multiple
  $\mathcal{PT}$-symmetry breaking transitions and reentrant localization
  transitions in non-{{Hermitian Floquet}} quasicrystals},\ }\href
  {https://doi.org/10.1103/PhysRevB.106.054307} {\bibfield  {journal} {\bibinfo
   {journal} {Phys. Rev. B}\ }\textbf {\bibinfo {volume} {106}},\ \bibinfo
  {pages} {054307} (\bibinfo {year} {2022})}\BibitemShut {NoStop}%
\bibitem [{\citenamefont {Aubry}\ and\ \citenamefont
  {Andr{\'e}}(1980)}]{Aubry_1980}%
  \BibitemOpen
  \bibfield  {author} {\bibinfo {author} {\bibfnamefont {S.}~\bibnamefont
  {Aubry}}\ and\ \bibinfo {author} {\bibfnamefont {G.}~\bibnamefont
  {Andr{\'e}}},\ }\bibfield  {title} {\bibinfo {title} {Analyticity breaking
  and {{Anderson}} localization in incommensurate lattices},\ }\href@noop {}
  {\bibfield  {journal} {\bibinfo  {journal} {Proceedings, VIII International
  Colloquium on Group-Theoretical Methods in Physics}\ }\textbf {\bibinfo
  {volume} {3}} (\bibinfo {year} {1980})}\BibitemShut {NoStop}%
\bibitem [{\citenamefont {Harper}(1955)}]{Harper_1955}%
  \BibitemOpen
  \bibfield  {author} {\bibinfo {author} {\bibfnamefont {P.~G.}\ \bibnamefont
  {Harper}},\ }\bibfield  {title} {\bibinfo {title} {Single {{Band Motion}} of
  {{Conduction Electrons}} in a {{Uniform Magnetic Field}}},\ }\href
  {https://doi.org/10.1088/0370-1298/68/10/304} {\bibfield  {journal} {\bibinfo
   {journal} {Proc. Phys. Soc. A}\ }\textbf {\bibinfo {volume} {68}},\ \bibinfo
  {pages} {874} (\bibinfo {year} {1955})}\BibitemShut {NoStop}%
\bibitem [{\citenamefont {Sokoloff}(1985)}]{Sokoloff_1985}%
  \BibitemOpen
  \bibfield  {author} {\bibinfo {author} {\bibfnamefont {J.~B.}\ \bibnamefont
  {Sokoloff}},\ }\bibfield  {title} {\bibinfo {title} {Unusual band structure,
  wave functions and electrical conductance in crystals with incommensurate
  periodic potentials},\ }\href {https://doi.org/10.1016/0370-1573(85)90088-2}
  {\bibfield  {journal} {\bibinfo  {journal} {Phys. Rep.}\ }\textbf {\bibinfo
  {volume} {126}},\ \bibinfo {pages} {189} (\bibinfo {year}
  {1985})}\BibitemShut {NoStop}%
\bibitem [{\citenamefont {Ma}\ \emph {et~al.}(1986)\citenamefont {Ma},
  \citenamefont {Halperin},\ and\ \citenamefont {Lee}}]{Ma_1986}%
  \BibitemOpen
  \bibfield  {author} {\bibinfo {author} {\bibfnamefont {M.}~\bibnamefont
  {Ma}}, \bibinfo {author} {\bibfnamefont {B.~I.}\ \bibnamefont {Halperin}},\
  and\ \bibinfo {author} {\bibfnamefont {P.~A.}\ \bibnamefont {Lee}},\
  }\bibfield  {title} {\bibinfo {title} {Strongly disordered superfluids:
  {{Quantum}} fluctuations and critical behavior},\ }\href
  {https://doi.org/10.1103/PhysRevB.34.3136} {\bibfield  {journal} {\bibinfo
  {journal} {Phys. Rev. B}\ }\textbf {\bibinfo {volume} {34}},\ \bibinfo
  {pages} {3136} (\bibinfo {year} {1986})}\BibitemShut {NoStop}%
\bibitem [{\citenamefont {Fisher}\ \emph {et~al.}(1989)\citenamefont {Fisher},
  \citenamefont {Weichman}, \citenamefont {Grinstein},\ and\ \citenamefont
  {Fisher}}]{Fisher_1989}%
  \BibitemOpen
  \bibfield  {author} {\bibinfo {author} {\bibfnamefont {M.~P.~A.}\
  \bibnamefont {Fisher}}, \bibinfo {author} {\bibfnamefont {P.~B.}\
  \bibnamefont {Weichman}}, \bibinfo {author} {\bibfnamefont {G.}~\bibnamefont
  {Grinstein}},\ and\ \bibinfo {author} {\bibfnamefont {D.~S.}\ \bibnamefont
  {Fisher}},\ }\bibfield  {title} {\bibinfo {title} {Boson localization and the
  superfluid-insulator transition},\ }\href
  {https://doi.org/10.1103/PhysRevB.40.546} {\bibfield  {journal} {\bibinfo
  {journal} {Phys. Rev. B}\ }\textbf {\bibinfo {volume} {40}},\ \bibinfo
  {pages} {546} (\bibinfo {year} {1989})}\BibitemShut {NoStop}%
\bibitem [{\citenamefont {Han}\ and\ \citenamefont {Zhou}(2022)}]{Han_2022}%
  \BibitemOpen
  \bibfield  {author} {\bibinfo {author} {\bibfnamefont {W.}~\bibnamefont
  {Han}}\ and\ \bibinfo {author} {\bibfnamefont {L.}~\bibnamefont {Zhou}},\
  }\bibfield  {title} {\bibinfo {title} {Dimerization-induced mobility edges
  and multiple reentrant localization transitions in non-{{Hermitian}}
  quasicrystals},\ }\href {https://doi.org/10.1103/PhysRevB.105.054204}
  {\bibfield  {journal} {\bibinfo  {journal} {Phys. Rev. B}\ }\textbf {\bibinfo
  {volume} {105}},\ \bibinfo {pages} {054204} (\bibinfo {year}
  {2022})}\BibitemShut {NoStop}%
\bibitem [{\citenamefont {Li}\ and\ \citenamefont
  {Das~Sarma}(2020)}]{Li_2020a}%
  \BibitemOpen
  \bibfield  {author} {\bibinfo {author} {\bibfnamefont {X.}~\bibnamefont
  {Li}}\ and\ \bibinfo {author} {\bibfnamefont {S.}~\bibnamefont {Das~Sarma}},\
  }\bibfield  {title} {\bibinfo {title} {Mobility edge and intermediate phase
  in one-dimensional incommensurate lattice potentials},\ }\href
  {https://doi.org/10.1103/PhysRevB.101.064203} {\bibfield  {journal} {\bibinfo
   {journal} {Physical Review B}\ }\textbf {\bibinfo {volume} {101}},\ \bibinfo
  {pages} {064203} (\bibinfo {year} {2020})}\BibitemShut {NoStop}%
\bibitem [{\citenamefont {Longhi}(2023)}]{Longhi_2023c}%
  \BibitemOpen
  \bibfield  {author} {\bibinfo {author} {\bibfnamefont {S.}~\bibnamefont
  {Longhi}},\ }\bibfield  {title} {\bibinfo {title} {Phase transitions and
  bunching of correlated particles in a non-{{Hermitian}} quasicrystal},\
  }\href {https://doi.org/10.1103/PhysRevB.108.075121} {\bibfield  {journal}
  {\bibinfo  {journal} {Phys. Rev. B}\ }\textbf {\bibinfo {volume} {108}},\
  \bibinfo {pages} {075121} (\bibinfo {year} {2023})}\BibitemShut {NoStop}%
\bibitem [{\citenamefont {Altshuler}\ \emph {et~al.}(1997)\citenamefont
  {Altshuler}, \citenamefont {Gefen}, \citenamefont {Kamenev},\ and\
  \citenamefont {Levitov}}]{Altshuler_1997}%
  \BibitemOpen
  \bibfield  {author} {\bibinfo {author} {\bibfnamefont {B.~L.}\ \bibnamefont
  {Altshuler}}, \bibinfo {author} {\bibfnamefont {Y.}~\bibnamefont {Gefen}},
  \bibinfo {author} {\bibfnamefont {A.}~\bibnamefont {Kamenev}},\ and\ \bibinfo
  {author} {\bibfnamefont {L.~S.}\ \bibnamefont {Levitov}},\ }\bibfield
  {title} {\bibinfo {title} {Quasiparticle {{Lifetime}} in a {{Finite System}}:
  {{A Nonperturbative Approach}}},\ }\href
  {https://doi.org/10.1103/PhysRevLett.78.2803} {\bibfield  {journal} {\bibinfo
   {journal} {Phys. Rev. Lett.}\ }\textbf {\bibinfo {volume} {78}},\ \bibinfo
  {pages} {2803} (\bibinfo {year} {1997})}\BibitemShut {NoStop}%
\bibitem [{\citenamefont {Alba}\ and\ \citenamefont
  {Calabrese}(2018)}]{Alba_2018}%
  \BibitemOpen
  \bibfield  {author} {\bibinfo {author} {\bibfnamefont {V.}~\bibnamefont
  {Alba}}\ and\ \bibinfo {author} {\bibfnamefont {P.}~\bibnamefont
  {Calabrese}},\ }\bibfield  {title} {\bibinfo {title} {Entanglement dynamics
  after quantum quenches in generic integrable systems},\ }\href
  {https://doi.org/10.21468/SciPostPhys.4.3.017} {\bibfield  {journal}
  {\bibinfo  {journal} {SciPost Phys.}\ }\textbf {\bibinfo {volume} {4}},\
  \bibinfo {pages} {017} (\bibinfo {year} {2018})}\BibitemShut {NoStop}%
\bibitem [{\citenamefont {Agarwal}\ \emph {et~al.}(2015)\citenamefont
  {Agarwal}, \citenamefont {Gopalakrishnan}, \citenamefont {Knap},
  \citenamefont {M{\"u}ller},\ and\ \citenamefont {Demler}}]{Agarwal_2015}%
  \BibitemOpen
  \bibfield  {author} {\bibinfo {author} {\bibfnamefont {K.}~\bibnamefont
  {Agarwal}}, \bibinfo {author} {\bibfnamefont {S.}~\bibnamefont
  {Gopalakrishnan}}, \bibinfo {author} {\bibfnamefont {M.}~\bibnamefont
  {Knap}}, \bibinfo {author} {\bibfnamefont {M.}~\bibnamefont {M{\"u}ller}},\
  and\ \bibinfo {author} {\bibfnamefont {E.}~\bibnamefont {Demler}},\
  }\bibfield  {title} {\bibinfo {title} {Anomalous {{Diffusion}} and
  {{Griffiths Effects Near}} the {{Many-Body Localization Transition}}},\
  }\href {https://doi.org/10.1103/PhysRevLett.114.160401} {\bibfield  {journal}
  {\bibinfo  {journal} {Phys. Rev. Lett.}\ }\textbf {\bibinfo {volume} {114}},\
  \bibinfo {pages} {160401} (\bibinfo {year} {2015})}\BibitemShut {NoStop}%
\bibitem [{\citenamefont {Orito}\ and\ \citenamefont
  {Imura}(2022)}]{Orito_2022}%
  \BibitemOpen
  \bibfield  {author} {\bibinfo {author} {\bibfnamefont {T.}~\bibnamefont
  {Orito}}\ and\ \bibinfo {author} {\bibfnamefont {K.-I.}\ \bibnamefont
  {Imura}},\ }\href@noop {} {\bibinfo {title} {Wave-packet and entanglement
  dynamics in a non-{{Hermitian}} many-body system}} (\bibinfo {year} {2022}),\
  \Eprint {https://arxiv.org/abs/2212.01001} {arxiv:2212.01001 [cond-mat,
  physics:quant-ph]} \BibitemShut {NoStop}%
\bibitem [{\citenamefont {Gornyi}\ \emph {et~al.}(2005)\citenamefont {Gornyi},
  \citenamefont {Mirlin},\ and\ \citenamefont {Polyakov}}]{Gornyi_2005}%
  \BibitemOpen
  \bibfield  {author} {\bibinfo {author} {\bibfnamefont {I.~V.}\ \bibnamefont
  {Gornyi}}, \bibinfo {author} {\bibfnamefont {A.~D.}\ \bibnamefont {Mirlin}},\
  and\ \bibinfo {author} {\bibfnamefont {D.~G.}\ \bibnamefont {Polyakov}},\
  }\bibfield  {title} {\bibinfo {title} {Interacting {{Electrons}} in
  {{Disordered Wires}}: {{Anderson Localization}} and {{Low-}}${{T}}$
  {{Transport}}},\ }\href {https://doi.org/10.1103/PhysRevLett.95.206603}
  {\bibfield  {journal} {\bibinfo  {journal} {Phys. Rev. Lett.}\ }\textbf
  {\bibinfo {volume} {95}},\ \bibinfo {pages} {206603} (\bibinfo {year}
  {2005})}\BibitemShut {NoStop}%
\bibitem [{\citenamefont {Basko}\ \emph {et~al.}(2006)\citenamefont {Basko},
  \citenamefont {Aleiner},\ and\ \citenamefont {Altshuler}}]{Basko_2006}%
  \BibitemOpen
  \bibfield  {author} {\bibinfo {author} {\bibfnamefont {D.~M.}\ \bibnamefont
  {Basko}}, \bibinfo {author} {\bibfnamefont {I.~L.}\ \bibnamefont {Aleiner}},\
  and\ \bibinfo {author} {\bibfnamefont {B.~L.}\ \bibnamefont {Altshuler}},\
  }\bibfield  {title} {\bibinfo {title} {Metal\textendash insulator transition
  in a weakly interacting many-electron system with localized single-particle
  states},\ }\href {https://doi.org/10.1016/j.aop.2005.11.014} {\bibfield
  {journal} {\bibinfo  {journal} {Ann. Phys.}\ }\textbf {\bibinfo {volume}
  {321}},\ \bibinfo {pages} {1126} (\bibinfo {year} {2006})}\BibitemShut
  {NoStop}%
\bibitem [{\citenamefont {Peruzzo}\ \emph {et~al.}(2010)\citenamefont
  {Peruzzo}, \citenamefont {Lobino}, \citenamefont {Matthews}, \citenamefont
  {Matsuda}, \citenamefont {Politi}, \citenamefont {Poulios}, \citenamefont
  {Zhou}, \citenamefont {Lahini}, \citenamefont {Ismail}, \citenamefont
  {W{\"o}rhoff}, \citenamefont {Bromberg}, \citenamefont {Silberberg},
  \citenamefont {Thompson},\ and\ \citenamefont {OBrien}}]{Peruzzo_2010}%
  \BibitemOpen
  \bibfield  {author} {\bibinfo {author} {\bibfnamefont {A.}~\bibnamefont
  {Peruzzo}}, \bibinfo {author} {\bibfnamefont {M.}~\bibnamefont {Lobino}},
  \bibinfo {author} {\bibfnamefont {J.~C.~F.}\ \bibnamefont {Matthews}},
  \bibinfo {author} {\bibfnamefont {N.}~\bibnamefont {Matsuda}}, \bibinfo
  {author} {\bibfnamefont {A.}~\bibnamefont {Politi}}, \bibinfo {author}
  {\bibfnamefont {K.}~\bibnamefont {Poulios}}, \bibinfo {author} {\bibfnamefont
  {X.-Q.}\ \bibnamefont {Zhou}}, \bibinfo {author} {\bibfnamefont
  {Y.}~\bibnamefont {Lahini}}, \bibinfo {author} {\bibfnamefont
  {N.}~\bibnamefont {Ismail}}, \bibinfo {author} {\bibfnamefont
  {K.}~\bibnamefont {W{\"o}rhoff}}, \bibinfo {author} {\bibfnamefont
  {Y.}~\bibnamefont {Bromberg}}, \bibinfo {author} {\bibfnamefont
  {Y.}~\bibnamefont {Silberberg}}, \bibinfo {author} {\bibfnamefont {M.~G.}\
  \bibnamefont {Thompson}},\ and\ \bibinfo {author} {\bibfnamefont {J.~L.}\
  \bibnamefont {OBrien}},\ }\bibfield  {title} {\bibinfo {title} {Quantum
  {{Walks}} of {{Correlated Photons}}},\ }\href
  {https://doi.org/10.1126/science.1193515} {\bibfield  {journal} {\bibinfo
  {journal} {Science}\ }\textbf {\bibinfo {volume} {329}},\ \bibinfo {pages}
  {1500} (\bibinfo {year} {2010})}\BibitemShut {NoStop}%
\bibitem [{\citenamefont {Esposito}\ \emph {et~al.}(2022)\citenamefont
  {Esposito}, \citenamefont {Barros}, \citenamefont {Dur{\'a}n~Hern{\'a}ndez},
  \citenamefont {Carvacho}, \citenamefont {Di~Colandrea}, \citenamefont
  {Barboza}, \citenamefont {Cardano}, \citenamefont {Spagnolo}, \citenamefont
  {Marrucci},\ and\ \citenamefont {Sciarrino}}]{Esposito_2022}%
  \BibitemOpen
  \bibfield  {author} {\bibinfo {author} {\bibfnamefont {C.}~\bibnamefont
  {Esposito}}, \bibinfo {author} {\bibfnamefont {M.~R.}\ \bibnamefont
  {Barros}}, \bibinfo {author} {\bibfnamefont {A.}~\bibnamefont
  {Dur{\'a}n~Hern{\'a}ndez}}, \bibinfo {author} {\bibfnamefont
  {G.}~\bibnamefont {Carvacho}}, \bibinfo {author} {\bibfnamefont
  {F.}~\bibnamefont {Di~Colandrea}}, \bibinfo {author} {\bibfnamefont
  {R.}~\bibnamefont {Barboza}}, \bibinfo {author} {\bibfnamefont
  {F.}~\bibnamefont {Cardano}}, \bibinfo {author} {\bibfnamefont
  {N.}~\bibnamefont {Spagnolo}}, \bibinfo {author} {\bibfnamefont
  {L.}~\bibnamefont {Marrucci}},\ and\ \bibinfo {author} {\bibfnamefont
  {F.}~\bibnamefont {Sciarrino}},\ }\bibfield  {title} {\bibinfo {title}
  {Quantum walks of two correlated photons in a {{2D}} synthetic lattice},\
  }\href {https://doi.org/10.1038/s41534-022-00544-0} {\bibfield  {journal}
  {\bibinfo  {journal} {npj Quantum Inf.}\ }\textbf {\bibinfo {volume} {8}},\
  \bibinfo {pages} {1} (\bibinfo {year} {2022})}\BibitemShut {NoStop}%
\bibitem [{\citenamefont {Jiao}\ \emph {et~al.}(2021)\citenamefont {Jiao},
  \citenamefont {Gao}, \citenamefont {Zhou}, \citenamefont {Wang},
  \citenamefont {Ren}, \citenamefont {Xu}, \citenamefont {Qiao}, \citenamefont
  {Wang},\ and\ \citenamefont {Jin}}]{Jiao_2021}%
  \BibitemOpen
  \bibfield  {author} {\bibinfo {author} {\bibfnamefont {Z.-Q.}\ \bibnamefont
  {Jiao}}, \bibinfo {author} {\bibfnamefont {J.}~\bibnamefont {Gao}}, \bibinfo
  {author} {\bibfnamefont {W.-H.}\ \bibnamefont {Zhou}}, \bibinfo {author}
  {\bibfnamefont {X.-W.}\ \bibnamefont {Wang}}, \bibinfo {author}
  {\bibfnamefont {R.-J.}\ \bibnamefont {Ren}}, \bibinfo {author} {\bibfnamefont
  {X.-Y.}\ \bibnamefont {Xu}}, \bibinfo {author} {\bibfnamefont {L.-F.}\
  \bibnamefont {Qiao}}, \bibinfo {author} {\bibfnamefont {Y.}~\bibnamefont
  {Wang}},\ and\ \bibinfo {author} {\bibfnamefont {X.-M.}\ \bibnamefont
  {Jin}},\ }\bibfield  {title} {\bibinfo {title} {Two-dimensional quantum walks
  of correlated photons},\ }\href {https://doi.org/10.1364/OPTICA.425879}
  {\bibfield  {journal} {\bibinfo  {journal} {Optica}\ }\textbf {\bibinfo
  {volume} {8}},\ \bibinfo {pages} {1129} (\bibinfo {year} {2021})}\BibitemShut
  {NoStop}%
\bibitem [{\citenamefont {Poulios}\ \emph {et~al.}(2014)\citenamefont
  {Poulios}, \citenamefont {Keil}, \citenamefont {Fry}, \citenamefont
  {Meinecke}, \citenamefont {Matthews}, \citenamefont {Politi}, \citenamefont
  {Lobino}, \citenamefont {Gr{\"a}fe}, \citenamefont {Heinrich}, \citenamefont
  {Nolte}, \citenamefont {Szameit},\ and\ \citenamefont
  {O'Brien}}]{Poulios_2014}%
  \BibitemOpen
  \bibfield  {author} {\bibinfo {author} {\bibfnamefont {K.}~\bibnamefont
  {Poulios}}, \bibinfo {author} {\bibfnamefont {R.}~\bibnamefont {Keil}},
  \bibinfo {author} {\bibfnamefont {D.}~\bibnamefont {Fry}}, \bibinfo {author}
  {\bibfnamefont {J.~D.~A.}\ \bibnamefont {Meinecke}}, \bibinfo {author}
  {\bibfnamefont {J.~C.~F.}\ \bibnamefont {Matthews}}, \bibinfo {author}
  {\bibfnamefont {A.}~\bibnamefont {Politi}}, \bibinfo {author} {\bibfnamefont
  {M.}~\bibnamefont {Lobino}}, \bibinfo {author} {\bibfnamefont
  {M.}~\bibnamefont {Gr{\"a}fe}}, \bibinfo {author} {\bibfnamefont
  {M.}~\bibnamefont {Heinrich}}, \bibinfo {author} {\bibfnamefont
  {S.}~\bibnamefont {Nolte}}, \bibinfo {author} {\bibfnamefont
  {A.}~\bibnamefont {Szameit}},\ and\ \bibinfo {author} {\bibfnamefont {J.~L.}\
  \bibnamefont {O'Brien}},\ }\bibfield  {title} {\bibinfo {title} {Quantum
  {{Walks}} of {{Correlated Photon Pairs}} in {{Two-Dimensional Waveguide
  Arrays}}},\ }\href {https://doi.org/10.1103/PhysRevLett.112.143604}
  {\bibfield  {journal} {\bibinfo  {journal} {Phys. Rev. Lett.}\ }\textbf
  {\bibinfo {volume} {112}},\ \bibinfo {pages} {143604} (\bibinfo {year}
  {2014})}\BibitemShut {NoStop}%
\bibitem [{\citenamefont {Greiner}\ \emph {et~al.}(2002)\citenamefont
  {Greiner}, \citenamefont {Mandel}, \citenamefont {Esslinger}, \citenamefont
  {H{\"a}nsch},\ and\ \citenamefont {Bloch}}]{Greiner_2002}%
  \BibitemOpen
  \bibfield  {author} {\bibinfo {author} {\bibfnamefont {M.}~\bibnamefont
  {Greiner}}, \bibinfo {author} {\bibfnamefont {O.}~\bibnamefont {Mandel}},
  \bibinfo {author} {\bibfnamefont {T.}~\bibnamefont {Esslinger}}, \bibinfo
  {author} {\bibfnamefont {T.~W.}\ \bibnamefont {H{\"a}nsch}},\ and\ \bibinfo
  {author} {\bibfnamefont {I.}~\bibnamefont {Bloch}},\ }\bibfield  {title}
  {\bibinfo {title} {Quantum phase transition from a superfluid to a {{Mott}}
  insulator in a gas of ultracold atoms},\ }\href
  {https://doi.org/10.1038/415039a} {\bibfield  {journal} {\bibinfo  {journal}
  {Nature}\ }\textbf {\bibinfo {volume} {415}},\ \bibinfo {pages} {39}
  (\bibinfo {year} {2002})}\BibitemShut {NoStop}%
\bibitem [{\citenamefont {Gommers}\ \emph {et~al.}(2006)\citenamefont
  {Gommers}, \citenamefont {Denisov},\ and\ \citenamefont
  {Renzoni}}]{Gommers_2006}%
  \BibitemOpen
  \bibfield  {author} {\bibinfo {author} {\bibfnamefont {R.}~\bibnamefont
  {Gommers}}, \bibinfo {author} {\bibfnamefont {S.}~\bibnamefont {Denisov}},\
  and\ \bibinfo {author} {\bibfnamefont {F.}~\bibnamefont {Renzoni}},\
  }\bibfield  {title} {\bibinfo {title} {Quasiperiodically {{Driven Ratchets}}
  for {{Cold Atoms}}},\ }\href {https://doi.org/10.1103/PhysRevLett.96.240604}
  {\bibfield  {journal} {\bibinfo  {journal} {Phys. Rev. Lett.}\ }\textbf
  {\bibinfo {volume} {96}},\ \bibinfo {pages} {240604} (\bibinfo {year}
  {2006})}\BibitemShut {NoStop}%
\bibitem [{\citenamefont {Roux}\ \emph {et~al.}(2008)\citenamefont {Roux},
  \citenamefont {Barthel}, \citenamefont {McCulloch}, \citenamefont {Kollath},
  \citenamefont {Schollw{\"o}ck},\ and\ \citenamefont {Giamarchi}}]{Roux_2008}%
  \BibitemOpen
  \bibfield  {author} {\bibinfo {author} {\bibfnamefont {G.}~\bibnamefont
  {Roux}}, \bibinfo {author} {\bibfnamefont {T.}~\bibnamefont {Barthel}},
  \bibinfo {author} {\bibfnamefont {I.~P.}\ \bibnamefont {McCulloch}}, \bibinfo
  {author} {\bibfnamefont {C.}~\bibnamefont {Kollath}}, \bibinfo {author}
  {\bibfnamefont {U.}~\bibnamefont {Schollw{\"o}ck}},\ and\ \bibinfo {author}
  {\bibfnamefont {T.}~\bibnamefont {Giamarchi}},\ }\bibfield  {title} {\bibinfo
  {title} {Quasiperiodic {{Bose-Hubbard}} model and localization in
  one-dimensional cold atomic gases},\ }\href
  {https://doi.org/10.1103/PhysRevA.78.023628} {\bibfield  {journal} {\bibinfo
  {journal} {Phys. Rev. A}\ }\textbf {\bibinfo {volume} {78}},\ \bibinfo
  {pages} {023628} (\bibinfo {year} {2008})}\BibitemShut {NoStop}%
\bibitem [{\citenamefont {Viebahn}\ \emph {et~al.}(2019)\citenamefont
  {Viebahn}, \citenamefont {Sbroscia}, \citenamefont {Carter}, \citenamefont
  {Yu},\ and\ \citenamefont {Schneider}}]{Viebahn_2019}%
  \BibitemOpen
  \bibfield  {author} {\bibinfo {author} {\bibfnamefont {K.}~\bibnamefont
  {Viebahn}}, \bibinfo {author} {\bibfnamefont {M.}~\bibnamefont {Sbroscia}},
  \bibinfo {author} {\bibfnamefont {E.}~\bibnamefont {Carter}}, \bibinfo
  {author} {\bibfnamefont {J.-C.}\ \bibnamefont {Yu}},\ and\ \bibinfo {author}
  {\bibfnamefont {U.}~\bibnamefont {Schneider}},\ }\bibfield  {title} {\bibinfo
  {title} {Matter-{{Wave Diffraction}} from a {{Quasicrystalline Optical
  Lattice}}},\ }\href {https://doi.org/10.1103/PhysRevLett.122.110404}
  {\bibfield  {journal} {\bibinfo  {journal} {Phys. Rev. Lett.}\ }\textbf
  {\bibinfo {volume} {122}},\ \bibinfo {pages} {110404} (\bibinfo {year}
  {2019})}\BibitemShut {NoStop}%
\bibitem [{\citenamefont {Ishiguro}\ \emph {et~al.}(2023)\citenamefont
  {Ishiguro}, \citenamefont {Sato},\ and\ \citenamefont
  {Nishinari}}]{Yuki2023}%
  \BibitemOpen
  \bibfield  {author} {\bibinfo {author} {\bibfnamefont {Y.}~\bibnamefont
  {Ishiguro}}, \bibinfo {author} {\bibfnamefont {J.}~\bibnamefont {Sato}},\
  and\ \bibinfo {author} {\bibfnamefont {K.}~\bibnamefont {Nishinari}},\
  }\bibfield  {title} {\bibinfo {title} {Asymmetry-induced delocalization
  transition in the integrable non-hermitian spin chain},\ }\href
  {https://doi.org/10.1103/PhysRevResearch.5.033102} {\bibfield  {journal}
  {\bibinfo  {journal} {Phys. Rev. Res.}\ }\textbf {\bibinfo {volume} {5}},\
  \bibinfo {pages} {033102} (\bibinfo {year} {2023})}\BibitemShut {NoStop}%
\bibitem [{\citenamefont {Li}\ \emph {et~al.}(2023{\natexlab{a}})\citenamefont
  {Li}, \citenamefont {Liu},\ and\ \citenamefont {Xu}}]{Li_2023b}%
  \BibitemOpen
  \bibfield  {author} {\bibinfo {author} {\bibfnamefont {K.}~\bibnamefont
  {Li}}, \bibinfo {author} {\bibfnamefont {Z.-C.}\ \bibnamefont {Liu}},\ and\
  \bibinfo {author} {\bibfnamefont {Y.}~\bibnamefont {Xu}},\ }\href
  {https://doi.org/10.48550/arXiv.2305.12342} {\bibinfo {title}
  {Disorder-{{Induced Entanglement Phase Transitions}} in {{Non-Hermitian
  Systems}} with {{Skin Effects}}}} (\bibinfo {year} {2023}{\natexlab{a}}),\
  \Eprint {https://arxiv.org/abs/2305.12342} {arxiv:2305.12342 [cond-mat,
  physics:quant-ph]} \BibitemShut {NoStop}%
\bibitem [{\citenamefont {Zhou}(2023{\natexlab{b}})}]{Zhou_2023a}%
  \BibitemOpen
  \bibfield  {author} {\bibinfo {author} {\bibfnamefont {L.}~\bibnamefont
  {Zhou}},\ }\href {https://doi.org/10.48550/arXiv.2309.00924} {\bibinfo
  {title} {Entanglement phase transitions in non-hermitian quasicrystals}}
  (\bibinfo {year} {2023}{\natexlab{b}}),\ \Eprint
  {https://arxiv.org/abs/2309.00924} {arxiv:2309.00924 [cond-mat,
  physics:quant-ph]} \BibitemShut {NoStop}%
\bibitem [{\citenamefont {Li}\ \emph {et~al.}(2023{\natexlab{b}})\citenamefont
  {Li}, \citenamefont {Yu},\ and\ \citenamefont {Li}}]{Li_2023c}%
  \BibitemOpen
  \bibfield  {author} {\bibinfo {author} {\bibfnamefont {S.-Z.}\ \bibnamefont
  {Li}}, \bibinfo {author} {\bibfnamefont {X.-J.}\ \bibnamefont {Yu}},\ and\
  \bibinfo {author} {\bibfnamefont {Z.}~\bibnamefont {Li}},\ }\href
  {https://doi.org/10.48550/arXiv.2309.03546} {\bibinfo {title} {Emergent
  entanglement phase transitions in non-{{Hermitian
  Aubry-Andr}}{\'e}-{{Harper}} chains}} (\bibinfo {year}
  {2023}{\natexlab{b}}),\ \Eprint {https://arxiv.org/abs/2309.03546}
  {arxiv:2309.03546 [cond-mat]} \BibitemShut {NoStop}%
\bibitem [{\citenamefont {Hamazaki}\ \emph {et~al.}(2019)\citenamefont
  {Hamazaki}, \citenamefont {Kawabata},\ and\ \citenamefont
  {Ueda}}]{Hamazaki_2019}%
  \BibitemOpen
  \bibfield  {author} {\bibinfo {author} {\bibfnamefont {R.}~\bibnamefont
  {Hamazaki}}, \bibinfo {author} {\bibfnamefont {K.}~\bibnamefont {Kawabata}},\
  and\ \bibinfo {author} {\bibfnamefont {M.}~\bibnamefont {Ueda}},\ }\bibfield
  {title} {\bibinfo {title} {Non-{{Hermitian Many-Body Localization}}},\ }\href
  {https://doi.org/10.1103/PhysRevLett.123.090603} {\bibfield  {journal}
  {\bibinfo  {journal} {Phys. Rev. Lett.}\ }\textbf {\bibinfo {volume} {123}},\
  \bibinfo {pages} {090603} (\bibinfo {year} {2019})}\BibitemShut {NoStop}%
\bibitem [{\citenamefont {Takahashi}(1977)}]{Takahashi_1977}%
  \BibitemOpen
  \bibfield  {author} {\bibinfo {author} {\bibfnamefont {M.}~\bibnamefont
  {Takahashi}},\ }\bibfield  {title} {\bibinfo {title} {Half-filled hubbard
  model at low temperature},\ }\href
  {https://doi.org/10.1088/0022-3719/10/8/031} {\bibfield  {journal} {\bibinfo
  {journal} {J. Phys. C: Solid State Phys.}\ }\textbf {\bibinfo {volume}
  {10}},\ \bibinfo {pages} {1289} (\bibinfo {year} {1977})}\BibitemShut
  {NoStop}%
\bibitem [{\citenamefont {Ke}\ \emph {et~al.}(2017)\citenamefont {Ke},
  \citenamefont {Qin}, \citenamefont {Kivshar},\ and\ \citenamefont
  {Lee}}]{Ke_2017}%
  \BibitemOpen
  \bibfield  {author} {\bibinfo {author} {\bibfnamefont {Y.}~\bibnamefont
  {Ke}}, \bibinfo {author} {\bibfnamefont {X.}~\bibnamefont {Qin}}, \bibinfo
  {author} {\bibfnamefont {Y.~S.}\ \bibnamefont {Kivshar}},\ and\ \bibinfo
  {author} {\bibfnamefont {C.}~\bibnamefont {Lee}},\ }\bibfield  {title}
  {\bibinfo {title} {Multiparticle {{Wannier}} states and {{Thouless}} pumping
  of interacting bosons},\ }\href {https://doi.org/10.1103/PhysRevA.95.063630}
  {\bibfield  {journal} {\bibinfo  {journal} {Phys. Rev. A}\ }\textbf {\bibinfo
  {volume} {95}},\ \bibinfo {pages} {063630} (\bibinfo {year}
  {2017})}\BibitemShut {NoStop}%
\end{thebibliography}%
	
	
\end{document}